\documentclass[useAMS,usenatbib,letterpaper]{mn2e}

\usepackage{amsmath}
\usepackage{relsize}
\usepackage{epstopdf}
\usepackage{graphicx}
\usepackage{graphics}
\usepackage{multirow}
\usepackage{booktabs}
\usepackage{supertabular}
\usepackage{lscape}
\usepackage{longtable}
\usepackage[figuresright]{rotating}
\usepackage{multicol}
\usepackage{pdflscape}
\usepackage{natbib}
\usepackage{rotating}
\usepackage{epsfig}
\usepackage{amsfonts}
\usepackage{threeparttable}
\usepackage[english]{babel}
\usepackage{textcomp}
\graphicspath{.}
\usepackage{natbib}
\usepackage{subfigure}
\usepackage{tabularx}
\usepackage[width=1.0\textwidth]{caption}
\usepackage{float}

\usepackage{url}  
\usepackage{amssymb}   

\newcolumntype{R}[1]{>{\raggedleft\arraybackslash}p{#1}}
\newcolumntype{C}[1]{>{\centering\arraybackslash}p{#1}}
\newcolumntype{L}[1]{>{\raggedright\arraybackslash}p{#1}}

\title[M-dwarf debris discs in young moving groups]{A $WISE$-based 
search for debris discs amongst M-dwarfs in nearby, young, moving groups}
\author[A. S. Binks and R. D. Jeffries]{A. S. Binks$^{1,2}$\thanks{E-mail: a.binks@crya.unam.mx} and R. D. Jeffries$^{2}$\\
$^{1}$Centro de Radioastronom\'ia y Astrof\'isica, Universidad Nacional Aut\'onoma de M\'exico, PO Box 3-72, 58090 Morelia, Michoac\'an, M\'exico\\
$^{2}$Astrophysics Group, School of Chemistry and Physics, Keele University, Keele, Staffordshire ST5 5BG}

\begin{document}

\date{Accepted. Received; in original form}

\pagerange{\pageref{firstpage}--\pageref{lastpage}} \pubyear{2016}

\maketitle

\label{firstpage}

\begin{abstract}
We present a search for debris discs amongst M-dwarf members of nearby,
young (5--150\,Myr) moving groups (MGs) using infrared (IR) photometry,
primarily from the Wide Infrared Survey Explorer ($WISE$). A catalogue
of 100 MG M-dwarfs that have suitable $WISE$ data is compiled and 19 of
these are found to have significant IR excess emission at 22\,$\mu$m. 
Our search is likely to be complete for
discs where the ratio of flux from the disc to flux from the star
$f_{\rm d}/f_{*} > 10^{-3}$. The spectral energy distributions are supplemented
with 2MASS photometry and data at longer wavelengths and fitted with
simple disc models to characterise the IR
excesses. There is a bimodal distribution -- twelve
targets have $W1-W4 > 3$, corresponding to $f_{\rm d}/f_{*} > 0.02$ and are likely
to be gas-rich, primordial discs. The remaining seven targets have
$W1-W4 < 1$ ($f_{\rm d}/f_{*} \lesssim 10^{-3}$) and include three objects with
previously known or suspected debris discs and four new debris disc
candidates that are all members of the Beta Pic MG.
All of the IR excesses are identified in stars that are likely members
of MGs with age $<30$\,Myr. The detected debris disc
frequency falls from $13 \pm 5$ per cent to $<7$ per cent (at 95 per
cent confidence) for objects younger or older than 30\,Myr
respectively. This provides evidence for the evolution of debris discs
on this timescale and does not support models where the maximum of debris
disc emission occurs much later in lower-mass stars.

\end{abstract}

\begin{keywords}
stars: discs -- stars: circumstellar matter -- stars: low-mass --
stars: pre-main-sequence
\end{keywords}

\section{Introduction}\label{S_Intro}
The formation of low-mass stars is thought to inevitably involve a
``primordial'' disc of gas and dust. Over the course of a few Myr, this
disc evolves in a poorly understood way: small dust particles form and
coagulate; some gas can accrete onto the star or is expelled in winds;
an inner hole may form. The progression from an optically thick
primordial disc, through transitional discs with inner holes to an
optically thin disc may take a few to 10\,Myr (\citealt{2001a_Haisch,
  2008a_Hernandez}). At the same time, grains may grow to form
planetesimals and then second generation, optically thin, debris discs
may form due to collisional cascades or collisions between rocky
protoplanets (\citealt{2008a_Wyatt}).

Much of the work on the evolution of debris discs around stars has
focused on coeval cohorts of known age in open clusters. The occurrence
of observed debris discs appears to depend on both age and
spectral-type; debris discs are more common around early-type stars and
around young stars \citep{2008a_Wyatt}. From 10 to 100\,Myr there is a
decline in debris disc frequency from $\sim 40$ to 10 per cent amongst
solar-type stars, although a large scatter remains present throughout
this range. A-type stars take $\sim 500$\,Myr for the same level of
decline \citep{2007a_Siegler}, in broad agreement with collisional cascade models (\citealt{2005a_Kenyon, 2008a_Thebault}). 

Whilst much observational work has focused on the debris disc
frequencies for solar-type stars, little is known about the evolution
of  M-dwarf debris discs, because most are too faint to be studied in
open clusters and young, nearby field M-dwarfs are rare
(\citealt{2014a_Theissen}). It has been suggested that the evolution of debris discs around lower mass M-type stars could be more rapid. The timescales for dust dissipation may be shorter, partly due to enhanced stellar wind drag forces (\citealt{2005a_Plavchan, 2008a_Trilling}) and partly due to stronger photoevaporation in the extreme ultraviolet (\citealt{2014a_Galvan-Madrid}).

The Wide-field Infrared Survey Explorer ({\it WISE},
\citealt{2010a_Wright}) has been successful in uncovering hundreds of
solar-type objects with an IR excess in the solar-neighbourhood
(\citealt{2014a_Patel}), however, very few young, nearby M-dwarfs have
been reported with IR excesses (\citealt{2012a_Avenhaus}). The
frequency with which debris discs are found (hereafter termed the disc
fraction) for young field M-dwarfs has been reported to be both smaller than FGK-types (\citealt{2009a_Lestrade}) and larger (\citealt{2008a_Forbrich}).

A potential alternative source of coeval lower mass stars are those
belonging to nearby ($< 100\,$pc), young ($< 150\,$Myr), moving groups
(MGs, \citealt{2004a_Zuckerman}). Some work has investigated the disc
fractions in MGs (\citealt{2012a_Simon, 2012a_Schneider, 2016a_Moor}), none has
been able to study large samples of M-dwarfs. Several recent surveys
have identified new M-dwarf members (or candidate members) of nearby
MGs (Shkolnik et al. 2009, 2012; Schlieder et al 2010, 2012; Malo et
al. 2013; Gagn\'e et al. 2014, 2015). To date, only three M-dwarf MG
members are known to have debris discs, identified via scattered light images and thermal emission; AU Mic in the Beta Pic MG (BPMG, \citealt{2004a_Kalas, 2006a_Augereau, 2014a_MacGregor}) and TWA 7 and 25 in the TW Hydrae MG (TWA, \citealt{2016a_Choquet}).

\nocite{2009a_Shkolnik}
\nocite{2012a_Shkolnik}
\nocite{2010a_Schlieder}
\nocite{2012a_Schlieder}
\nocite{2013a_Malo}
\nocite{2014a_Gagne}
\nocite{2015a_Gagne}

Motivated by the swathe of new M-dwarfs discovered in MGs, this work
makes use of IR photometry to search for debris disc candidates among a
large number of MG M-dwarfs. Near-IR photometry is available from
$2MASS$ (\citealt{2003a_Cutri}), the main source of mid-IR photometry
is $WISE$ (\citealt{2010a_Wright}) and fluxes measured at longer
wavelengths are available for some
objects. $\S$\ref{S_Target_Selection} describes how the initial target
catalog of M-dwarfs is compiled. The collated $WISE$ photometry is
presented in $\S$\ref{S_Identifying_IR_Excess} with a description of
the photometric criteria required for the detection of an IR excess. In
$\S$\ref{S_SED} SED models and simple blackbody fits are
made to stars qualifying from the photometric selection in
$\S$\ref{S_Identifying_IR_Excess}. A discussion of the sensitivity
limits for $WISE$ detection of debris discs is presented in $\S$\ref{S_Sensitivity}. In $\S$\ref{S_Nature} we make assessments of the nature of the IR excess for each target with significant IR excess and we discuss previously confirmed M-dwarf debris disc systems in MGs. In $\S$\ref{S_Discussion} we present the results from our IR search and discuss its limitations. A summary is provided in $\S$\ref{S_Summary}.

\section{Target selection}\label{S_Target_Selection}
\subsection{Moving group membership confirmation}\label{S_MG_Confirmation}
Recent large-scale kinematic surveys have revealed hundreds of new
candidate MG M-dwarfs, many of which require follow-up data to confirm
their membership status. Some searches (e.g.,
\citealt{2012a_Schlieder}) use proper motions to identify possible
comoving stars with a MG and then perform follow-up spectroscopy to
measure ages and kinematics, identifying likely coeval and comoving
objects as MG candidates. Another approach is to identify candidate
young stars based on general youth-indicators such as UV and/or X-ray activity (\citealt{2012a_Shkolnik}) or rapid rotation (\citealt{2015a_Binks}) and then acquire spectroscopy to confirm youth and test their kinematics for membership with known MGs. An alternative method based on Bayesian inference techniques compares the proper motions, positions and colours of a sample of H$\alpha$-emitting stars with the distributions observed in known MGs (e.g., \citealt{2013a_Malo}). Table~\ref{T_MGs} lists the 9 MGs considered in this work and the source paper(s) from which candidates for this paper are selected. Ages are adopted from \cite{2015a_Bell}, unless otherwise stated in Table~\ref{T_MGs}.

All 286 of the selected targets have proper-motions that are reported
to be consistent with MG membership in the source paper and we also
require that an object has an M-type spectral classification. If a
candidate has a measured radial velocity (RV), MG membership is tested
by comparing the RV of a candidate and the RV it would be expected to
have were it a member -- given by $V_{\rm T}\cos\lambda$,
where $V_{\rm T}$ is the total speed of the MG and $\lambda$ is the
angle between the target's position and the MG convergent point. Convergent points are extracted from table 7 in \cite{2015a_Binks}. If $|{\rm RV} - V_{\rm T}\cos\lambda| > 5\,{\rm km\,s}^{-1}$ then the
target is rejected as a MG candidate. The sample is inevitably inhomogeneous, but where available, all kinematic information is consistent with MG membership.

{\tiny
\begin{table}
\begin{center}
\setlength{\tabcolsep}{0.0cm}
\begin{tabular}{p{1.8cm}p{1.5cm}p{1.3cm}p{3.4cm}}
\toprule
\toprule
MG name		& Age (Myr)		& $N_{\rm initial}$	& Source reference		\\
\toprule
$\epsilon$ Cha	& 3--5$^{\rm a}$	& 16			& 1				\\
TWA		& $10 \pm 3$		& 33			& 2, 3, 4			\\
$\eta$ Cha	& $11 \pm 3$		& 12			& 5				\\
BPMG		& $24 \pm 3$		& 45			& 3, 6, 7, 8, 9, 10, 11, 12, 13	\\
Argus		& 30--50$^{\rm b}$	& 12			& 3, 14				\\
Columba		& $42^{+6}_{-4}$	& 17			& 3				\\
Carina		& $45^{+11}_{-7}$	& 4			& 3				\\
Tuc-Hor		& $45 \pm 4$		& 116			& 3, 15				\\
ABDMG		& $149^{+51}_{-19}$	& 31			& 3, 9, 10, 13, 16			\\
\bottomrule
\end{tabular}
\end{center}
\caption{MGs considered in this work. Ages are from \protect\cite{2015a_Bell} except for a: \protect\cite{2013a_Murphy} and b: \protect\cite{2013a_Malo}. $N_{\rm initial}$ gives the number of M-dwarfs in each MG which satisfied all criteria for inclusion provided in $\S$\ref{S_MG_Confirmation}. Sources references: 1) \protect\cite{2013a_Murphy}, 2) \protect\cite{2012a_Nakajima}, 3) \protect\cite{2013a_Malo}, 4) \protect\cite{2014a_Ducourant}, 5) \protect\cite{2004a_Luhman}, 6) \protect\cite{1999a_Barrado_y_Navascues}, 7) \protect\cite{2001a_Zuckerman}, 8) \protect\cite{2010a_Schlieder}, 9) \protect\cite{2012a_Schlieder}, 10) \protect\cite{2012a_Shkolnik}, 11) \protect\cite{2014a_Malo}, 12) \protect\cite{2014a_Binks}, 13) \protect\cite{2016a_Binks} 14) \protect\cite{2013a_De_Silva}, 15) \protect\cite{2014a_Kraus}, 16) \protect\cite{2004b_Zuckerman}.}
\label{T_MGs}
\end{table}}

\subsection{Selection criteria using $WISE$}\label{S_Selection_Criteria}
$WISE$ \citep{2010a_Wright} is a satellite-based IR telescope which
mapped $> 99$ per cent of the sky at 3.4, 4.6, 12 and 22\,$\mu$m ($W1,
W2, W3, W4$) with angular resolutions of 6.1", 6.4", 6.5" and 12.0",
respectively. It is capable of providing 5$\sigma$ point-source
sensitivities of 0.068, 0.098, 0.86 and 5.4\,mJy, which is two
orders-of-magnitude deeper than previous satellite-based mid-IR surveys such as $IRAS$.

To obtain a reliable sample of mid-IR photometry for disc identification,
targets were only retained if they had signal-to-noise ratios (SNRs)
greater than 5.0 in the $WISE$ $W4$ band (hereafter, SNR refers
exclusively to the $W4$ band). Whilst an IR excess can be indicative of
a debris disc, there are a number of false positive signals that we
sought to eliminate following guidelines in
\cite{2014a_Patel}. Firstly, since interstellar cirrus near the
Galactic plane can strongly contaminate $WISE$ images, any objects with
$l \leq 5^{\circ}$ are excluded. Searching the SIMBAD database, we
ensured that no objects in our sample matched any of the following
object descriptions that could be false-positive sources of IR-excess:
post-AGB stars, white dwarfs, carbon stars, novae, Cepheids,
cataclysmic variables, high-mass X-ray binaries, planetary nebulae or
Wolf-Rayet stars. We then removed objects that had projected companions
with $\Delta K_{s} \leq 5$ mag within 16'' in $2MASS$; this is a strict
criterion but one deemed necessary to exclude contamination by
extremely red, unresolved sources in $WISE$. In order to include only
objects with non-variable photometry, which could affect our
determination of $WISE$ colours, we removed 8 stars whose $AllWISE$
photometry deviates from the mean of the single exposure measurements
(from the $AllWISE$ L1b catalog\footnote{available at \url{http://irsa.ipac.caltech.edu/holdings/catalogs.html}}) by more than $2\sigma$, where $\sigma$
represents one standard deviation in $W1$ or $W2$ from the single frame
exposures. In one case (WISE\,J112117.15$-$344645.4), the difference in $W1$
and $W2$ magnitudes between $AllWISE$ and the mean of the L1b images is
$\sim 0.8$ mag ($\sim 6\sigma$). In addition the sample was limited to objects with $W1 > 4.5$ mag and $W2 > 2.8$ to avoid saturation. To avoid contamination by known $2MASS$ extended sources, we include only stars with $WISE$ {\tt ext\_flg} = 0 or 1. We include only $WISE$ sources indicative of unconfused photometry: i.e., only stars with {\tt cc\_flg[Wi]} = 0 (or flagged with a lower-case initial). Finally, in an attempt to exclude variable sources, we include only objects with consistent variability detections in $W1$ and $W2$, excluding stars whose {\tt var\_flag[W1]} $>$ 8 and {\tt var\_flag[W2]} $<$ 5 or {\tt var\_flag[W1]} $<$ 5 and {\tt var\_flag[W2]} $>$ 8 (see \citealt{2014a_Patel} for an extensive description of this criterion).

Of the 286 candidate M-dwarf members in the nine MGs, 133 have SNR $>
5.0$; 10 in $\epsilon$~Cha, 8 in $\eta$~Cha, 22 in TWA, 33 in BPMG, 2
in Carina, 5 in Columba, 26 in Tuc-Hor, 9 in Argus and 18 in ABDMG. Of
these objects, 100 satisfy all the additional criteria described in
this section, and these are listed in Table~2.

\subsection{Flux conversion}\label{S_Flux_Conversion}
$JHK$ photometry is from the $2MASS$ catalog \citep{2003a_Cutri} and
magnitudes were converted to fluxes (all in units of
erg/s/cm$^{2}$/\AA) using the $2MASS$ isophotal bandpasses and
zero-point magnitude fluxes from table 2 in \cite{2003a_Cohen}. For the
$WISE$ data, zero-magnitude fluxes are taken from \cite{2011a_Jarrett}
and colour corrections are available in \cite{2010a_Wright}. Following
the work in \cite{2014a_Theissen} and the advisory notes in
\cite{2010a_Wright}, $W4$ fluxes are reduced by 10 per cent. For
objects with a significant $W4$ excess (see $\S$\ref{S_Identifying_IR_Excess}),
additional photometry/flux values were searched for in the $IRAS$, $AKARI$,
$Spitzer$ $IRAC$ and $Spitzer$ $MIPS$, $Herschel$ $SPIRE$ and $Herschel$ $PACS$ and $SCUBA2$ catalogs and any additional fluxes are
listed in Table~\ref{T_Disc_Type}. The fluxes for $IRAS$ sources
were extracted directly from the $IRAS$ point source catalog
\citep{1988a_Helou}. For the $AKARI$ $IRC$ bands (9 and 18\,$\mu$m), zero
magnitude fluxes are from table 8 in \cite{2008a_Tanabe} and the $AKARI$ $FIS$ (90\,$\mu$m), the $Spitzer$ $IRAC$ and $Spitzer$ $MIPS$ zero magnitude fluxes were extracted from the calibration tables available at \url{http://irsa.ipac.caltech.edu/}. For $Herschel$ $PACS$ and $SPIRE$ and $SCUBA2$ data, fluxes are taken directly from the source papers listed in the footnotes of Table~\ref{T_Disc_Type}.. All fluxes redward of the $K$ band have a 10 per cent calibration error added in quadrature to the photometric uncertainties.

{\voffset = +2.0in

{
\begin{table*}
\setlength{\tabcolsep}{0.0cm}
\begin{center}
\footnotesize
\caption{RV and photometric data for the 100 objects that satisfy both the RV criteria described in $\S$\ref{S_Target_Selection} and additional criteria in $\S$\ref{S_Selection_Criteria} and have SNR $> 5.0$ in the $W4$ band. The M-dwarf spectral-type (SpT) sub-classes are from the source paper, radial velocities (RVs) are extracted from the literature source (see $\S$\ref{S_Target_Selection}) and ${\rm RV}_{\rm p} = {\rm RV} - V_{\rm T}\cos{\lambda}$, unless there is no published RV, in which case ${\rm RV}_{\rm p} = V_{\rm T}\cos{\lambda}$. Errors for $J, H, K, W1, W2$ and $W3$ magnitudes are usually $0.01-0.03$ and are always $< 0.10$. $E_{W3}$ and $E_{W4}$ are the significance of any photospheric excess in the $W3$ and $W4$ bands, respectively. Objects in bold have $W1-W4 > 1.0$.}
\begin{tabular}{p{35mm}p{7mm}p{19mm}p{13mm}p{35mm}p{10mm}p{10mm}p{10mm}p{25mm}p{9mm}p{9mm}}
\toprule
\toprule
Name                    & SpT    & RV                   & ${\rm RV}_{\rm p}$    & Name & $W1$        & $W2$       & $W3$        & $W4$                         & $E_{W3}$   & $E_{W4}$          \\ 
(2MASS- )               & M-     & (${\rm km\,s}^{-1}$) & (${\rm km\,s}^{-1}$)  & (WISE-)   & (mag)  & (mag)      & (mag)       & (mag, SNR)                    &              &             \\
\toprule
\multicolumn{11}{c}{$\epsilon$\,Cha ($3-5$\,Myr)} \\ 
{\bf J11183572$-$7935548} &  {\bf 4.5}  &  {\bf $+19.3 \pm 1.6$}  &  {\bf $+4.2$}  & {\bf J111835.64$-$793554.8}  &  {\bf 9.44}  &  {\bf 9.14}  &  {\bf 7.66}  &  {$\bf 4.47 \pm 0.02, 45.1$}  &  {\bf 27.2} &  {\bf 46.5}  \\
{\bf J11432669$-$7804454} &  {\bf 4.7}  &  {\bf $+15.6 \pm 1.0$}  &  {\bf $+0.9$}  & {\bf J114326.57$-$780445.5}  &  {\bf 10.22} &  {\bf 9.83}  &  {\bf 8.68}  &  {$\bf 7.05 \pm 0.07, 15.3$}  &  {\bf 24.8} &  {\bf 14.4}  \\
J11474812$-$7841524 &  3.0  &  $+16.1 \pm 0.9$  &  $+1.5$  & J114748.00$-$784152.6  &  8.47  &  8.33  &  8.20  &  $8.09 \pm 0.18, 6.2$    &  2.3  &  0  \\
{\bf J11493184$-$7851011} &  {\bf 0.0}  &  {\bf $+13.4 \pm 1.3$}  &  {\bf $-1.2$}  & {\bf J114931.74$-$785101.0}  &  {\bf 8.17}  &  {\bf 7.61}  &  {\bf 4.54}  &  {$\bf 1.83 \pm 0.01, 88.8$}  &  {\bf 34.7} &  {\bf 98.3}  \\
{\bf J11550485$-$7919108} &  {\bf 3.0}  &  {\bf $+14.0 \pm 1.3$}  &  {\bf $-0.5$}  & {\bf J115504.71$-$791911.0}  &  {\bf 9.89}  &  {\bf 9.66}  &  {\bf 9.31}  &  {$\bf 7.16 \pm 0.10, 10.9$}  &  {\bf 10.7} &  {\bf 18.1}  \\
{\bf J12005517$-$7820296} &  {\bf 5.8}  &  {\bf $+10.7 \pm 1.3$}  &  {\bf $-3.6$}  & {\bf J120055.08$-$782029.5}  &  {\bf 10.64} &  {\bf 10.17} &  {\bf 8.51}  &  {$\bf 6.58 \pm 0.05, 20.8$}  &  {\bf 29.1} &  {\bf 22.2}  \\
J12020369$-$7853012 &  0.0  &  $+11.0 \pm 6.0$  &  $-3.3$  & J120203.59$-$785301.3  &  8.11  &  8.04  &  7.91  &  $7.79 \pm 0.15, 7.5$    &  1.9  &  0  \\
J12202177$-$7407393 &  0.0  &  $+12.3 \pm 0.4$  &  $13.7$  & J122021.70$-$740739.5  &  8.28  &  8.15  &  8.03  &  $7.97 \pm 0.18, 6.0$    &  3.5  &  0  \\
\toprule
\multicolumn{11}{c}{TWA ($10 \pm 3$\,Myr)} \\
J02224418$-$6022476 &  4.0  &  $+13.1 \pm 0.9$  &  $+1.3$  & J022244.32$-$602247.7  &  7.95  &  7.80  &  7.68  &  $7.50 \pm 0.12, 9.2$    &  1.0  &  0.2  \\
J10423011$-$3340162 &  2.0  &  $+11.4 \pm 0.0$  &  $-2.9$  & J104230.01$-$334016.4  &  6.79  &  6.66  &  6.60  &  $6.00 \pm 0.05, 21.4$   &  0.7  &  16.1  \\
J11132622$-$4523427 &  0.0  &  $+15.8 \pm 2.0$  &  $-2.7$  & J111326.18$-$452342.8  &  8.37  &  8.27  &  8.18  &  $8.13 \pm 0.19, 5.7$    &  1.6  &  0.3  \\
J11315526$-$3436272 &  2.0  &  $+12.7 \pm 3.8$  &  $-1.3$  & J113155.20$-$343627.3  &  6.66  &  6.44  &  6.41  &  $6.24 \pm 0.06, 19.1$   &  2.6  &  0.2  \\
{\bf J11321831$-$3019518} &  {\bf 5.0}  &  {\bf $+15.8 \pm 2.0$}  &  {\bf $+2.7$}  & {\bf J113218.24$-$301952.0}  &  {\bf 8.82}  &  {\bf 8.44}  &  {\bf 7.07}  &  {$\bf 5.14 \pm 0.03, 39.3$}  &  {\bf 26.7} &  {\bf 36.8}  \\
J12072738$-$3247002 &  3.0  &   $+8.5 \pm 1.2$  &  $-0.3$  & J120727.32$-$324700.4  &  7.61  &  7.49  &  7.40  &  $7.24 \pm 0.09, 12.1$   &  0.3  &  0.7  \\
{\bf J12073346$-$3932539} &  {\bf 8.0}  & {\bf $+11.2 \pm 2.0$}  &  {\bf $+1.6$}  & {\bf J120733.42$-$393254.2}  &  {\bf 11.57} &  {\bf 11.02} &  {\bf 9.47}  &  {$\bf 8.12 \pm 0.17, 6.2$}   &  {\bf 28.1} &  {\bf 7.6}  \\
J12313807$-$4558593 &  3.0  &   $+8.1 \pm 4.0$  &  $-0.5$  & J123138.03$-$455859.6  &  8.34  &  8.20  &  8.07  &  $7.79 \pm 0.13, 8.7$    &  2.3  &  1.9  \\
J12345629$-$4538075 &  1.5  &   $+9.0 \pm 0.4$  &  $+0.4$  & J123456.26$-$453807.7  &  7.94  &  7.86  &  7.75  &  $7.54 \pm 0.12, 9.1$    &  1.0  &  0.8  \\
J22440873$-$5413183 &  4.0  &   $+1.6 \pm 1.6$  &  $+1.0$  & J224408.79$-$541319.0  &  8.30  &  8.14  &  8.01  &  $7.77 \pm 0.18, 6.2$    &  1.6  &  0  \\
J23261069$-$7323498 &  0.0  &   $+8.0 \pm 1.9$  &  $+0.7$  & J232610.84$-$732350.5  &  7.86  &  7.81  &  7.71  &  $7.61 \pm 0.13, 8.5$    &  0.3  &  0  \\
\toprule
\multicolumn{11}{c}{$\eta$\,Cha ($11 \pm 3$\,Myr)} \\
{\bf J08413030$-$7853064} &  {\bf 4.8}  &                 {\bf 0} &    {\bf 17.3}  & {\bf J084130.24$-$785306.3}  &  {\bf 10.72} &  {\bf 10.35} &  {\bf 8.99}  &  {$\bf 7.37 \pm 0.10, 11.4$}  &  {\bf 26.6} &  {\bf 14.7}  \\
{\bf J08422710$-$7857479} &  {\bf 4.0}  &                 {\bf 0} &    {\bf 17.3}  & {\bf J084227.02$-$785747.7}  &  {\bf 9.72}  &  {\bf 9.47}  &  {\bf 7.86}  &  {$\bf 5.19 \pm 0.03, 37.7$}  &  {\bf 28.0} &  {\bf 39.3}  \\
J08422372$-$7904030 &  1.8  &                 0 &    17.3  & J084223.64$-$790402.7  &  8.52  &  8.44  &  8.32  &  $7.77 \pm 0.12, 8.9$    &  5.2  &  12.3  \\
{\bf J08431857$-$7905181} &  {\bf 3.3}  &                 {\bf 0} &    {\bf 17.3}  & {\bf J084318.52$-$790518.0}  &  {\bf 8.60}  &  {\bf 7.88}  &  {\bf 5.52}  &  {$\bf 3.41 \pm 0.02, 66.8$}  &  {\bf 33.6} &  {\bf 67.0}  \\
{\bf J08440914$-$7833457} &  {\bf 5.8}  &                 {\bf 0} &    {\bf 17.4}  & {\bf J084409.09$-$783345.6}  &  {\bf 11.21} &  {\bf 10.70} &  {\bf 9.04}  &  {$\bf 7.21 \pm 0.07, 15.0$}  &  {\bf 29.4} &  {\bf 15.1}  \\
{\bf J08441637$-$7859080} &  {\bf 4.5}  &                 {\bf 0} &    {\bf 17.3}  & {\bf J084416.33$-$785907.8}  &  {\bf 9.11}  &  {\bf 8.75}  &  {\bf 7.23}  &  {$\bf 5.47 \pm 0.03, 36.6$}  &  {\bf 28.0} &  {\bf 38.2}  \\
J08443188$-$7846311 &  1.0  &  $+15.0 \pm 1.1$  &  $-2.3$  & J084431.82$-$784630.9  &  8.60  &  8.59  &  8.45  &  $8.25 \pm 0.18, 6.2$    &  0    &  0  \\
J08475676$-$7854532 &  3.3  &                 0 &    17.3  & J084756.68$-$785452.9  &  8.31  &  8.16  &  8.01  &  $8.03 \pm 0.15, 7.4$    &  2.8  &  0  \\
\toprule
\multicolumn{11}{c}{BPMG ($24 \pm 3$\,Myr)} \\
J00172353$-$6645124 &  2.5  &  $+10.7 \pm 0.2$  &  $-0.2$  & J001723.69$-$664512.4  &  7.59  &  7.50  &  7.40  &  $7.34 \pm 0.13, 8.2$    &  0.2  &  0  \\
J01112542+1526214   &  5.0  &   $+3.1 \pm 1.6$  &  $-0.3$  & J011125.54+152620.7    &  8.02  &  7.79  &  7.63  &  $7.47 \pm 0.12, 8.9$    &  2.9  &  0  \\
J01132817$-$3821024 &  3.0  &  $+14.3 \pm 0.5$  &  $+2.4$  & J011328.27$-$382102.9  &  7.46  &  7.46  &  7.36  &  $7.19 \pm 0.11, 10.2$   &  0    &  0  \\
J01351393$-$0712517 &  4.0  &   $+6.5 \pm 1.8$  &  $-2.7$  & J013513.98$-$071251.9  &  7.97  &  7.80  &  7.68  &  $7.46 \pm 0.14, 8.0$    &  1.6  &  1.1  \\
J01535076$-$1459503 &  4.0  &  $+10.5 \pm 0.4$  &  $-1.5$  & J015350.81$-$145950.6  &  6.79  &  6.72  &  6.67  &  $6.59 \pm 0.05, 20.1$   &  0    &  0  \\
J04593483+0147007   &  0.0  &  $+19.8 \pm 0.0$  &  $+1.5$  & J045934.85+014659.7    &  6.21  &  6.06  &  6.06  &  $6.01 \pm 0.05, 22.0$   &  0.2  &  0  \\
J05004714$-$5715255 &  0.5  &  $+19.4 \pm 0.3$  &  $+0.4$  & J050047.16$-$571524.7  &  6.16  &  6.04  &  6.06  &  $5.92 \pm 0.03, 31.7$   &  0    &  0  \\
J05335981$-$0221325 &  3.0  &  $+22.0 \pm 1.3$  &  $-3.2$  & J053359.82$-$022132.9  &  7.54  &  7.43  &  7.34  &  $7.41 \pm 0.19, 5.6$    &  0    &  0  \\
J06131330$-$2742054 &  4.0  &  $+22.5 \pm 0.2$  &  $+0.9$  & J061313.30$-$274205.6  &  7.01  &  6.82  &  6.77  &  $6.64 \pm 0.06, 17.5$   &  0    &  0.1  \\
J08173943$-$8243298 &  4.5  &  $+15.6 \pm 1.5$  &  $+2.8$  & J081738.97$-$824328.8  &  6.48  &  6.27  &  6.22  &  $6.05 \pm 0.04, 29.2$   &  0.7  &  3.3  \\
J10172689$-$5354265 &  6.0  &  $+13.6 \pm 0.3$  &  $+0.3$  & J101726.70$-$535426.5  &  7.44  &  7.27  &  7.15  &  $7.18 \pm 0.09, 11.6$   &  0    &  4.1  \\
J13545390$-$7121476 &  2.5  &   $+5.7 \pm 0.2$  &  $-1.6$  & J135453.61$-$712148.9  &  7.61  &  7.49  &  7.38  &  $7.39 \pm 0.12, 9.3$    &  1.9  &  0  \\
J16572029$-$5343316 &  3.0  &   $+1.4 \pm 0.2$  &  $+3.4$  & J165720.25$-$534332.4  &  7.68  &  7.57  &  7.48  &  $7.33 \pm 0.13, 8.2$    &  0    &  0  \\
J17173128$-$6657055 &  3.0  &   $+2.7 \pm 1.8$  &  $-0.1$  & J171731.26$-$665706.8  &  7.53  &  7.36  &  7.20  &  $6.94 \pm 0.09, 11.7$   &  4.1  &  3.1  \\
J17292067$-$5014529 &  3.0  &   $-0.4 \pm 0.0$  &  $-3.3$  & J172920.64$-$501453.4  &  7.81  &  7.68  &  7.56  &  $7.58 \pm 0.18, 6.2$    &  4.3  &  4.1  \\
J18420694$-$5554254 &  3.0  &   $+0.3 \pm 0.5$  &  $+1.5$  & J184206.97$-$555426.2  &  8.49  &  8.33  &  8.25  &  $7.81 \pm 0.16, 6.8$    &  1.3  &  0  \\
J18465255$-$6210366 &  1.0  &   $+2.4 \pm 0.1$  &  $+1.2$  & J184652.56$-$621037.3  &  7.75  &  7.71  &  7.62  &  $7.52 \pm 0.15, 7.4$    &  0    &  0  \\
J19560294$-$3207186 &  4.0  &   $-3.7 \pm 2.2$  &  $+4.1$  & J195602.95$-$320719.3  &  7.92  &  7.76  &  7.67  &  $7.47 \pm 0.17, 6.4$    &  0.3  &  0  \\
J19560438$-$3207376 &  0.0  &   $-7.2 \pm 0.5$  &  $+0.6$  & J195604.39$-$320738.3  &  7.71  &  7.74  &  7.66  &  $7.31 \pm 0.12, 9.1$    &  0    &  1.1  \\
J20100002$-$2801410 &  2.5  &   $-5.8 \pm 0.6$  &  $+2.7$  & J201000.06$-$280141.6  &  7.61  &  7.45  &  7.36  &  $7.18 \pm 0.14, 7.6$    &  2.6  &  0.4  \\
J20333759$-$2556521 &  3.0  &   $-7.6 \pm 0.4$  &  $+0.5$  & J203337.63$-$255652.8  &  8.68  &  8.44  &  8.32  &  $7.76 \pm 0.17, 6.5$    &  4.1  &  0  \\
J20434114$-$2433534 &  3.7  &   $-5.8 \pm 0.6$  &  $+2.1$  & J204341.18$-$243353.8  &  7.58  &  7.44  &  7.39  &  $7.07 \pm 0.11, 9.8$    &  0    &  0.8  \\
\bottomrule
\end{tabular}
\end{center}
\end{table*}

\begin{table*}
\setlength{\tabcolsep}{0.0cm}
\begin{center}
\footnotesize
		\contcaption{}
\begin{tabular}{p{35mm}p{7mm}p{19mm}p{13mm}p{35mm}p{10mm}p{10mm}p{10mm}p{25mm}p{9mm}p{9mm}}
\toprule
\toprule
Name                    & SpT    & RV                   & ${\rm RV}_{\rm p}$    & Name & $W1$        & $W2$       & $W3$        & $W4$                         & $E_{W3}$   & $E_{W4}$          \\ 
(2MASS- )               & M-     & (${\rm km\,s}^{-1}$) & (${\rm km\,s}^{-1}$)  & (WISE-)   & (mag)  & (mag)      & (mag)       & (mag, SNR)                    &              &             \\
\toprule
J20450949$-$3120266 &  1.0  &   $-4.1 \pm 0.0$  &  $+1.9$  & J204509.76$-$312030.9  &  4.50  &  4.01  &  4.31  &  $4.14 \pm 0.03, 42.3$   &  1.3  &  3.4  \\
J21100535$-$1919573 &  2.0  &   $-5.7 \pm 0.4$  &  $+2.2$  & J211005.41$-$191958.4  &  7.02  &  7.00  &  6.93  &  $6.78 \pm 0.10, 10.9$   &  0    &  0  \\
J22004158+2715135   &  0.0  &  $-13.3 \pm 2.4$  &  $-0.5$  & J220041.64+271513.4    &  7.60  &  7.60  &  7.54  &  $7.28 \pm 0.11, 9.8$    &  0    &  0.5  \\
J22450004$-$3315258 &  5.0  &   $+2.0 \pm 0.0$  &  $+0.5$  & J224500.20$-$331527.2  &  7.63  &  7.44  &  7.30  &  $7.18 \pm 0.10, 10.6$   &  1.0  &  0.4  \\
J23172807+1936469   &  3.5  &   $-3.7 \pm 0.0$  &  $+2.7$  & J231728.40+193645.7    &  6.96  &  6.85  &  6.77  &  $6.58 \pm 0.06, 17.6$   &  0    &  0  \\
J23323085$-$1215513 &  0.0  &   $+1.2 \pm 0.6$  &  $+0.4$  & J233230.95$-$121552.0  &  6.51  &  6.37  &  6.39  &  $6.21 \pm 0.06, 19.2$   &  0    &  0  \\
\toprule
\multicolumn{11}{c}{Argus ($30-50$\,Myr)} \\
J00503319+2449009   &  3.5  &   $+6.0 \pm 1.1$  &  $+2.0$  & J005033.39+244900.3    &  6.87  &  6.70  &  6.64  &  $6.47 \pm 0.07, 16.6$   &  0.3  &  0.5  \\
J03033668$-$2535329 &  0.0  &  $+20.1 \pm 0.8$  &  $+4.0$  & J030336.86$-$253531.6  &  7.00  &  7.00  &  6.93  &  $6.81 \pm 0.06, 17.8$   &  0    &  0  \\
J05090356$-$4209199 &  3.5  &  $+16.8 \pm 1.7$  &  $-1.7$  & J050903.58$-$420919.2  &  8.60  &  8.43  &  8.32  &  $8.19 \pm 0.16, 6.8$    &  1.9  &  0  \\
J06134539$-$2352077 &  3.5  &  $+22.9 \pm 0.2$  &  $-0.5$  & J061345.36$-$235206.3  &  7.36  &  7.16  &  7.04  &  $6.51 \pm 0.06, 18.8$   &  3.2  &  0.8  \\
J15553178+3512028   &  4.0  &  $-15.5 \pm 0.7$  &  $+2.1$  & J155531.60+351204.3    &  7.83  &  7.69  &  7.54  &  $7.37 \pm 0.09, 12.6$   &  1.6  &  0  \\
J18450097$-$1409053 &  5.0  &  $-23.0 \pm 0.3$  &  $+1.1$  & J184500.95$-$140905.9  &  7.05  &  6.94  &  6.87  &  $6.58 \pm 0.07, 15.2$   &  0    &  0.6  \\
J19312434$-$2134226 &  2.5  &  $-25.6 \pm 1.5$  &  $+3.8$  & J193124.38$-$213423.8  &  7.71  &  7.59  &  7.52  &  $7.37 \pm 0.15, 7.1$    &  0.2  &  0  \\
J20163382$-$0711456 &  0.0  &  $-23.0 \pm 0.2$  &  $+1.8$  & J201633.88$-$071145.5  &  7.59  &  7.59  &  7.53  &  $7.49 \pm 0.16, 6.8$    &  0    &  0  \\
\toprule
\multicolumn{11}{c}{Columba ($42^{+6}_{-4}$\,Myr)} \\
J03413724+5513068   &  0.5  &   $-3.2 \pm 0.6$  &  $-1.4$  & J034137.39+551305.7    &  7.43  &  7.44  &  7.35  &  $7.27 \pm 0.11, 10.0$   &  0    &  0  \\
J04515303$-$4647309 &  0.0  &  $+24.0 \pm 0.8$  &  $+2.0$  & J045153.05$-$464730.8  &  8.79  &  8.72  &  8.60  &  $8.25 \pm 0.19, 5.8$    &  1.6  &  0  \\
J05100488$-$2340148 &  2.0  &  $+24.4 \pm 0.2$  &  $+1.4$  & J051004.90$-$234015.1  &  8.38  &  8.23  &  8.14  &  $8.10 \pm 0.19, 5.7$    &  2.3  &  0  \\
J07065772$-$5353463 &  0.0  &  $+22.4 \pm 0.6$  &  $-0.5$  & J070657.72$-$535345.9  &  7.56  &  7.57  &  7.48  &  $7.33 \pm 0.08, 13.7$   &  0    &  0  \\
\toprule
\multicolumn{11}{c}{Carina ($45^{+11}_{-7}$\,Myr)} \\
J06112997$-$7213388 &  4.0  &   $+6.0 \pm 1.1$  &  $+2.0$  & J061130.01$-$721338.2  &  8.55  &  8.36  &  8.23  &  $8.17 \pm 0.11, 10.3$   &  2.6  &  0  \\
J09032434$-$6348330 &  0.5  &  $+20.7 \pm 0.4$  &  $-0.1$  & J090324.30$-$634832.9  &  8.57  &  8.53  &  8.42  &  $8.19 \pm 0.14, 8.1$    &  0.2  &  0  \\
\toprule
\multicolumn{11}{c}{Tuc$-$Hor ($45 \pm 4$\,Myr)} \\
J00152752$-$6414545 &  1.8  &   $+6.7 \pm 0.3$  &  $+0.5$  & J001527.62$-$641455.2  &  8.70  &  8.51  &  8.36  &  $8.17 \pm 0.25, 5.3$    &  5.4  &  0  \\
J00493566$-$6347416 &  1.7  &   $+8.1 \pm 0.3$  &  $+0.2$  & J004935.79$-$634742.0  &  8.33  &  8.23  &  8.14  &  $8.00 \pm 0.18, 6.0$    &  0.9  &  0  \\
J01521830$-$5950168 &  1.6  &  $+10.3 \pm 0.3$  &  $+0.1$  & J015218.43$-$595016.9  &  7.96  &  7.88  &  7.78  &  $7.65 \pm 0.09, 11.9$   &  0.6  &  0  \\
J02001277$-$0840516 &  2.1  &   $+4.5 \pm 0.4$  &  $-1.0$  & J020012.84$-$084052.4  &  7.77  &  7.68  &  7.59  &  $7.54 \pm 0.14, 7.7$    &  0.2  &  0  \\
J02125819$-$5851182 &  1.9  &   $+9.1 \pm 0.8$  &  $+2.1$  & J021258.28$-$585118.3  &  8.31  &  8.21  &  8.10  &  $8.03 \pm 0.15, 7.4$    &  1.4  &  0  \\
J02205139$-$5823411 &  3.2  &  $+12.1 \pm 0.6$  &  $+0.5$  & J022051.50$-$582341.3  &  8.67  &  8.53  &  8.38  &  $8.04 \pm 0.16, 6.9$    &  2.6  &  1.5  \\
J02474639$-$5804272 &  1.8  &  $+13.1 \pm 0.5$  &  $+0.3$  & J024746.49$-$580427.4  &  8.34  &  8.24  &  8.12  &  $8.09 \pm 0.19, 5.7$    &  1.8  &  0  \\
J02564708$-$6343027 &  3.6  &  $+16.7 \pm 4.7$  &  $-3.5$  & J025647.15$-$634302.5  &  8.80  &  8.63  &  8.49  &  $8.25 \pm 0.18, 6.2$    &  2.8  &  0.9  \\
J03050976$-$3725058 &  1.4  &  $+14.2 \pm 0.5$  &  $-1.4$  & J030509.79$-$372505.8  &  8.60  &  8.46  &  8.33  &  $8.30 \pm 0.18, 5.9$    &  3.5  &  0  \\
J04133314$-$5231586 &  2.4  &  $+18.4 \pm 0.2$  &  $-1.7$  & J041333.21$-$523158.5  &  9.01  &  8.88  &  8.74  &  $8.67 \pm 0.19, 5.7$    &  2.8  &  0  \\
J04240094$-$5512223 &  2.0  &  $+19.0 \pm 0.7$  &  $-2.1$  & J042400.99$-$551222.2  &  8.80  &  8.67  &  8.51  &  $8.21 \pm 0.13, 8.2$    &  3.8  &  1.3  \\
J04365738$-$1613065 &  3.3  &  $+16.6 \pm 1.9$  &  $+0.1$  & J043657.44$-$161306.7  &  8.14  &  7.98  &  7.88  &  $7.46 \pm 0.14, 7.7$    &  1.6  &  0.3  \\
J04440099$-$6624036 &  0.0  &  $+16.0 \pm 0.5$  &  $+0.3$  & J044401.08$-$662403.2  &  8.50  &  8.47  &  8.37  &  $8.13 \pm 0.17, 6.4$    &  0    &  0  \\
J05392505$-$4245211 &  1.7  &  $+21.7 \pm 0.2$  &  $-1.2$  & J053925.08$-$424521.0  &  8.47  &  8.38  &  8.26  &  $8.14 \pm 0.19, 5.8$    &  1.5  &  0.1  \\
J23124644$-$5049240 &  3.9  &   $+4.1 \pm 1.9$  &  $-1.9$  & J231246.53$-$504924.8  &  8.09  &  7.90  &  7.77  &  $7.74 \pm 0.15, 7.5$    &  2.7  &  0  \\
J23285763$-$6802338 &  2.3  &   $+8.0 \pm 1.5$  &  $-1.4$  & J232857.75$-$680234.5  &  8.27  &  8.16  &  8.03  &  $7.96 \pm 0.18, 6.0$    &  2.0  &  0  \\
J23474694$-$6517249 &  1.0  &   $+6.1 \pm 0.3$  &  $+0.4$  & J234747.06$-$651725.3  &  8.09  &  8.02  &  7.91  &  $7.96 \pm 0.18, 6.0$    &  1.0  &  0  \\
\toprule
\multicolumn{11}{c}{ABDMG ($149^{+51}_{-19}$\,Myr)} \\
J01034210+4051158   &  2.6  &  $-10.9 \pm 0.4$  &  $+0.6$  & J010342.25+405114.2    &  8.11  &  7.94  &  7.87  &  $7.62 \pm 0.11, 10.1$   &  1.7  &  0.3  \\
J01484087$-$4830519 &  6.0  &  $+21.5 \pm 0.2$  &  $-1.6$  & J014840.98$-$483052.3  &  8.26  &  8.19  &  8.08  &  $8.16 \pm 0.20, 5.4$    &  0    &  0  \\
J03472333$-$0158195 &  2.5  &  $+16.0 \pm 1.7$  &  $-1.8$  & J034723.45$-$015822.7  &  6.81  &  6.66  &  6.62  &  $6.45 \pm 0.07, 16.6$   &  0.2  &  0  \\
J04522441$-$1649219 &  3.0  &  $+26.7 \pm 1.5$  &  $+0.9$  & J045224.49$-$164924.0  &  6.78  &  6.58  &  6.53  &  $6.39 \pm 0.06, 19.5$   &  1.6  &  0  \\
J04571728$-$0621564 &  0.5  &  $+23.4 \pm 0.3$  &  $+0.9$  & J045717.30$-$062157.5  &  8.53  &  8.51  &  8.40  &  $8.08 \pm 0.21, 5.1$    &  0    &  1.0  \\
J05254166$-$0909123 &  3.5  &  $+26.3 \pm 0.3$  &  $+2.2$  & J052541.69$-$090914.4  &  7.45  &  7.31  &  7.21  &  $7.06 \pm 0.09, 12.8$   &  0.7  &  0  \\
J12383713$-$2703348 &  1.5  &   $+7.8 \pm 1.2$  &  $+0.0$  & J123837.00$-$270336.9  &  7.66  &  7.57  &  7.47  &  $7.30 \pm 0.11, 9.6$    &  1.0  &  0.1  \\
J12574030+3513306   &  4.0  &  $-14.1 \pm 1.6$  &  $-2.8$  & J125740.02+351328.7    &  6.21  &  6.25  &  6.31  &  $6.16 \pm 0.05, 22.7$   &  0    &  0  \\
J15244849$-$4929473 &  2.0  &  $+10.3 \pm 0.2$  &  $+2.9$  & J152448.37$-$492949.9  &  7.14  &  7.02  &  6.98  &  $6.99 \pm 0.12, 9.5$    &  0    &  0  \\
J16334161$-$0933116 &  0.5  &  $-15.0 \pm 0.4$  &  $+0.0$  & J163341.57$-$093313.6  &  7.46  &  7.45  &  7.37  &  $7.30 \pm 0.13, 8.2$    &  0    &  0  \\
J17383964+6114160   &  0.0  &  $-26.7 \pm 0.1$  &  $+3.7$  & J173839.62+611416.5    &  6.64  &  6.68  &  6.64  &  $6.52 \pm 0.05, 21.2$   &  0    &  0  \\
J21464282$-$8543046 &  3.5  &  $+23.5 \pm 0.7$  &  $+1.3$  & J214644.83$-$854306.3  &  7.82  &  7.65  &  7.51  &  $7.25 \pm 0.09, 11.5$   &  2.9  &  2.3  \\
J21521039+0537356   &  2.0  &  $-15.1 \pm 0.2$  &  $-0.8$  & J215210.48+053734.4    &  7.14  &  7.07  &  7.00  &  $6.84 \pm 0.08, 14.0$   &  0.7  &  0  \\
J23320018$-$3917368 &  3.0  &  $+11.1 \pm 0.2$  &  $-0.8$  & J233200.34$-$391738.9  &  7.88  &  7.75  &  7.63  &  $7.61 \pm 0.15, 7.4$    &  1.6  &  0  \\
\bottomrule
\label{T_WISE_Phot_All4}
\end{tabular}
\end{center}
\end{table*}
}

}

\section{Identifying IR-excess}\label{S_Identifying_IR_Excess}

In this section we discuss only how an IR excess was identified, delaying discussion of the possible cause of an IR excess until $\S$\ref{S_Nature}. We define two quantities $E_{W3}$ and $E_{W4}$ that define the significance of any IR excess over that expected from the photosphere in the respective $WISE$ wavebands. These are calculated by subtracting the photospheric flux expected for a star of that spectral-type from the observed flux and then dividing by the uncertainty in the observed flux. The photospheric flux was estimated by interpolating the $W_{1-4}$ photometry in table 5 from \cite{2013a_Pecaut} for candidates younger than 30\,Myr, or from their table 4 for candidates older than 30\,Myr. The flux error is derived as the quadrature sum of the $W_{x}$ photometric uncertainty and the scatter in $W1-Wx$ for a disc-less star at a given spectral-type (from a sample of field M-dwarfs from the third catalogue of nearby stars; \citealt{1991a_Gliese}). Following \cite{2012a_Kennedy}, objects with $E_{W4} > 3.0$ are classed as having significant excess and these are discussed in more detail in $\S$\ref{S_Nature}.

\cite{2012a_Schneider} apply a criteria of $W1-W4 > 1.0$\,mag to
identify circumstellar discs around M-dwarf TWA members. From a sample of M-dwarfs from the
third catalogue of nearby stars \citep{1991a_Gliese} a typical
colour for an M-dwarf field star without an excess is $W1-W4 \sim 0$
with a range of $\sim 0.3$ mag (see below); therefore $W1-W4 = 1.0$
represents a $\sim 3\sigma$ separation from the field star sample. The
range in $W1-W3$ for M-dwarf field stars is $\sim 0.15$ mag, centered
about 0. 

\begin{figure*}
 \begin{center}
\includegraphics[width=1.0\textwidth]{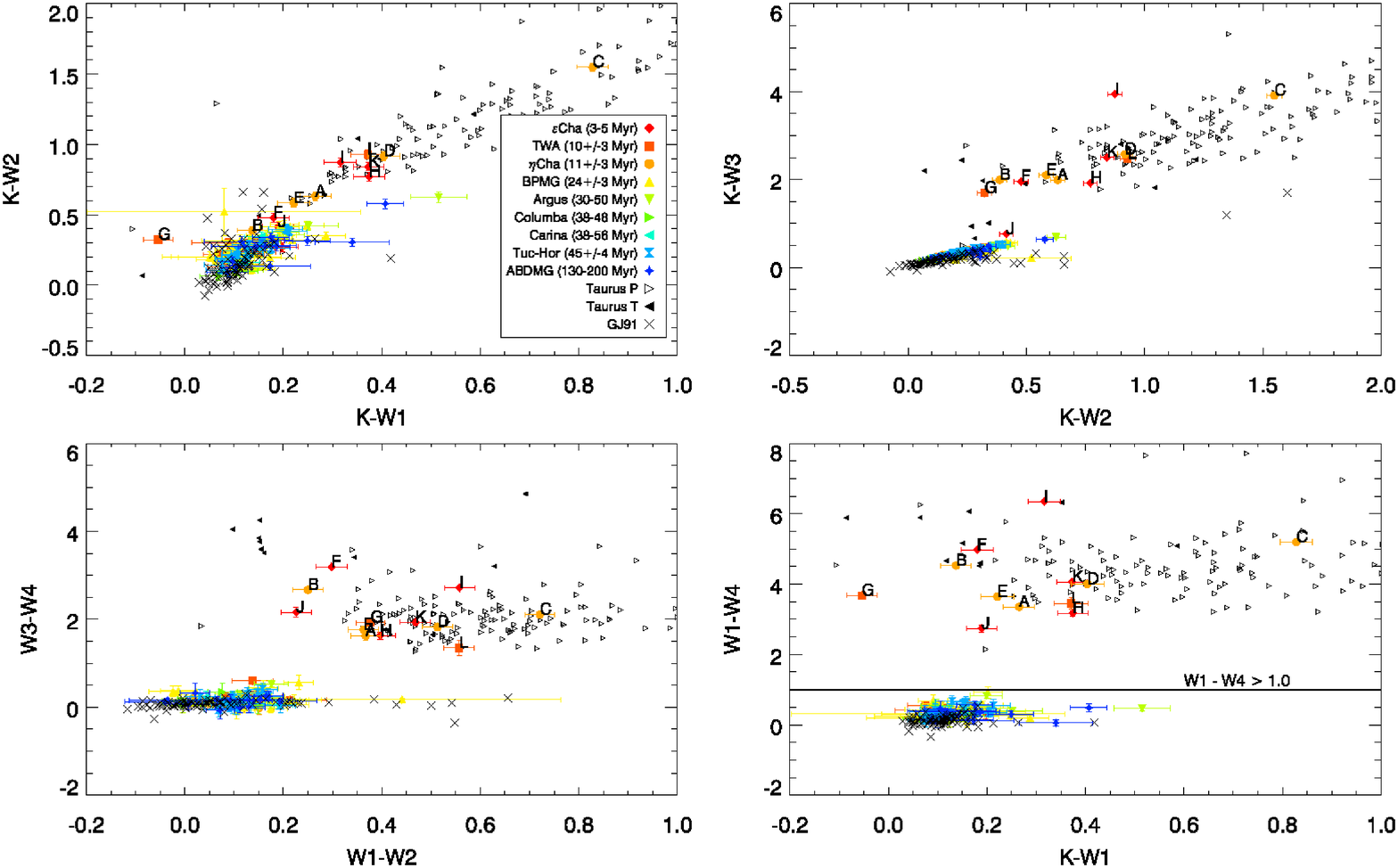}
 \end{center}
 \flushleft
 \caption{Colour-colour diagrams for the entire sample of M-dwarfs in this analysis with SNR values $> 5.0$. Taurus P and Taurus T refer to the primordial and transitional disc sample in \protect\cite{2014a_Esplin}. Black crosses represent M-type field stars from the \protect\cite{1991a_Gliese} catalog. Objects labelled with initials in the $W1-W4$ versus $K-W1$ plot correspond to the 12 targets with $W1-W4 > 1.0$, listed in Table~\ref{T_SED_Params}.}
  \label{F_Colours}
\end{figure*}

Figure~\ref{F_Colours} shows a variety of colour-colour diagrams for
the MG sample. A sample of primordial and transitional discs in the
Taurus star-forming region \citep{2014a_Esplin} is also shown and can
be used to judge the efficacy of the diagrams and whether they are
capable of distinguishing transitional from primordial
discs. Figure~\ref{F_Colours} shows that $W1-W4$ is likely to be the
best colour indicator to pick out objects with discs. This work will
utilise the same photometric colour cut as \cite{2012a_Schneider}, however should an object with
$W1-W4 < 1.0$ have an $E_{W4} > 3.0$, then they are also assumed to be
debris disc candidates, unless there is evidence to suggest otherwise. 

Twelve objects survive the $W1-W4 > 1.0$ cut: 5 in $\epsilon$~Cha, 5 in
$\eta$~Cha and 2 in TWA (see Table~\ref{T_Disc_Type}) and their nature
is discussed in $\S$\ref{S_Disc_Criteria}. A further seven objects
have $W1-W4 < 1.0$ and $E_{W4} > 3.0$ and these are discussed in
$\S$\ref{S_Other}. As a further check on the reliability of these
sources, postage stamp $AllWISE$ images in the $W4$ band were examined
with the {\sc daophot} photometry package in IRAF\footnote{IRAF is distributed by the National Optical Astronomy Observatories, which are operated by the Association of Universities for Research 
in Astronomy, Inc., under cooperative agreement with the National 
Science Foundation.}. All of the
detections showed no significant confusion with other sources within
the $WISE$ point spread function and all bar one showed no evidence of any non-circularity
that might betray the presence of multiple components, an extended
extragalactic source or cirrus. The exception was target R (see
Table~\ref{T_SED_Params}), which showed marginal evidence for
non-circularity, but was our IR-excess source with the lowest detection
significance in $W4$.
There were also 5 objects with $W1-W4 > 1.0$, SNR $>
5.0$, but which were excluded by other criteria described in
$\S$\ref{S_Selection_Criteria} -- these are briefly discussed in Appendix~A. 

\section{SED models}\label{S_SED}

For those stars suspected of having an IR excess, SED models were
generated to provide a best-fit star plus single-temperature (or two-temperature, if appropriate) disc to the
available photometry. BT-Settl model atmospheres were used to simulate
photospheres with $T_{\rm eff} > 2700$\,K \citep{2014a_Allard} and the
AMES-DUSTY models were used for $T_{\rm eff} \leq 2700$\,K
\citep{2000a_Chabrier}. All SEDs were fit using atmospheres with [M/H]
= 0.0 and $\log g = 4.5$. The flux profiles of the discs were generated from Planck functions. Model fluxes for each photometry point were calculated by integrating the relative spectral response (RSR) curve with the combined stellar and disc flux over the photometric bandwidth (i.e., between ${\lambda_{1}}$ and ${\lambda_{2}}$) and dividing this by the integral of the RSR curve.

\begin{figure*}
    \begin{minipage}[b]{\textwidth}
    \centering
      \begin{tabular}{ccc}
        {\bf A}\includegraphics[width=0.31\textwidth]{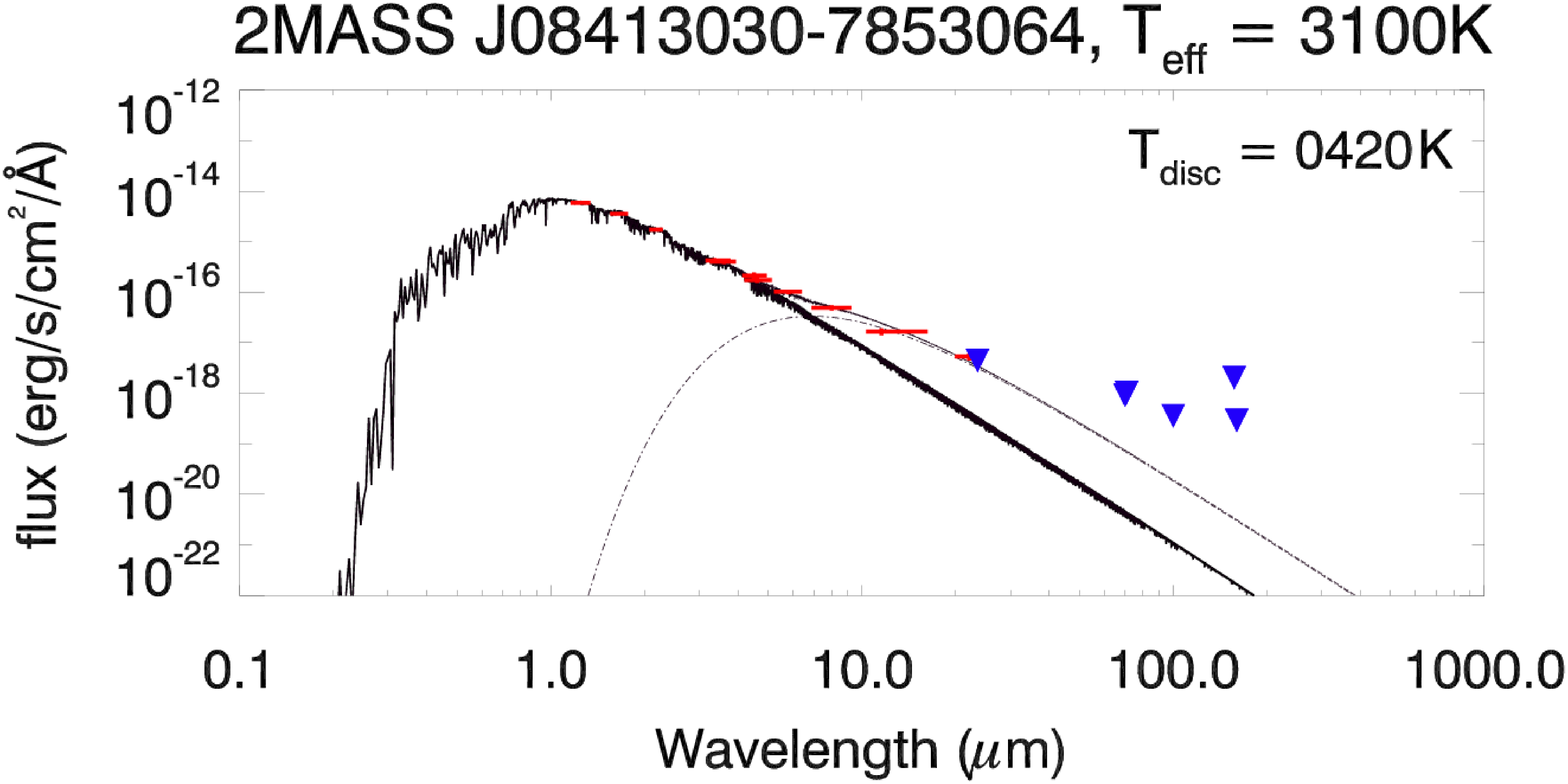} & {\bf B}\includegraphics[width=0.31\textwidth]{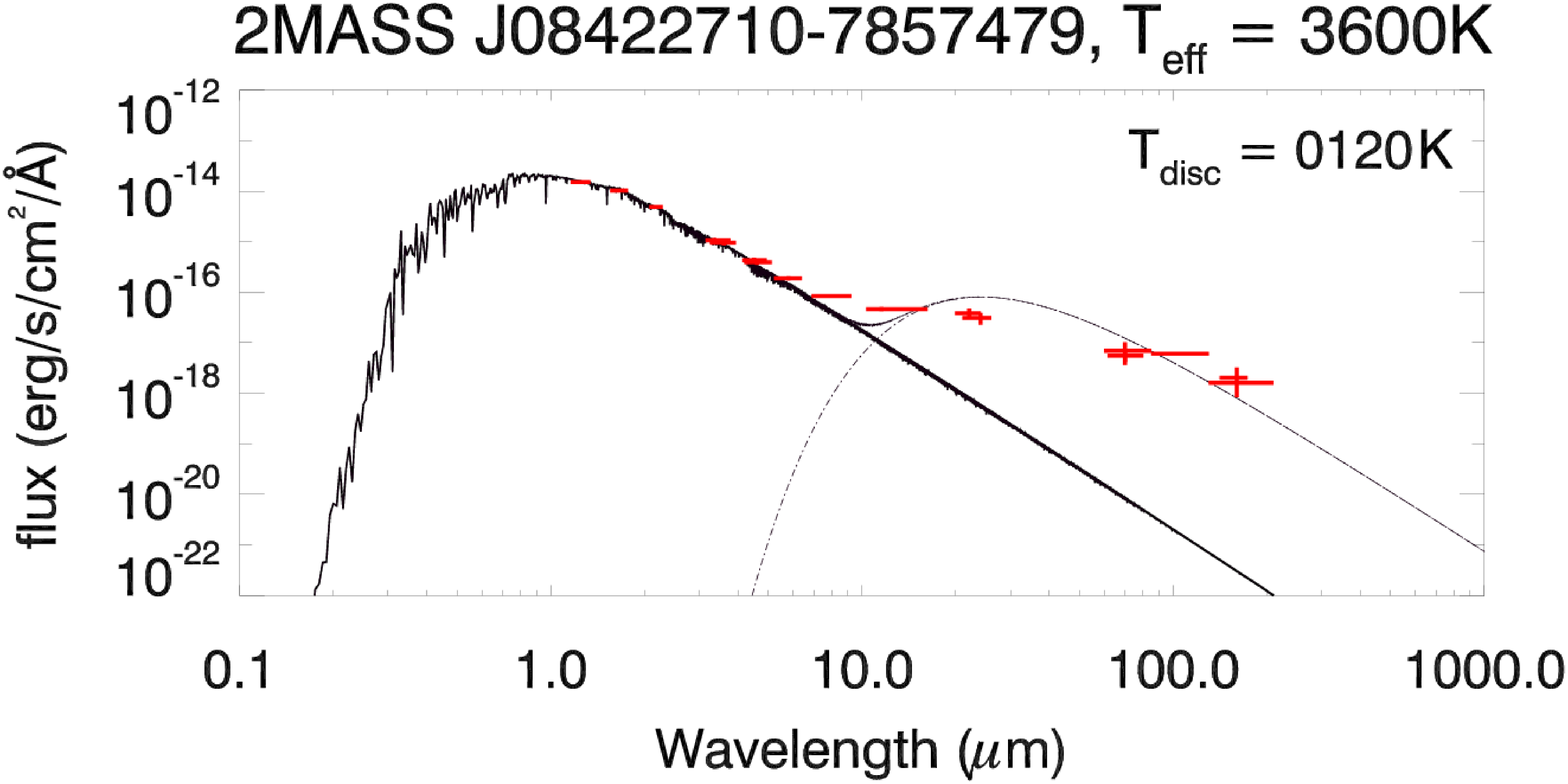} & {\bf C}\includegraphics[width=0.31\textwidth]{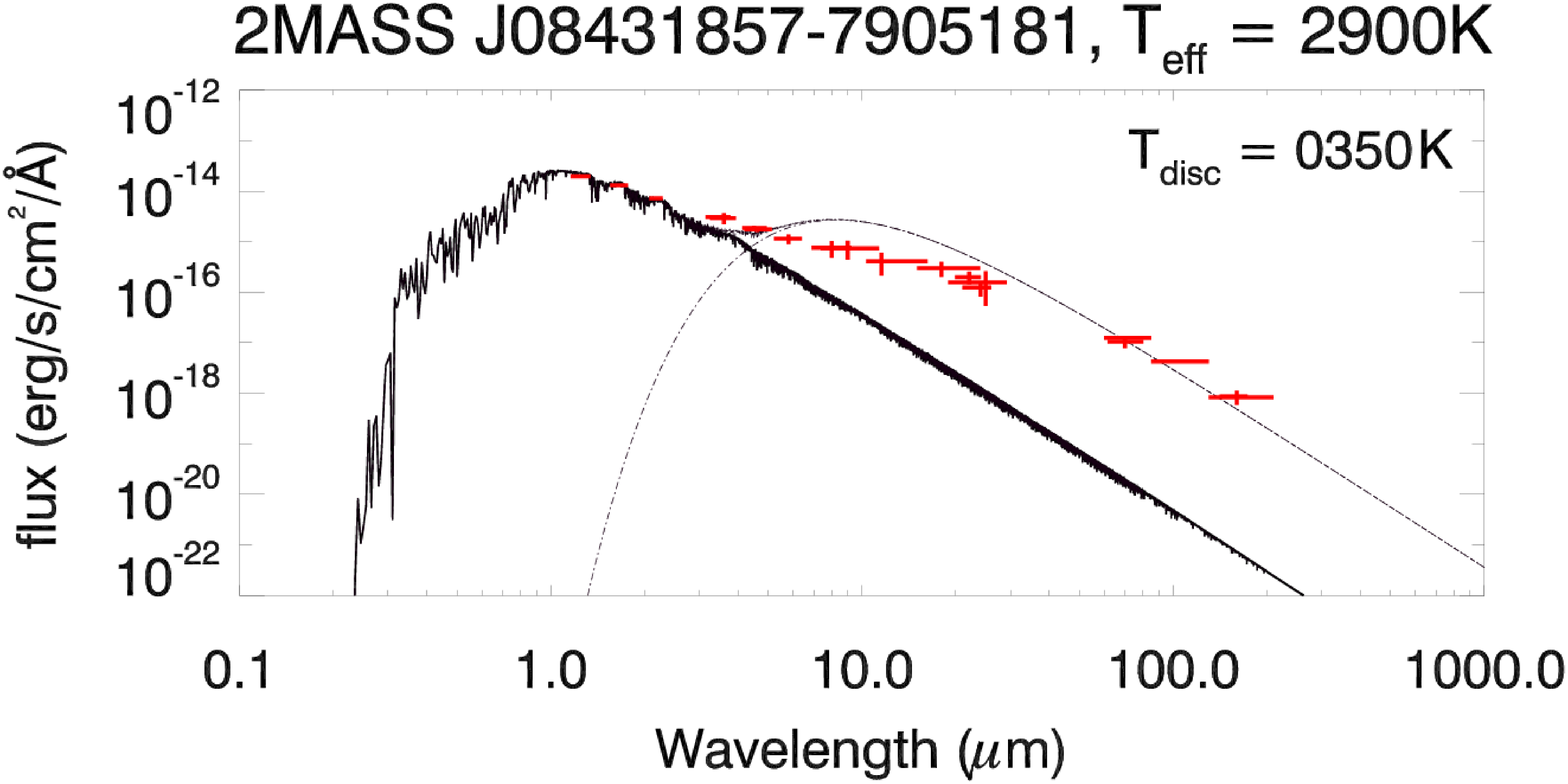} \\
        {\bf D}\includegraphics[width=0.31\textwidth]{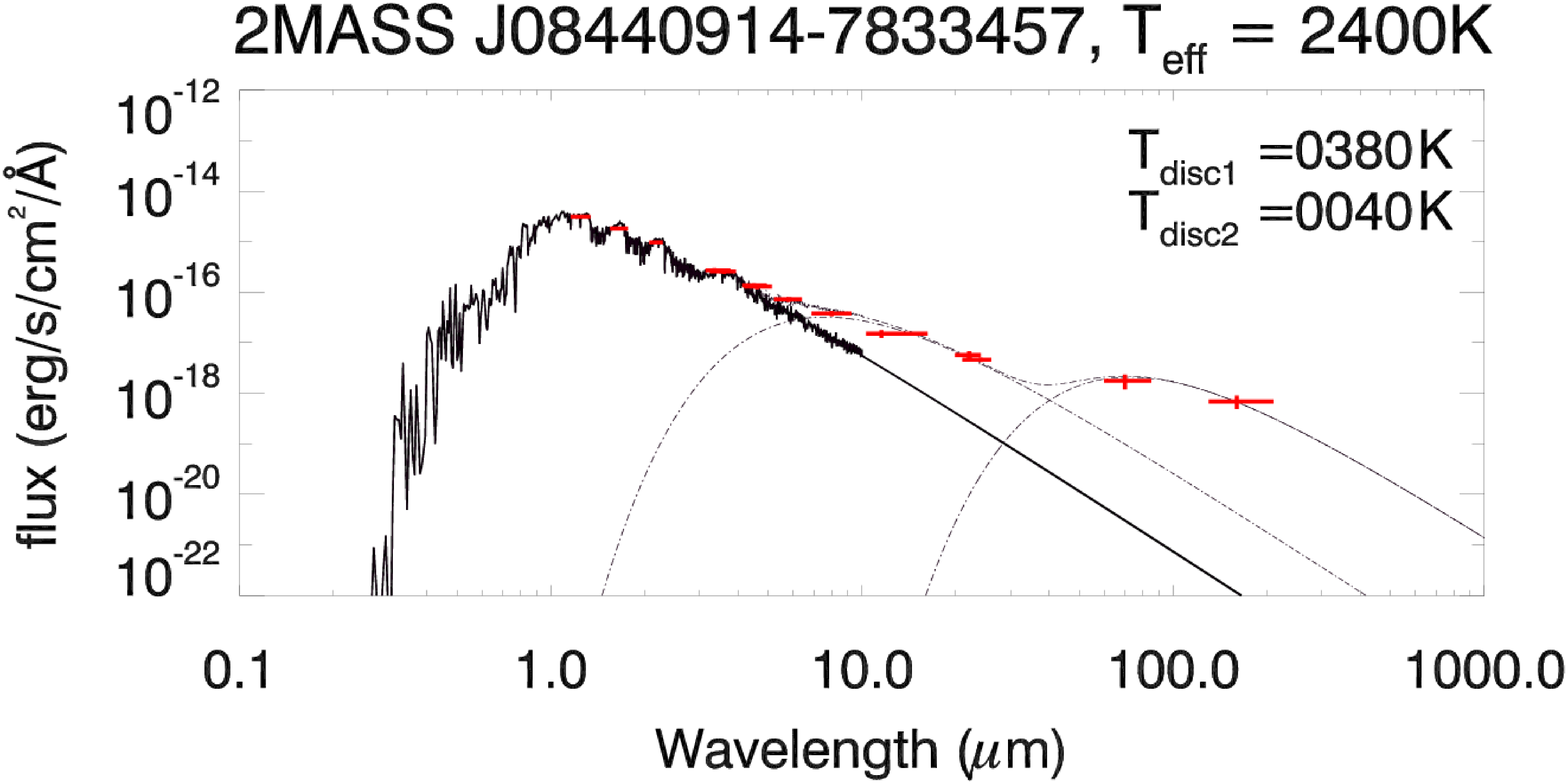} & {\bf E}\includegraphics[width=0.31\textwidth]{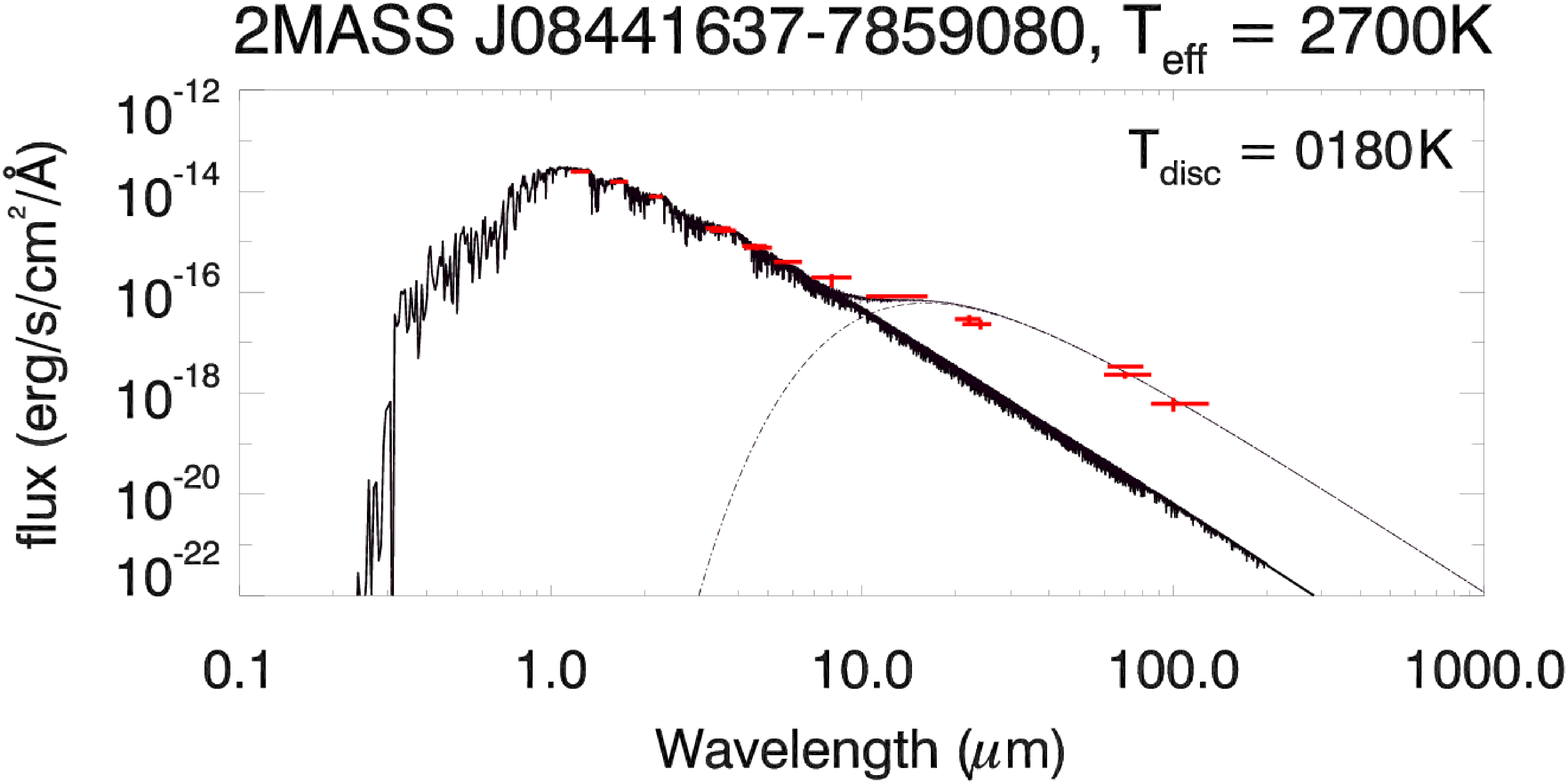} & {\bf F}\includegraphics[width=0.31\textwidth]{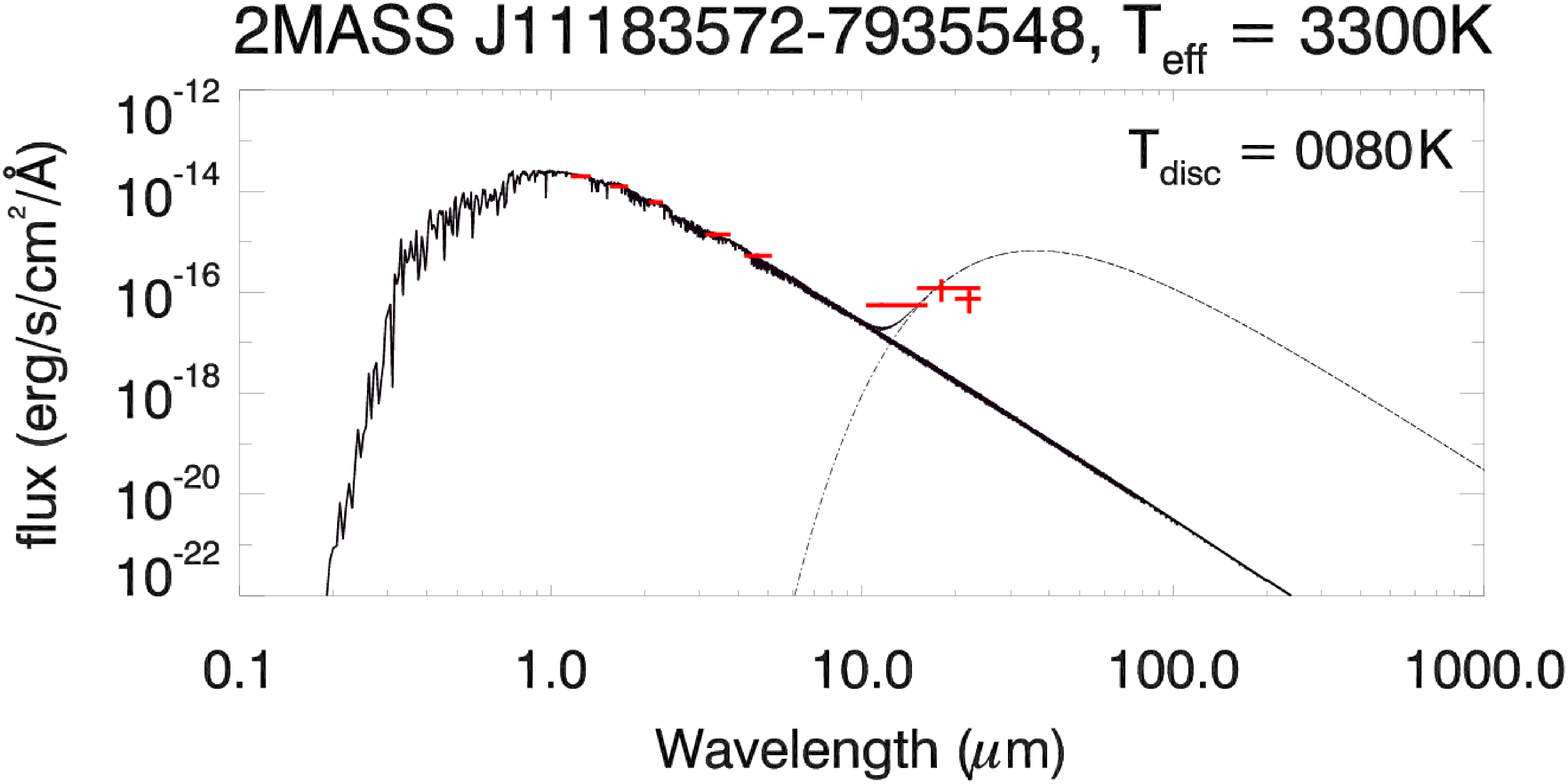} \\
        {\bf G}\includegraphics[width=0.31\textwidth]{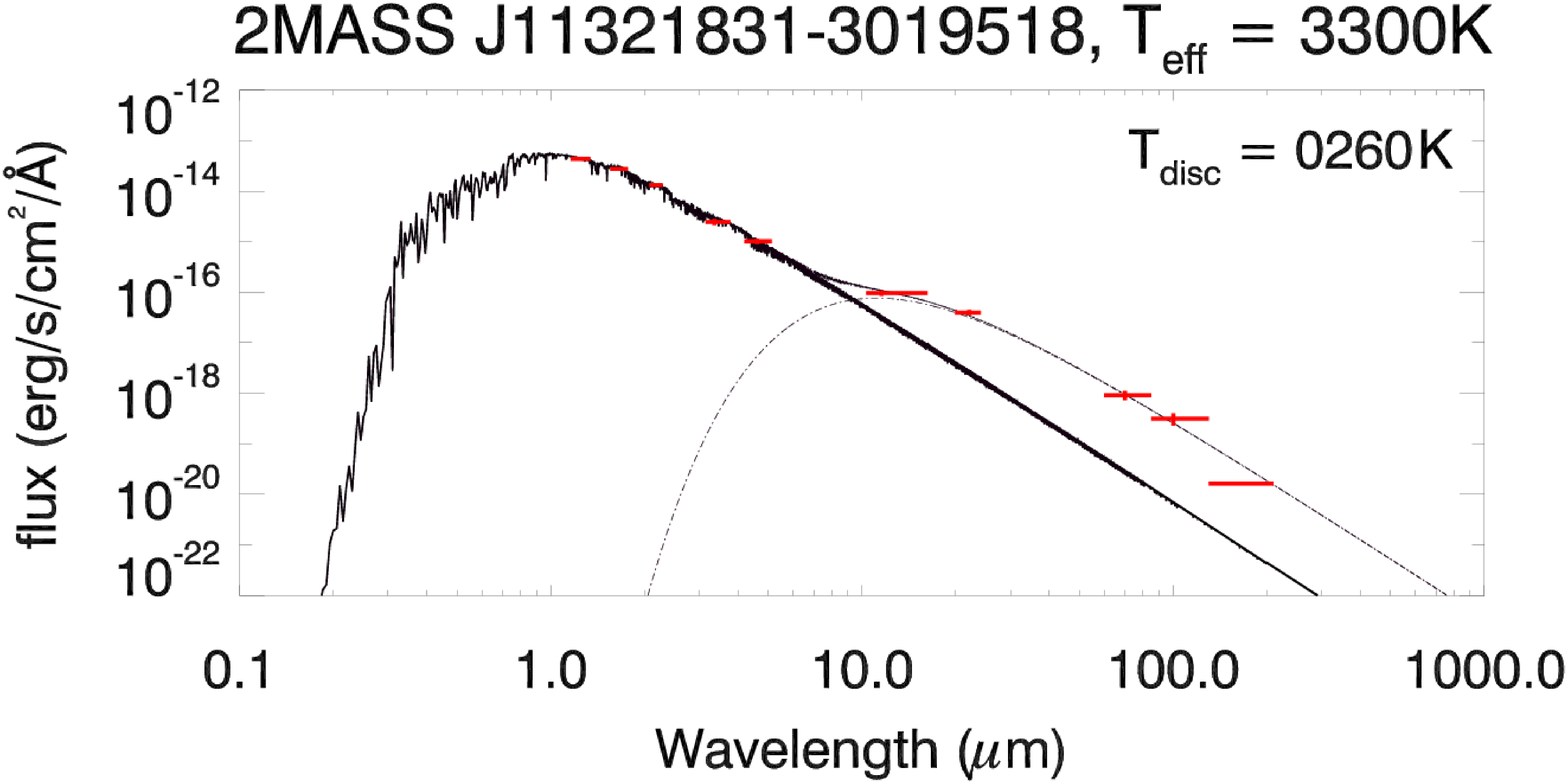} & {\bf H}\includegraphics[width=0.31\textwidth]{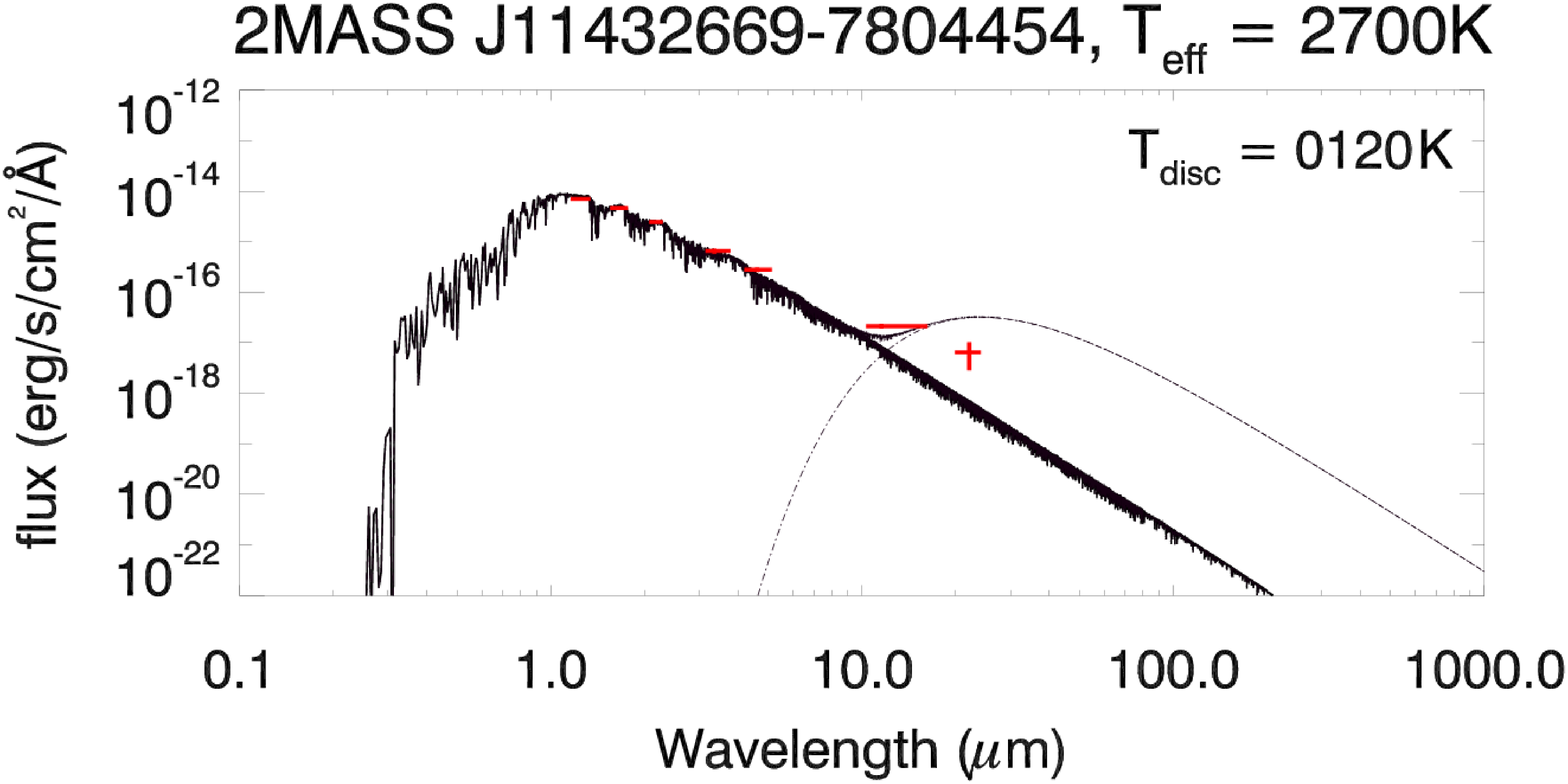} & {\bf I}\includegraphics[width=0.31\textwidth]{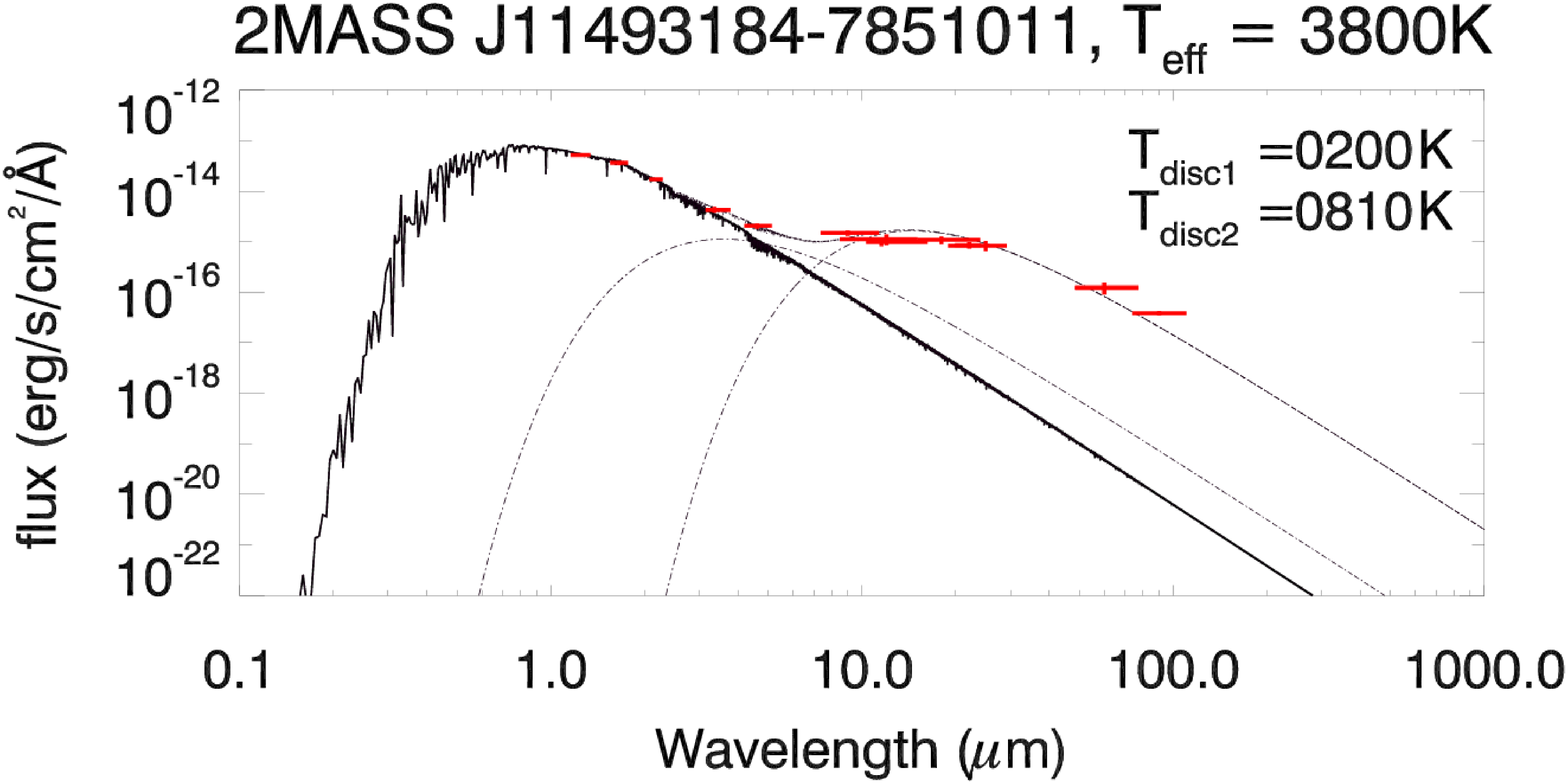} \\
        {\bf J}\includegraphics[width=0.31\textwidth]{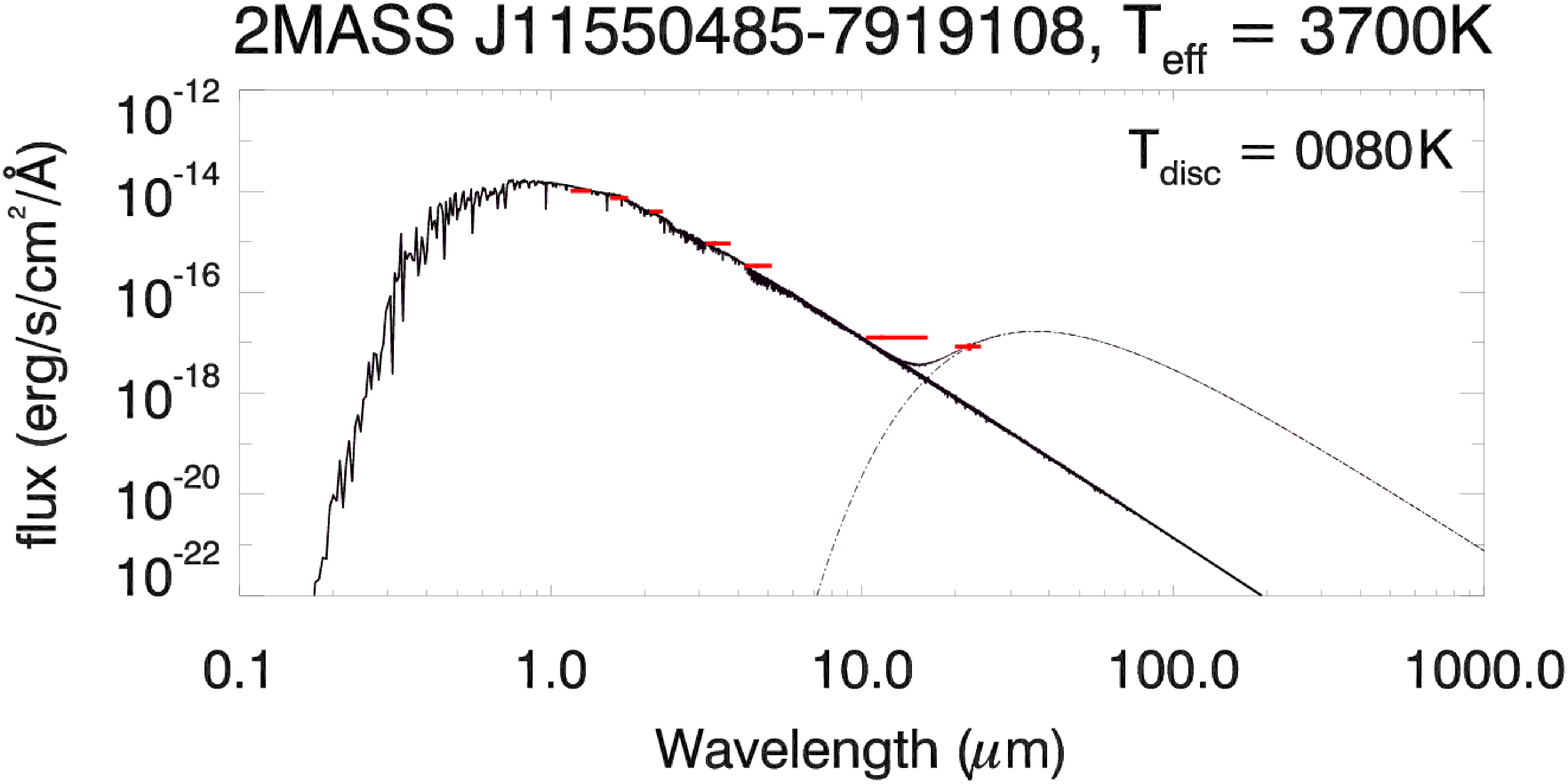} & {\bf K}\includegraphics[width=0.31\textwidth]{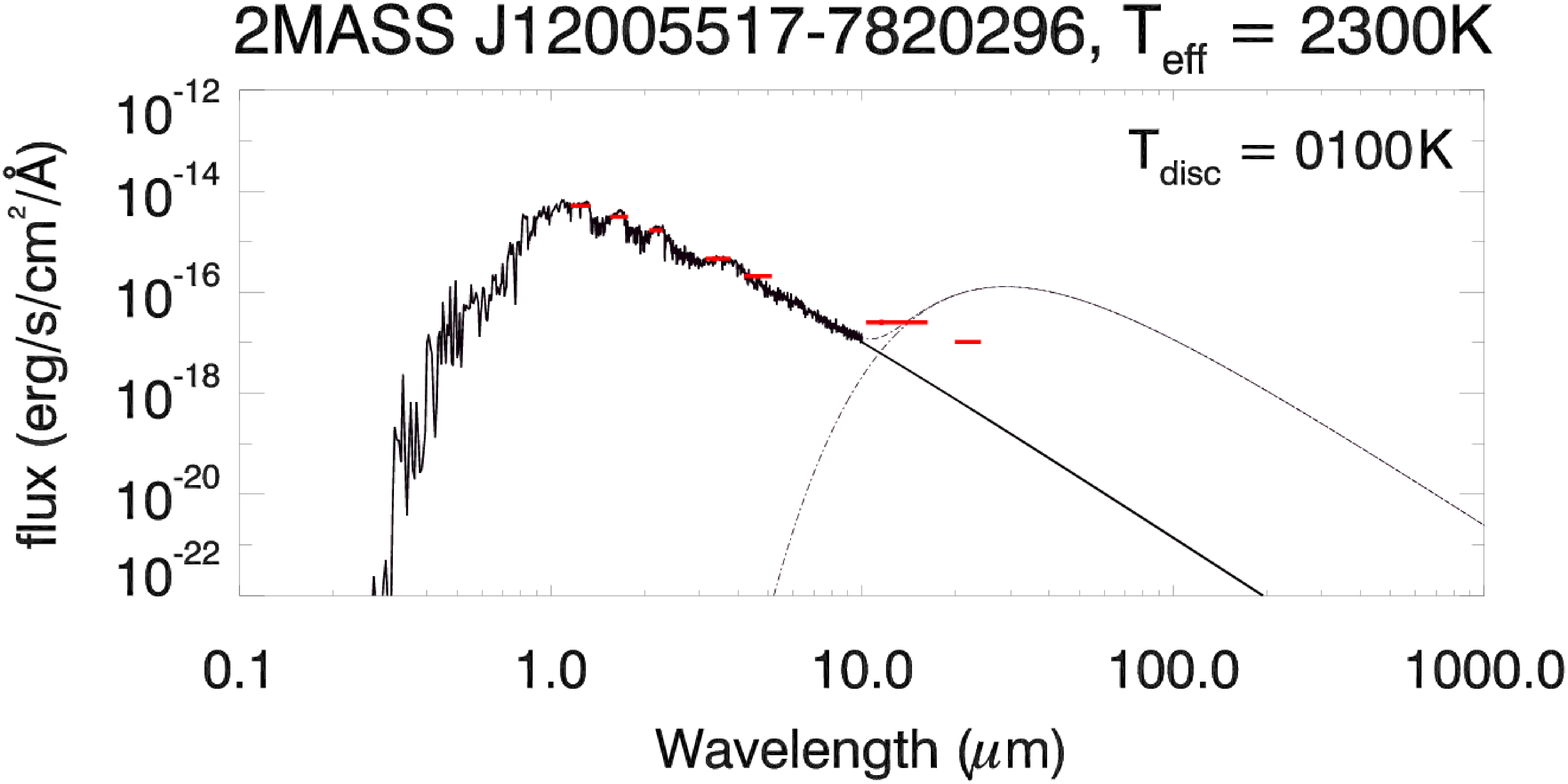} & {\bf L}\includegraphics[width=0.31\textwidth]{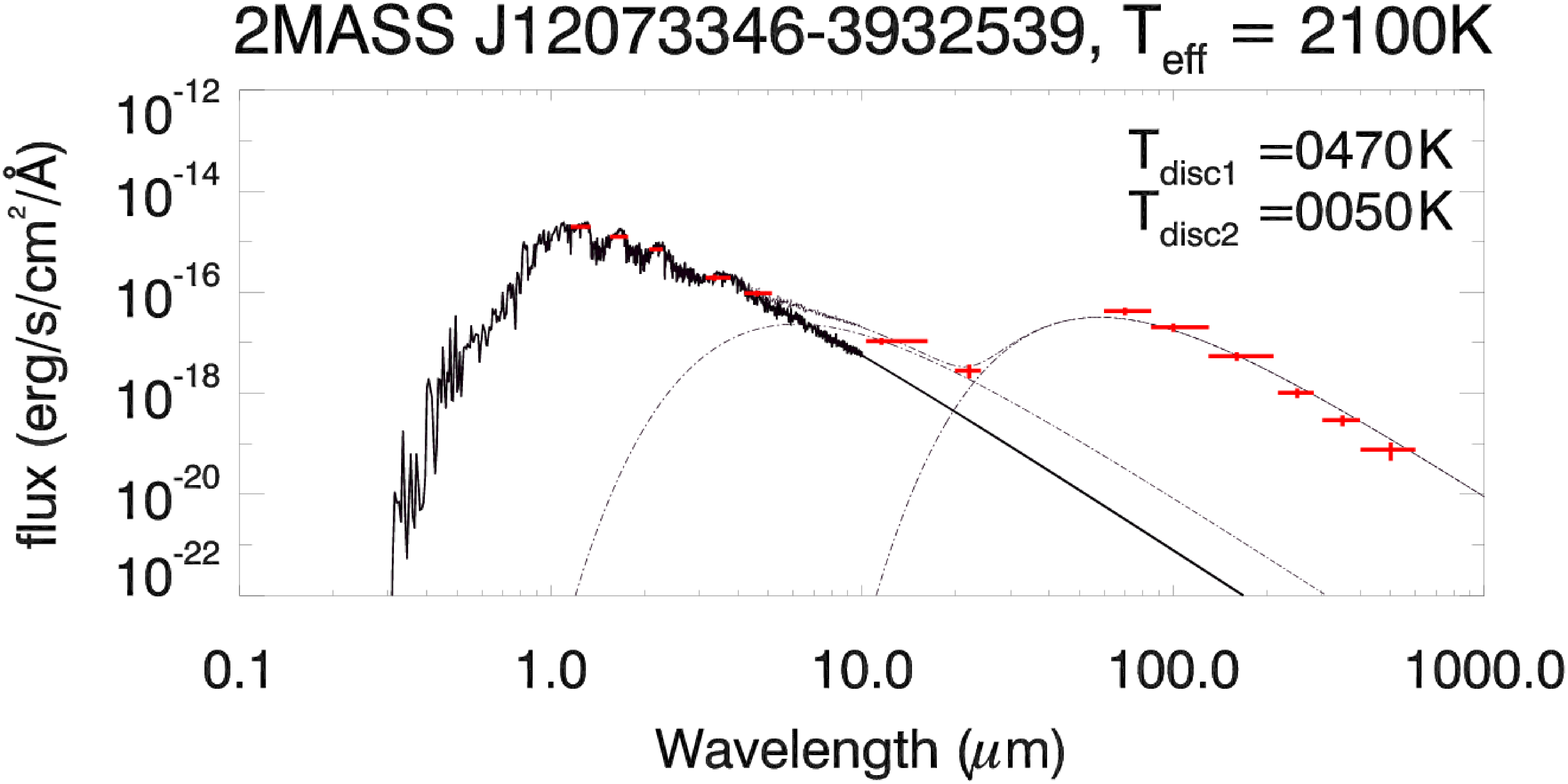} \\
      \end{tabular}
      \caption{SED models for the 12 objects identified with $W1-W4>1.0$.  Red symbols represent the flux data and the best-fit stellar SED, single-temperature (or two-temperature) black-body and their sum are overplotted.  The fitting parameters for each SED are provided in Table~\ref{T_SED_Params}. Blue downwards-facing triangles represent values for 3$\sigma$ upper limits.}
       \label{F_SED}
    \end{minipage}
\end{figure*}

\begin{figure*}
    \begin{minipage}[b]{\textwidth}
    \centering
      \begin{tabular}{ccc}
        {\bf M}\includegraphics[width=0.31\textwidth]{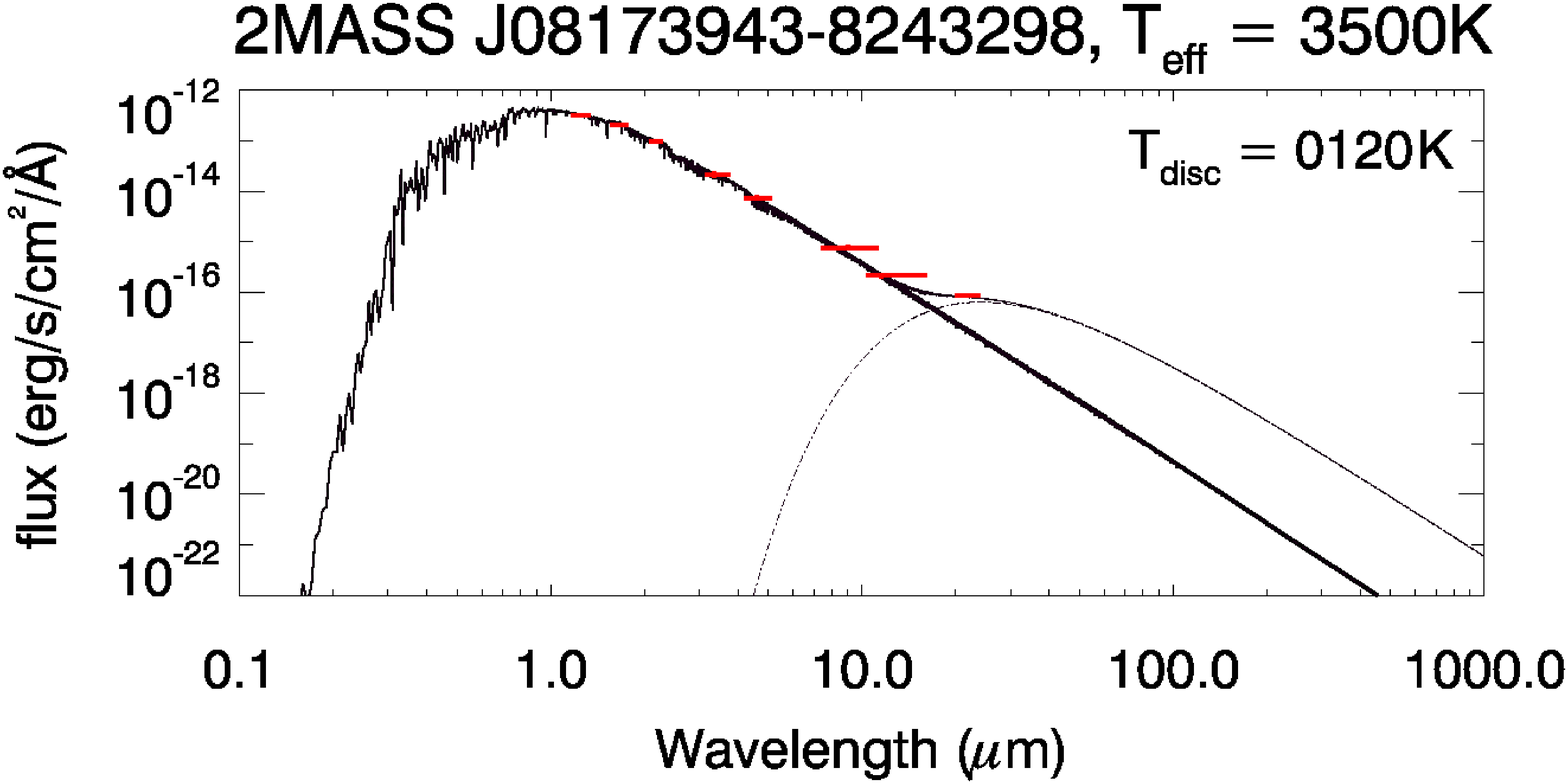} & {\bf N}\includegraphics[width=0.31\textwidth]{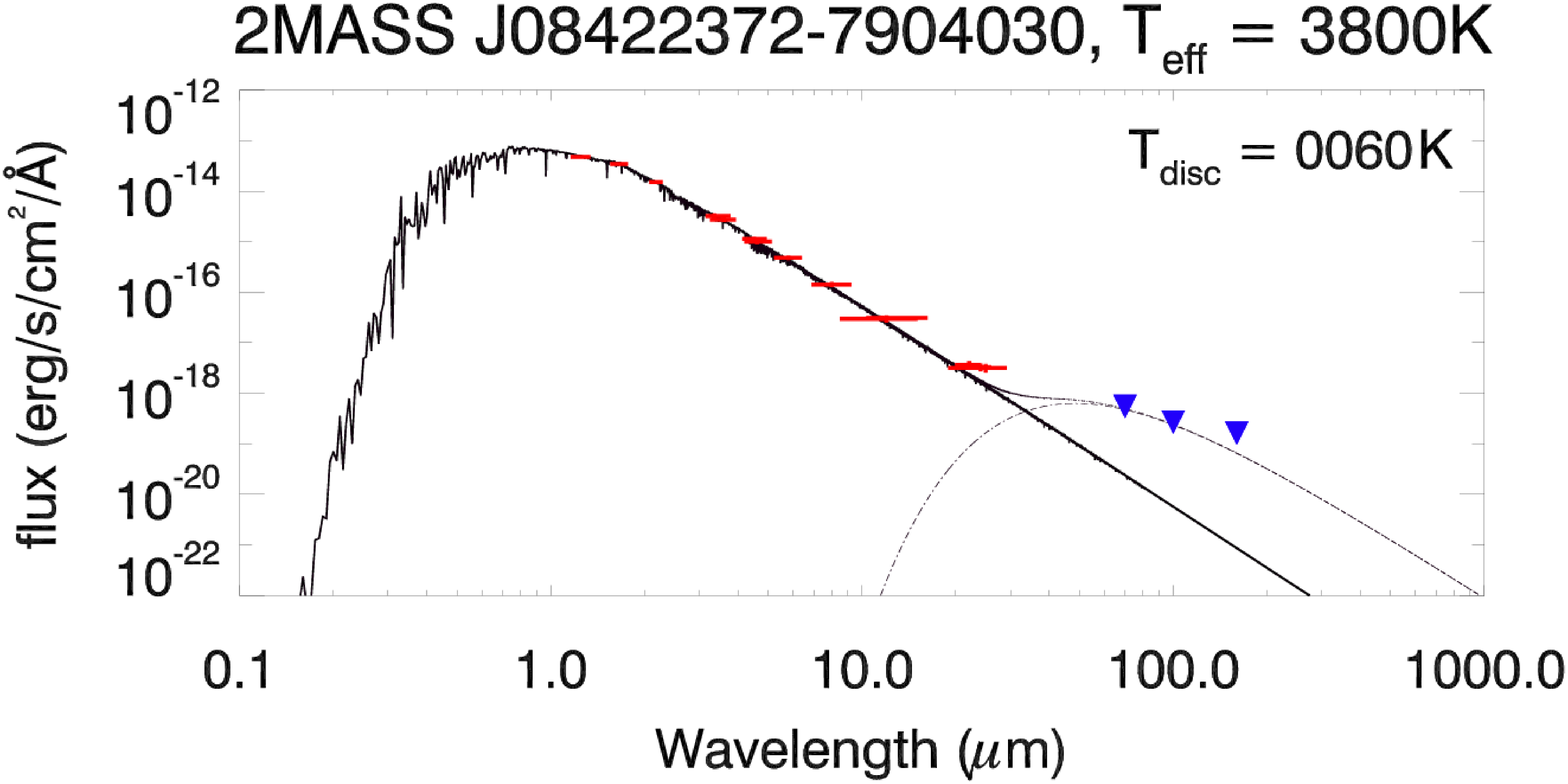} & {\bf O}\includegraphics[width=0.31\textwidth]{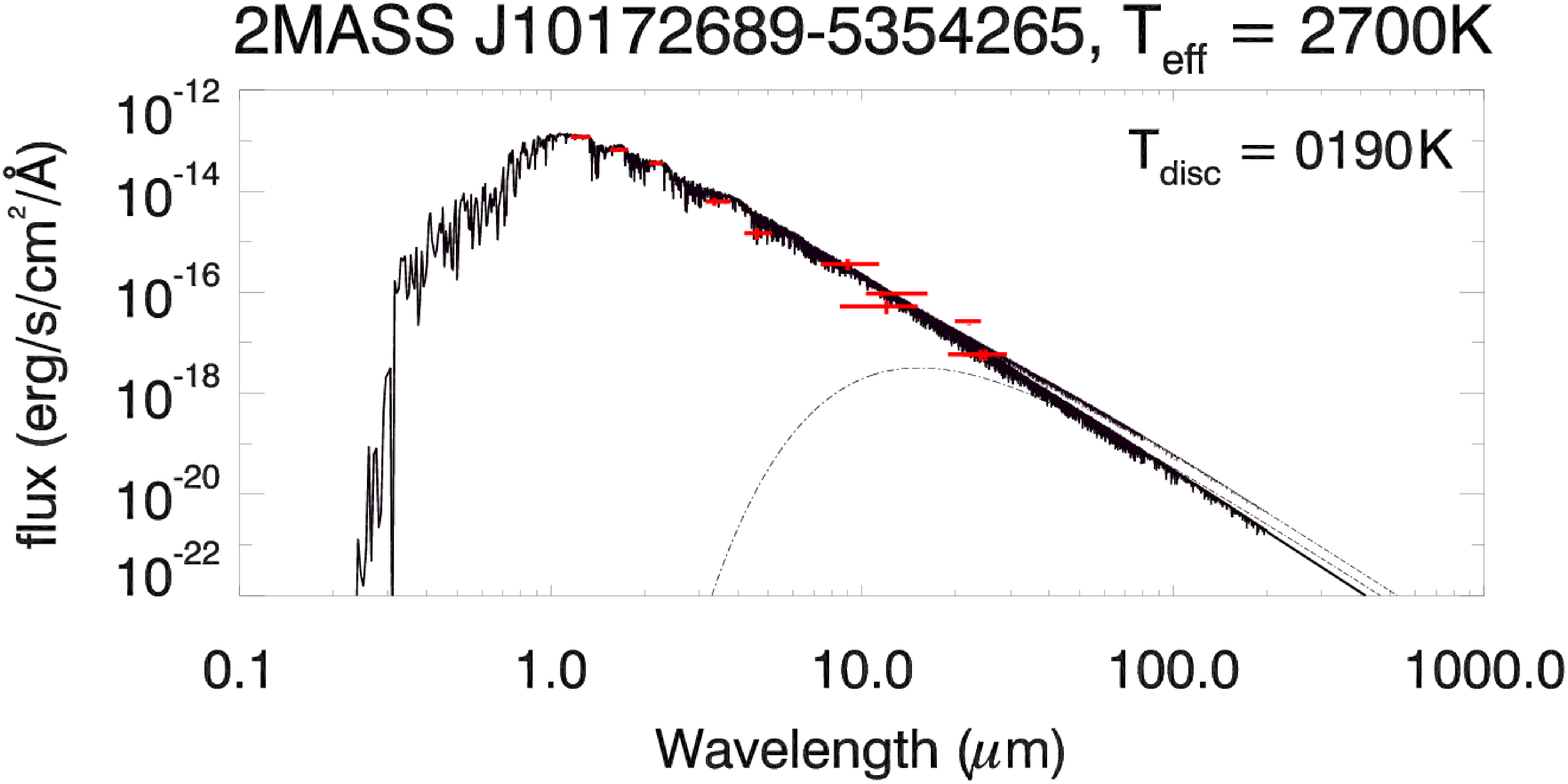} \\
        {\bf P}\includegraphics[width=0.31\textwidth]{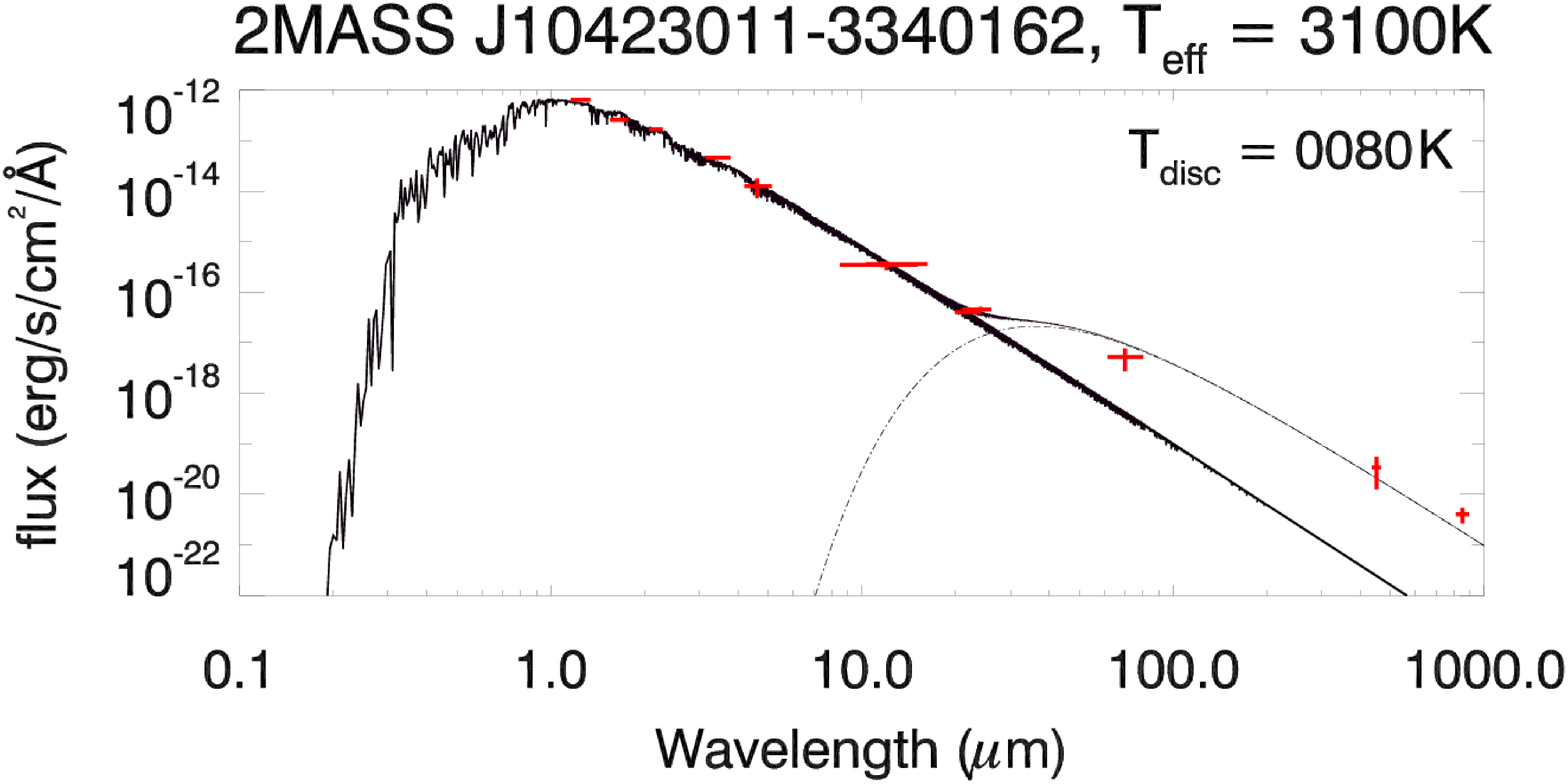} & {\bf Q}\includegraphics[width=0.31\textwidth]{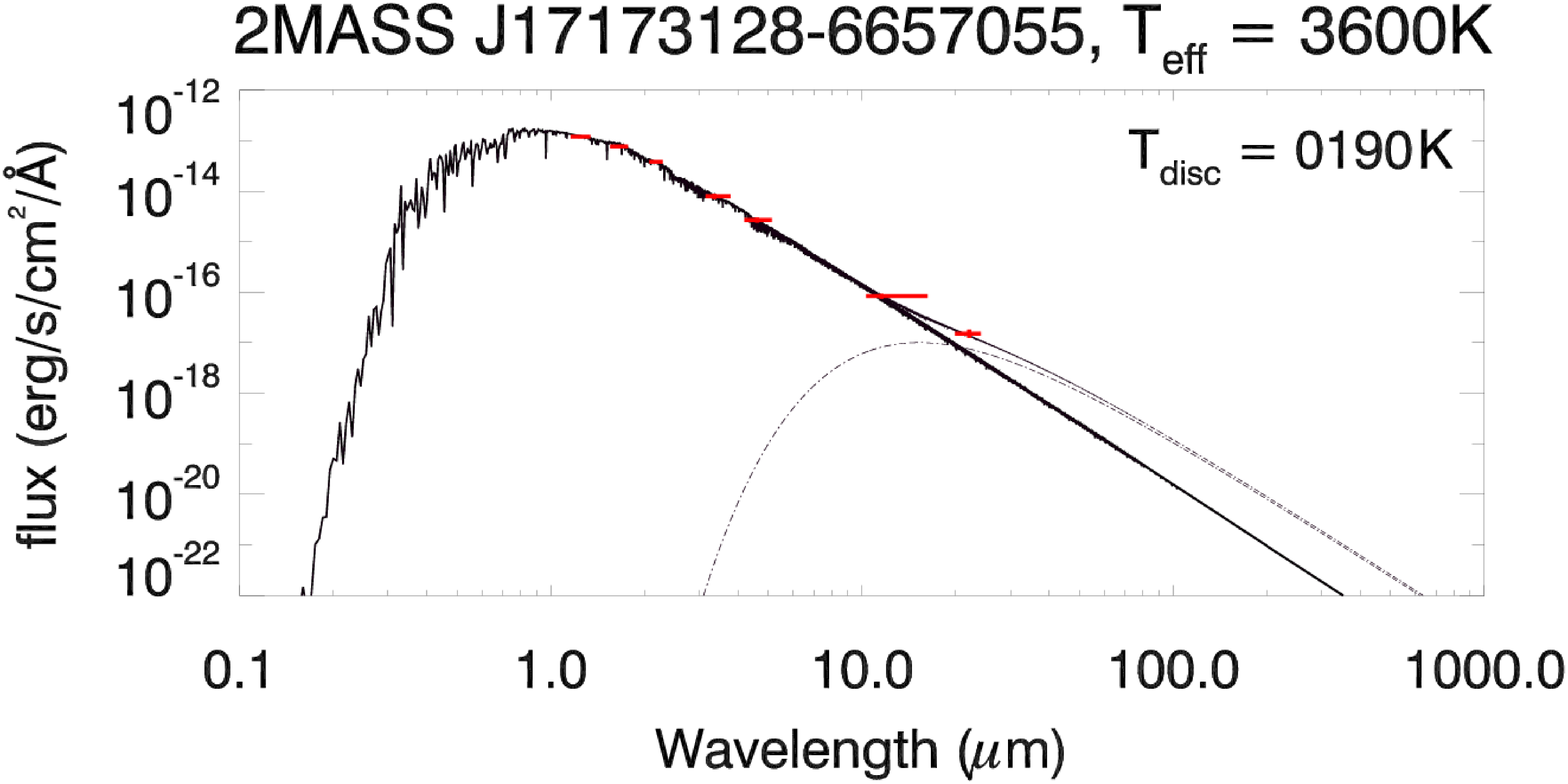} & {\bf R}\includegraphics[width=0.31\textwidth]{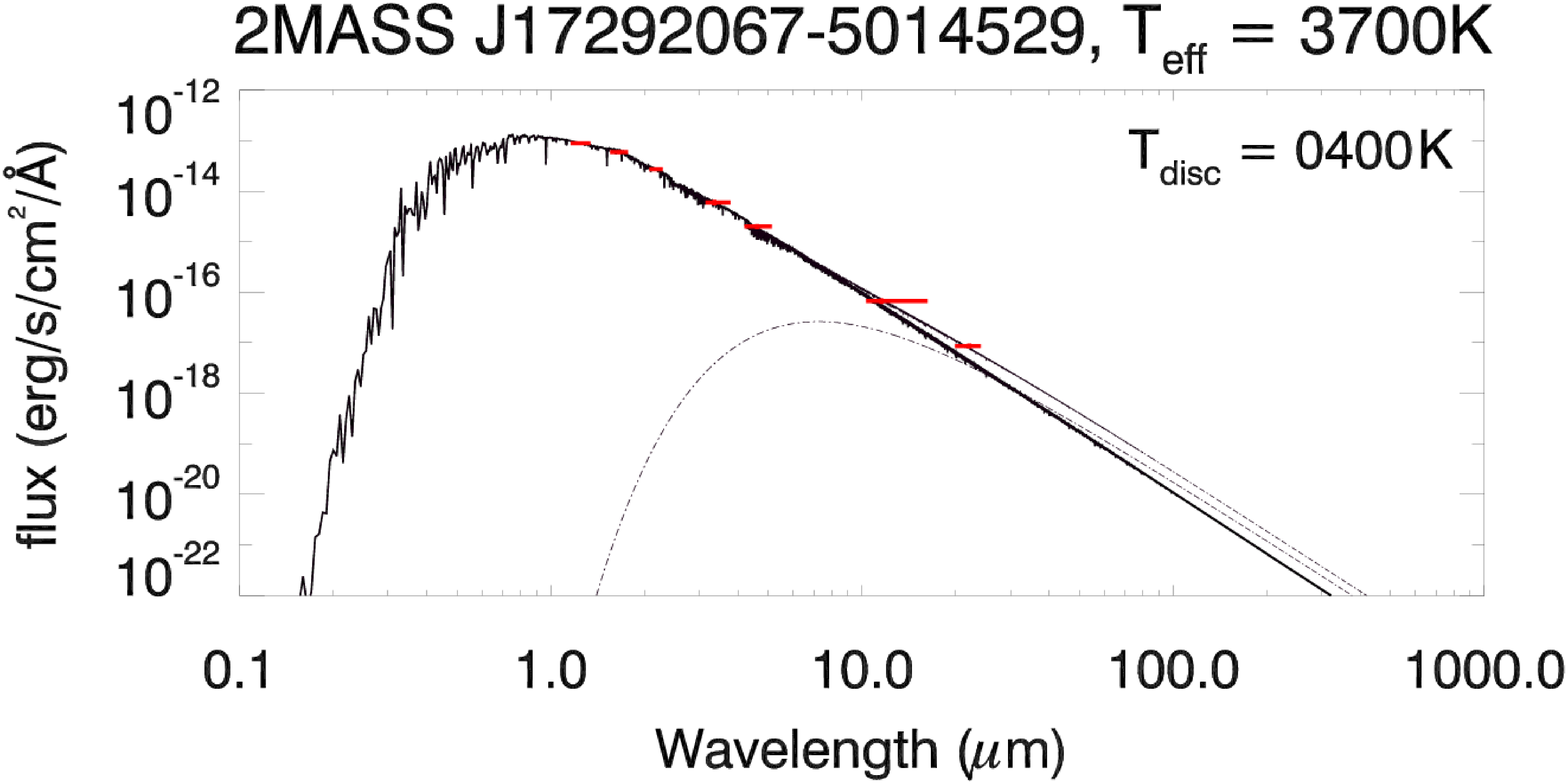} \\
        {\bf S}\includegraphics[width=0.31\textwidth]{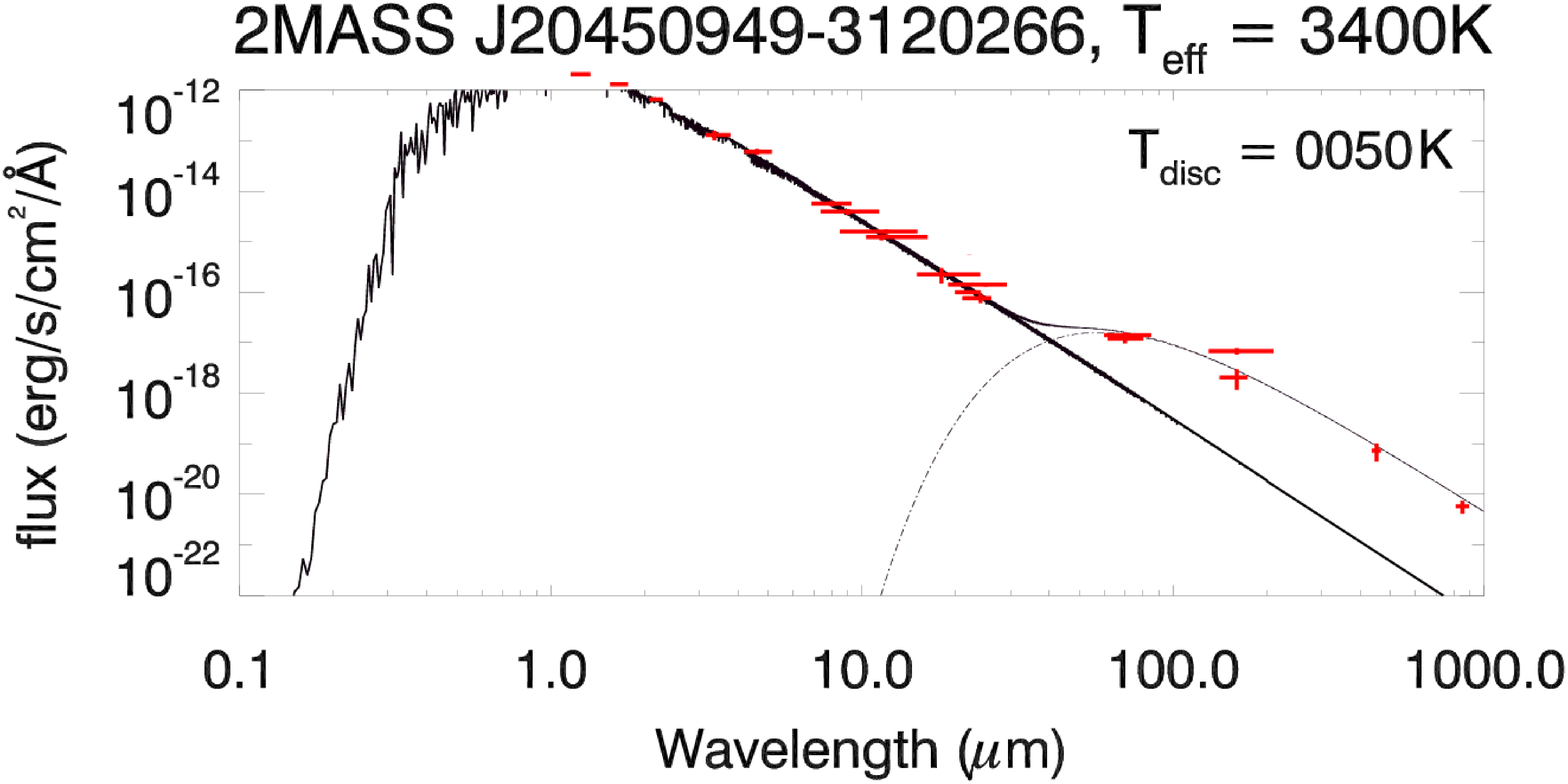} &  & \\
      \end{tabular}
      \caption[SEDs for objects with $W1-W4 < 1.0$ and $E_{W4} >
        3.0$.]{SEDs for objects with $W1-W4 < 1.0$ and $E_{W4} >
        3.0$. Only in the cases of TWA~7 (target P) and AU~Mic (target S) are data available beyond 25\,$\mu$m, although EI~Cha (target N) is constrained by upper-limits from Herschel PACS 70, 100 and 160\,$\mu$m.}
       \label{F_EW4}
    \end{minipage}
\end{figure*}

\begin{figure*}
    \begin{minipage}[b]{\textwidth}
    \centering
      \begin{tabular}{ccc}
        {\bf A}\includegraphics[width=0.31\textwidth]{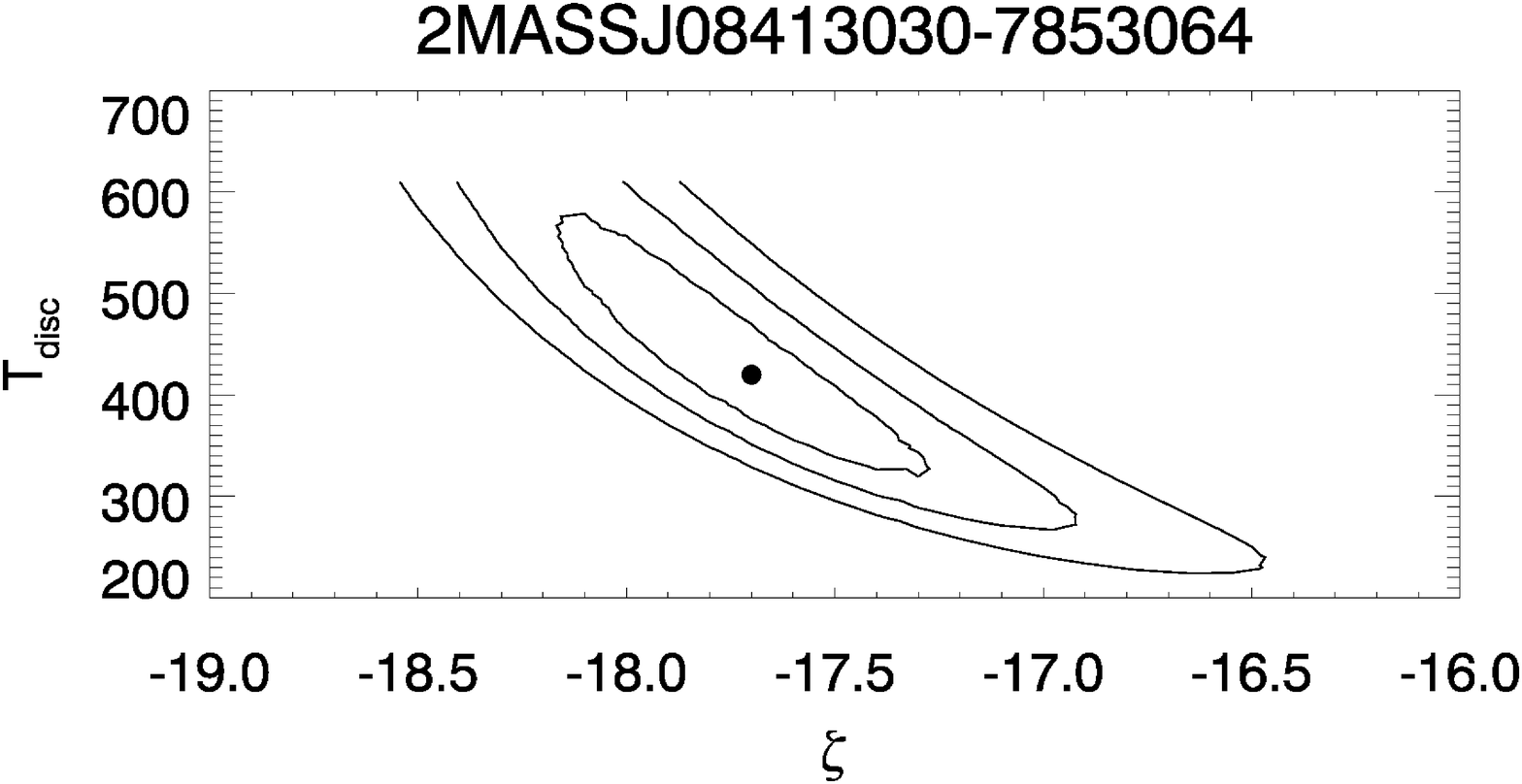} & {\bf B}\includegraphics[width=0.31\textwidth]{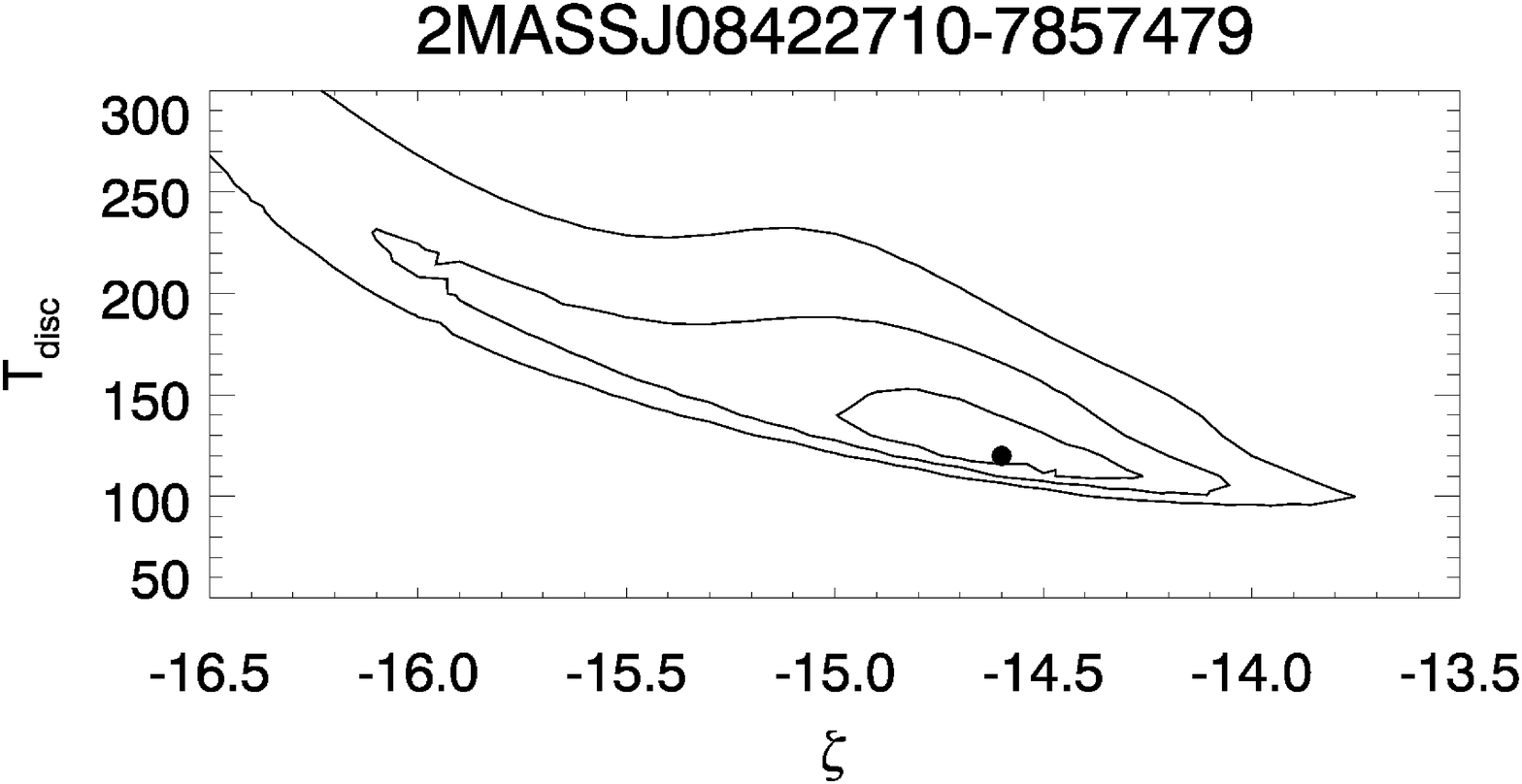} & {\bf C}\includegraphics[width=0.31\textwidth]{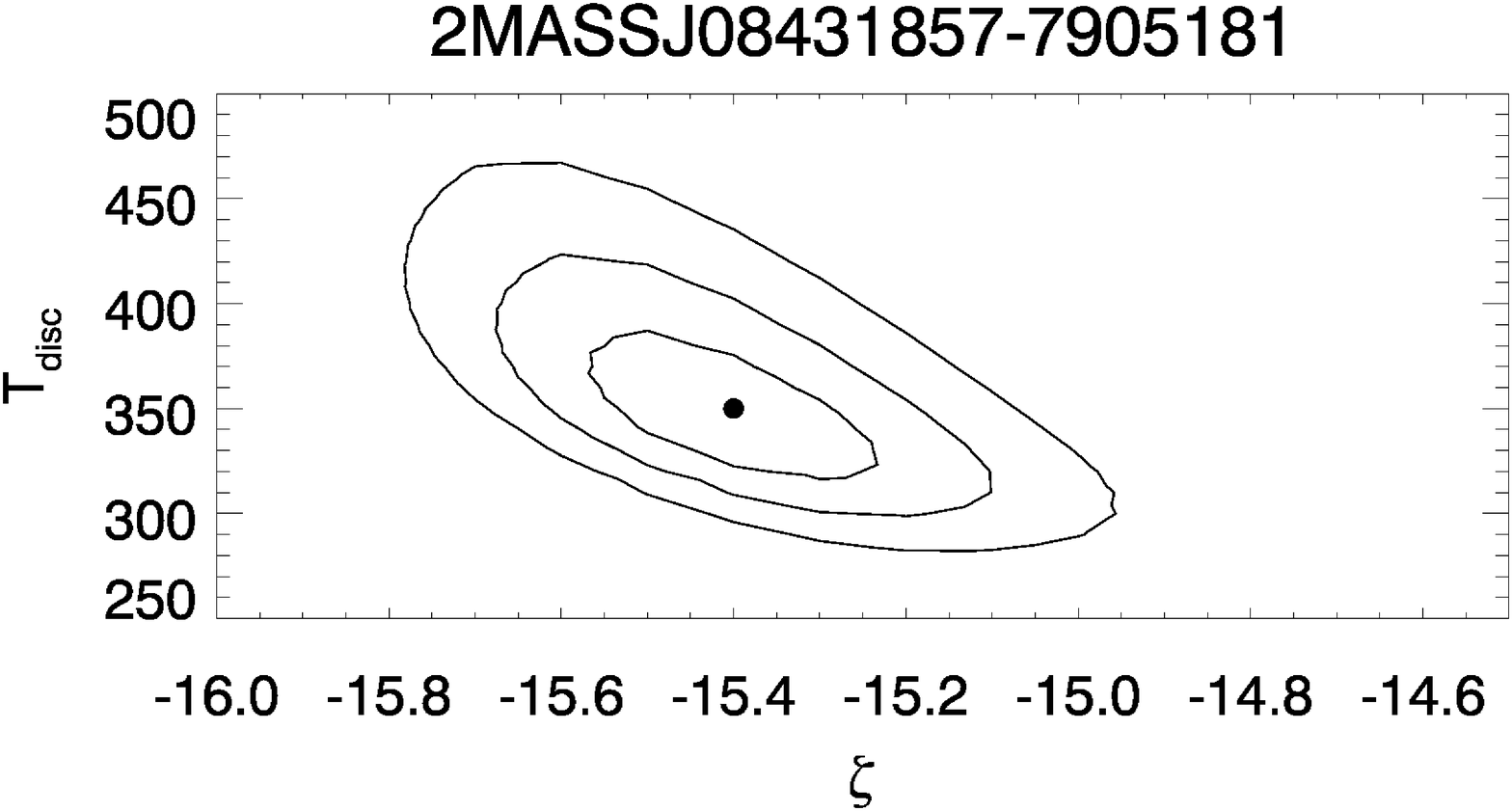} \\
        {\bf E}\includegraphics[width=0.31\textwidth]{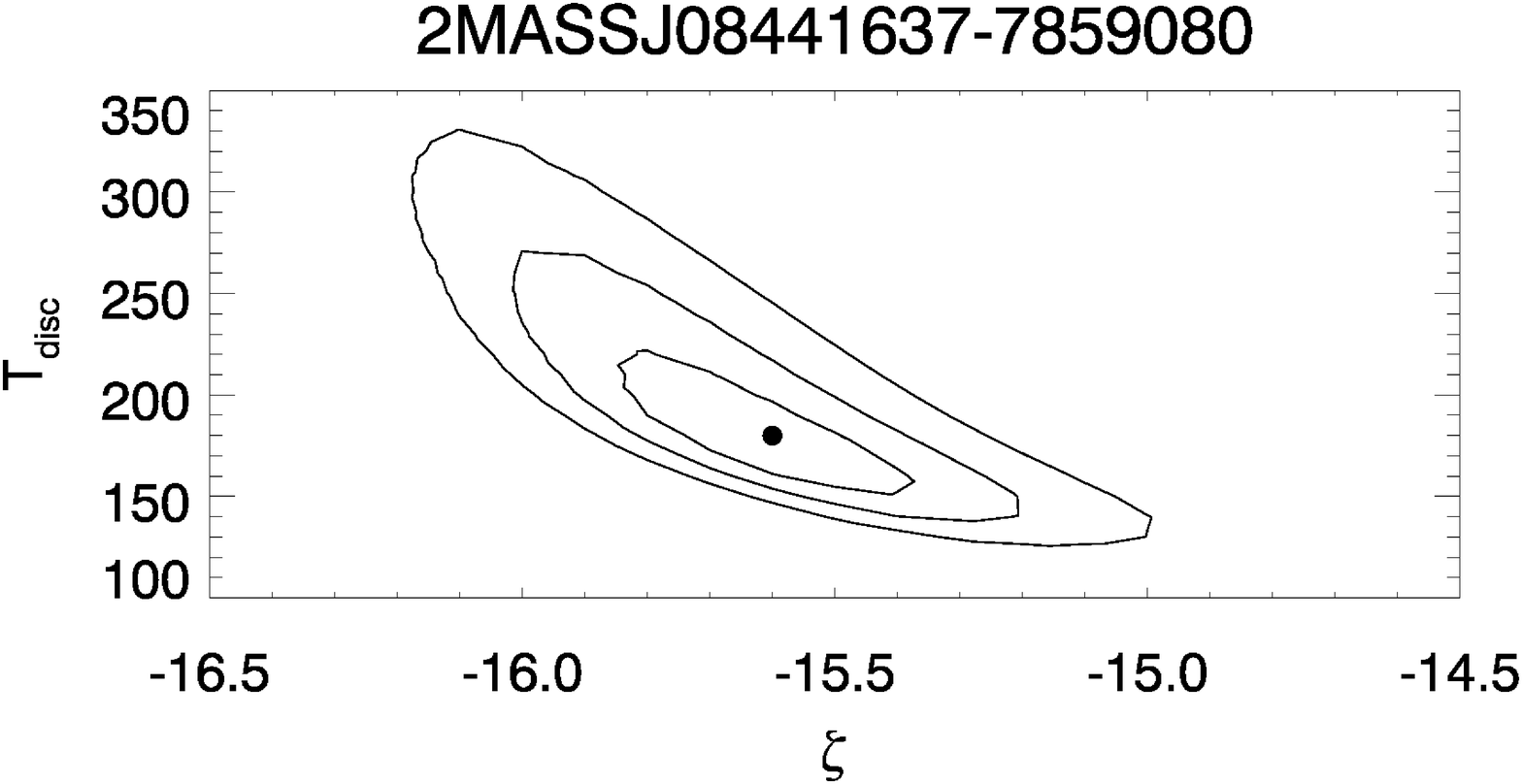} & {\bf G}\includegraphics[width=0.31\textwidth]{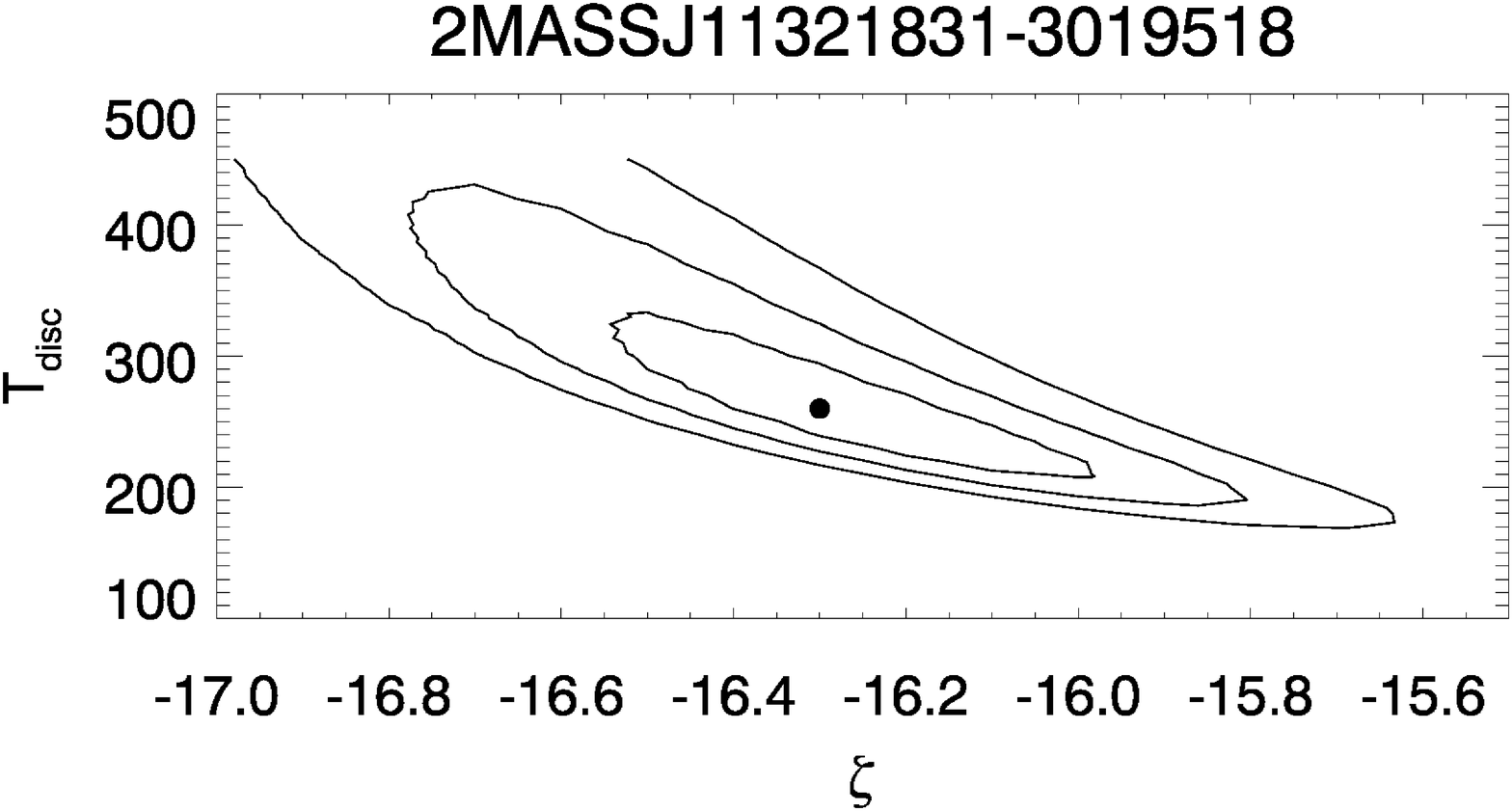} & {\bf M}\includegraphics[width=0.31\textwidth]{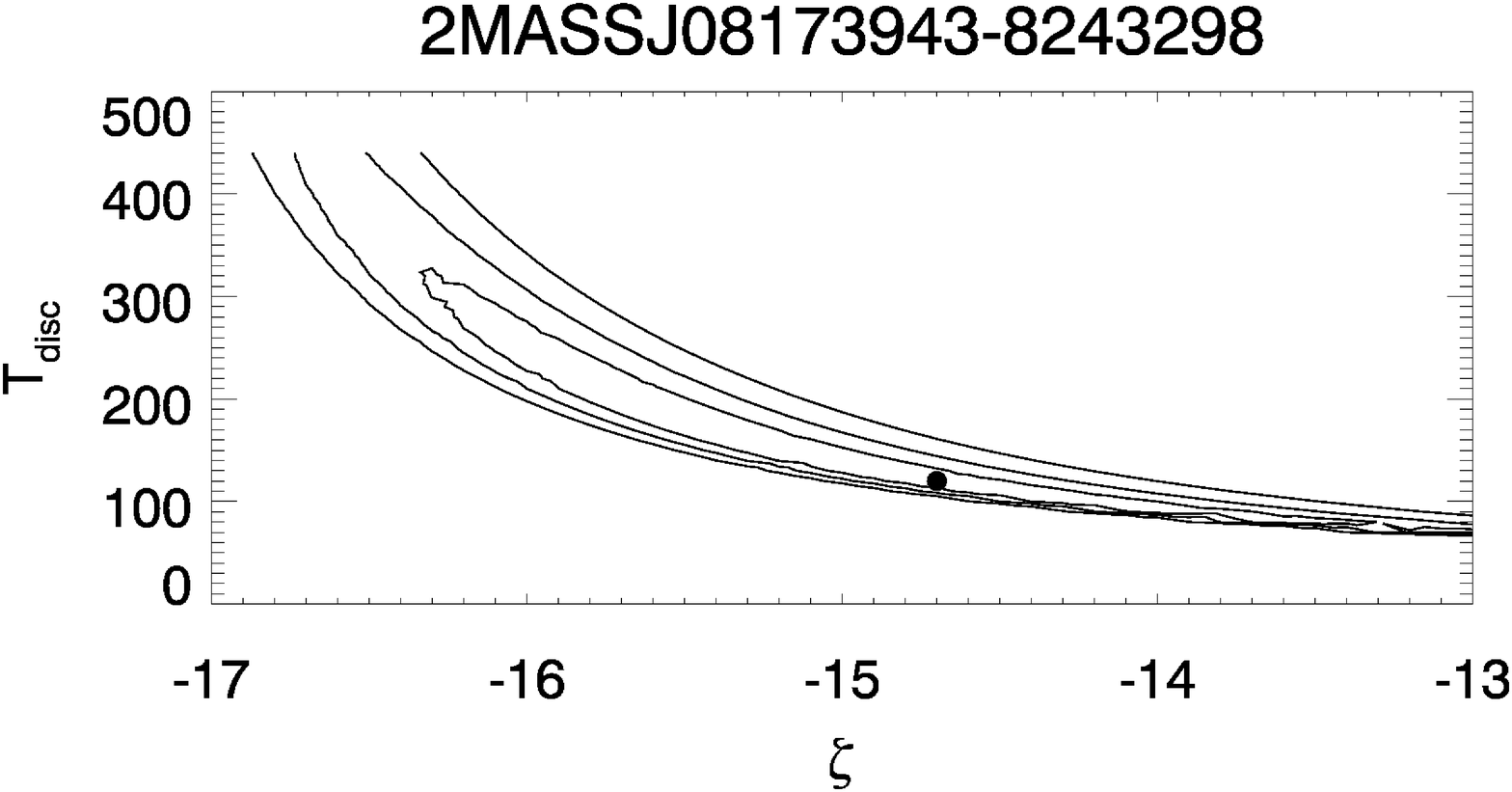} \\
{\bf P}\includegraphics[width=0.31\textwidth]{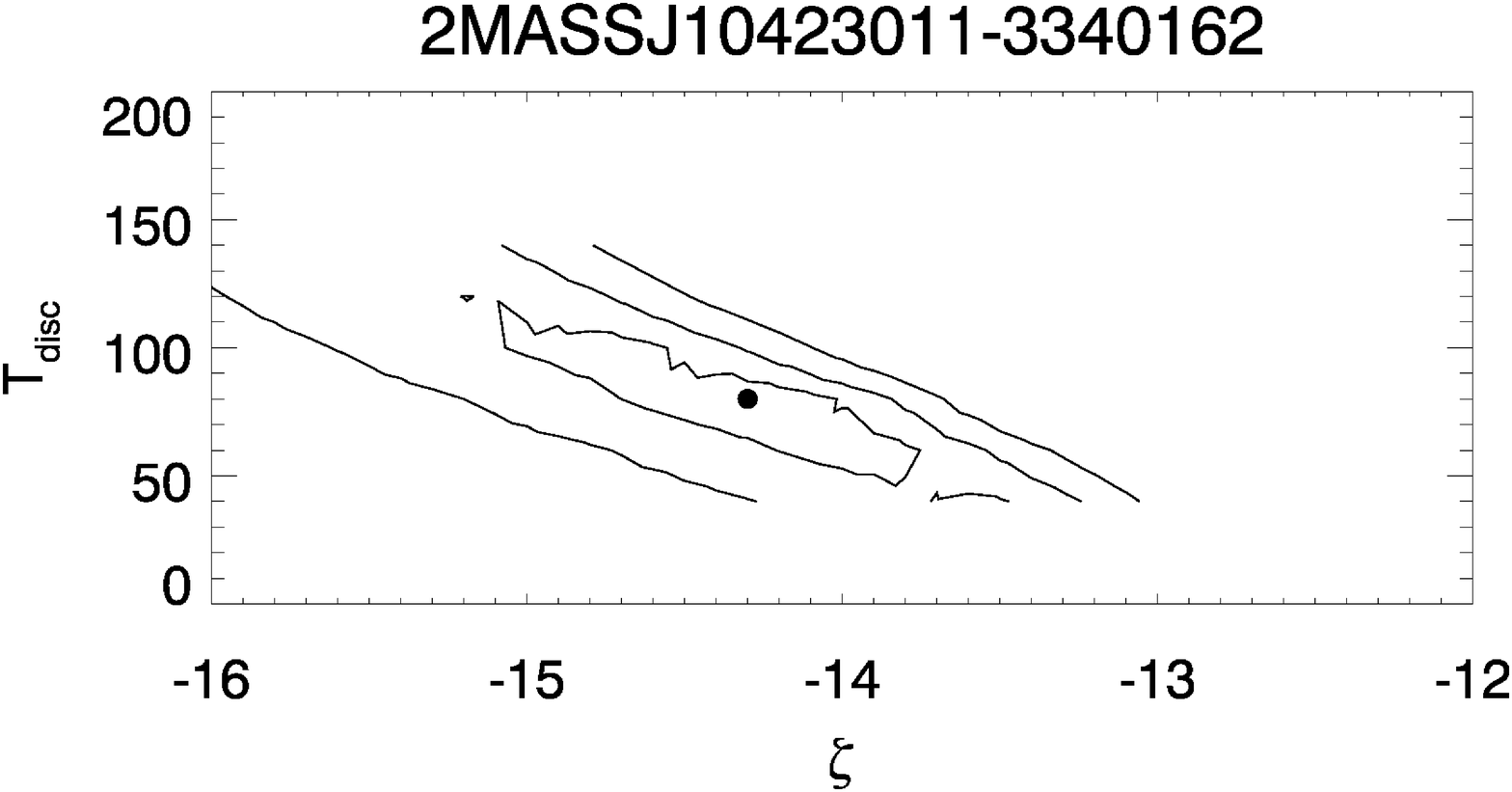} & {\bf S}\includegraphics[width=0.31\textwidth]{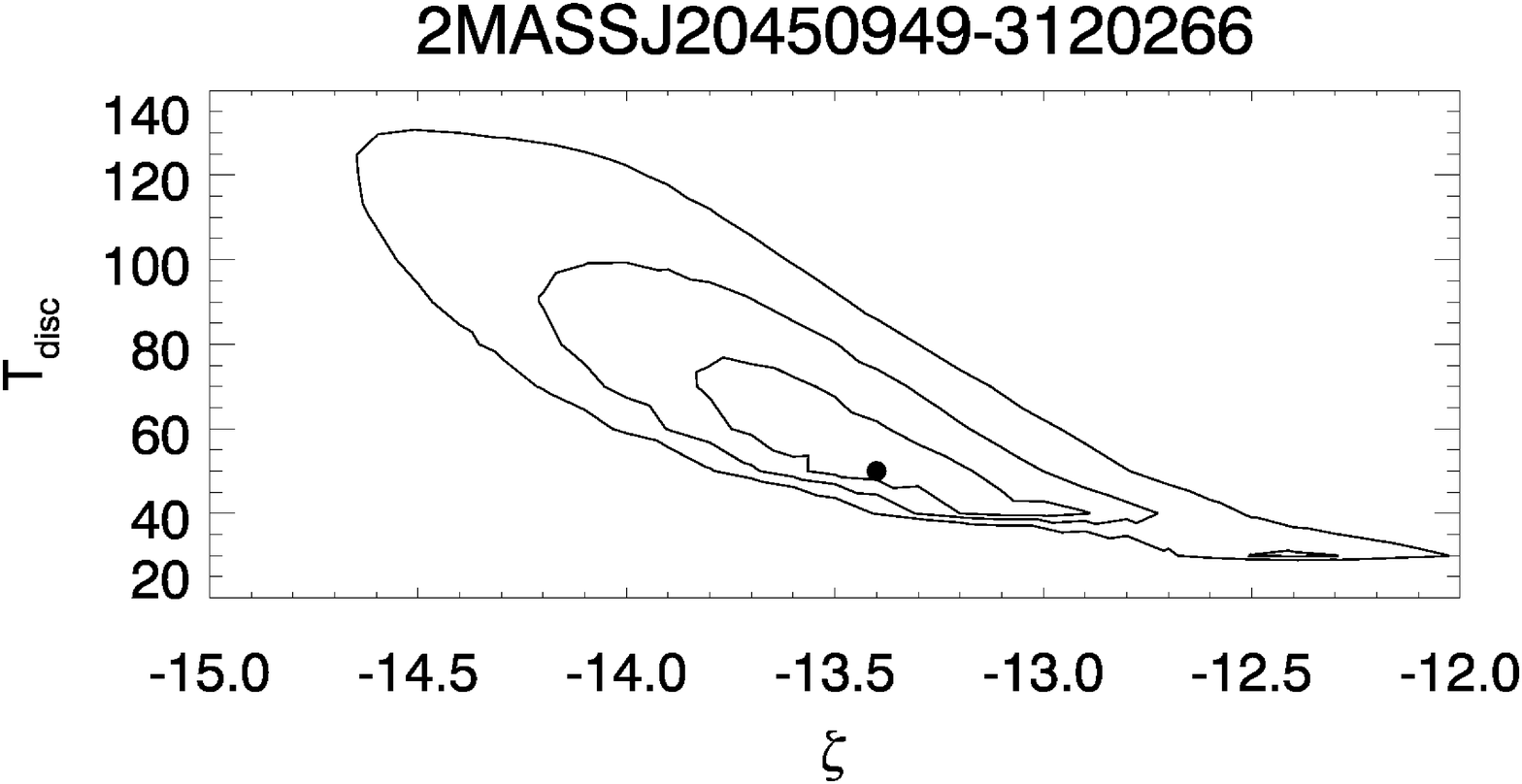} & \\
      \end{tabular}
      \caption{Contour plots indicating the 1, 2 and 3$\sigma$ errors
        from simultaneously fitting $\zeta$ and $T_{\rm eff}$ in the
        SED models ($\chi^2_{\rm min} +$ 1, 2 or 3$\sigma$,
        respectively). These were only available for 8 objects because
        in all other cases there was not enough far-IR data to
        constrain the disc model.}
       \label{F_Contour}
    \end{minipage}
\end{figure*}

The photospheric model is constrained using $2MASS$ $JHK$ photometry,
assuming that a disc component would contribute negligble flux at these
wavelengths, and then this is used as a baseline upon which the disc
flux is added. $T_{\rm eff}$ was estimated to the nearest 100\,K by
comparing spectral-types with the colour-$T_{\rm eff}$ tables in
\cite{2013a_Pecaut}, leaving three free parameters to constrain: the
disc temperature ($T_{\rm d}$), the surface area of the star and the
surface area of the disc. For the cases where we fit a two temperature disc we constrain two disc areas and two disc temperatures. Because the fitted SEDs were calibrated for a
source at 10\,pc, the flux profiles were multiplied by a normalisation
parameter of $10^{\eta}$ to characterise the stellar surface area. The
value of $\eta$ was altered until the minimum ${\chi}^{2}_{*}$ value
was obtained with respect to the $JHK$ photometry. Error bars for $\eta$ were calculated using a parabolic interpolation of ${\chi}^{2}_{\rm *}$ versus $\eta$. The $\eta$ and $T_{\rm eff}$ values are listed in Table~\ref{T_SED_Params} and all $\eta$ errors were $\sim 0.01$.

A ``star plus disc'' model was then used and the parameters varied 
to minimise a chi-squared value labelled as $\chi^2_{\rm d}$ using the best photospheric model, $WISE$
photometry and any supplementary IR data available (see
Table~\ref{T_Disc_Type}). A single-temperature Planck function
was used with the effective disc area and $T_{\rm d}$ as free parameters. Flux upper limits were not included in the chi-square minimisation, however we require that the model flux is less than the upper limits where an upper limit is present. The Planck
function normalisation was multiplied by $10^{\zeta}$, altering
$\zeta$ in steps of 0.05 and $T_{\rm d}$ in steps of 10\,K to provide
the lowest $\chi^{2}_{\rm d}$. We then froze into the model the best-fitting star and single-disc parameters and introduce a second disc using the same $\zeta$ and $T_{\rm d}$ steps to obtain a new $\chi^{2}_{\rm d}$. We use the F-test with two degrees of freedom at the 90 per cent confidence interval to determine whether to use the 1 temperature or 2 temperature model ($\Delta\chi^{2} = 9.00$) -- for targets D, I and L (see Figure~\ref{F_SED}) the 2 temperature model is applied.

Uncertainties in $\zeta$ and $T_{\rm d}$
were calculated only for single-temperature discs that had IR data either side of the
peak of the Planck function (5/12 of the objects with $W1-W4 > 1.0$ and 3/7 of the objects with $W1-W4 < 1.0$). If no IR data were available redward of the Planck function peak then only
upper limits to $T_{\rm d}$ could be estimated (see below). In Figure~\ref{F_SED}, the best-fit star plus disc models are displayed for all 12 objects that had
$W1-W4 > 1.0$ (see $\S$\ref{S_Identifying_IR_Excess}) and in
Figure~\ref{F_Contour} contour plots are provided for 68, 95 and 99.7
per cent confidence intervals for the simultaneous fits of $\zeta$ and
$T_{\rm d}$. The maximum/minimum values of $\zeta$ and $T_{\rm d}$
which provide 1$\sigma$ errors in the fit are used as estimates of the
error bar on both of these parameters. Figure~\ref{F_EW4} displays
equivalent SED models for the seven objects that have $W1-W4 <
1.0$ and $E_{W4} > 3.0$ (see $\S$\ref{S_Other}). We cannot provide errors for the two-temperature discs as the fitting process was not performed simultaneously for both discs, but as the aim of this paper is to identify disc candidates, we only require an estimate of the total flux from the disc and are not overly concerned with the detailed parameters for probably over-simplistic disc models.

Table~\ref{T_SED_Params} lists the reduced ${\chi}^{2}$ values for the photospheric fit and the best disc fit, the corresponding $\eta$ and
$\zeta$ values and the corresponding $T_{\rm eff}$ and $T_{\rm d}$
values. The amount of flux from the disc compared to the star ($f_{\rm
  d}/f_{*}$, hereafter referred to as the ``disc flux fraction'') is
calculated by integrating the best-fit flux profiles for the disc and
star between 0.1\,$\mu$m and 1.0\,mm. In the situation where
only $WISE$ data were available and the SED did not feature any
wavelength points redward of the disc flux peak, the fit was much less
constrained and the models more subject to degeneracies. Temperature
upper limits (and hence $f_{\rm d}/f_{*}$ lower limits) for these
objects are set at the point where the fit would be excluded with
2$\sigma$ confidence. There is a strong correlation in the parameter uncertainties. However because lower temperatures goes with higher $\zeta$, the effect on the estimated $f_{\rm d}/f_{*}$ is not serious.

{\scriptsize
\begin{table*}
\begin{tabular}{lrrrrrrrrr}
\toprule
\toprule
Name/Label              & MG             & $\chi_{\rm *}^{2}$ & $\eta$   & $T_{\rm eff}$ & $N_{\rm IR}$  & $\chi_{\rm d}^{2}$ & $\zeta$  & $T_{\rm d}$ & $f_{\rm d}/f_{*}$         \\
2MASS-                  &                &                    &          & (K)           &               &                    &          & (K)         &                           \\
\toprule
\multicolumn{10}{c}{$W1-W4 > 1.0$} \\
\toprule
J08413030$-$7853064, A  & $\eta$~Cha     &  0.21              & $-$19.80 &  3100         &  8$^{\rm a}$  &    1.36            &       $-17.7 \pm 0.5$ & 420$^{+160}_{-100}$ & 0.043$^{+0.037}_{-0.014}$ \\
J08422710$-$7857479, B  & $\eta$~Cha     &  2.14              & $-$19.58 &  3600         & 14            &   13.13            & $-14.6^{+0.3}_{-0.4}$ & 120$^{+30}_{-10}$   & 0.118$^{+0.093}_{-0.042}$ \\  
J08431857$-$7905181, C  & $\eta$~Cha     & 16.66              & $-$19.14 &  2900         & 17            &   24.65            &       $-15.4 \pm 0.2$ & 350$^{+40}_{-30}$   & 1.170$^{+0.448}_{-0.323}$ \\
J08440914$-$7833457, D  & $\eta$~Cha     &  0.37              & $-$19.74 &  2400         & 11            &    2.58            & $-$17.5, $-$13.8      & 380, 40             & 0.175 \\
J08441637$-$7859080, E  & $\eta$~Cha     &  2.61              & $-$18.98 &  2700         & 12            &    1.95            & $-15.4^{+0.1}_{-0.4}$ & 180$^{+130}_{-40}$   & 0.047$^{+0.028}_{-0.016}$ \\
J11183572$-$7935548, F  & $\epsilon$~Cha &  1.49              & $-$19.36 &  3300         &  5*           &    1.51            & $-$15.1               & $< 250$             & $> 0.083$ \\
J11321831$-$3019518, G  & TWA            &  1.60              & $-$19.02 &  3300         &  7            &    0.87            & $-16.3^{+0.3}_{-0.2}$ & 260$^{+70}_{-50}$   & 0.020$^{+0.016}_{-0.006}$ \\
J11432669$-$7804454, H  & $\epsilon$~Cha & 11.02              & $-$19.50 &  2700         &  4*           &    2.78            & $-$17.7               & $< 710$             & $> 0.075$ \\
J11493184$-$7851011, I  & $\epsilon$~Cha &  3.58              & $-$19.11 &  3800         & 10            &    6.14            & $-$14.4, $-$17.6      & 200, 810            & 0.455 \\
J11550485$-$7919108, J  & $\epsilon$~Cha & 25.40              & $-$19.75 &  3700         &  4*           &    6.60            & $-$16.2               & $< 270$             & $> 0.013$ \\
J12005517$-$7820296, K  & $\epsilon$~Cha &  0.60              & $-$19.44 &  2300         &  4*           &    2.26            & $-$16.9               & $< 490$             & $> 0.094$ \\
J12073346$-$3932539, L  & TWA            &  0.82              & $-$19.69 &  2100         & 10            &    4.39            & $-$18.1, $-$13.1      & 470, 50             & 1.351 \\
\toprule
\multicolumn{10}{c}{$W1-W4 < 1.0$ and $E_{W4} > 3.0$} \\
\toprule
J08173943$-$8243298, M  & BPMG           &  0.79              & $-$18.19 &  3500         &  5            &    1.73            & $-14.7^{+1.4}_{-1.6}$ & 120$^{+210}_{-40}$             & 0.005$^{+0.015}_{-0.002}$ \\
J08422372$-$7904030, N  & $\eta$~Cha     &  1.53              & $-$19.14 &  3800         & 10*$^{\rm b}$ &    2.71            & $-$15.2               & $< 60$              & $> 4 \times 10^{-4}$ \\
J10172689$-$5354265, O  & BPMG           &  0.53              & $-$18.32 &  2700         &  8*           &   16.50            & $-$17.0               & $< 380$             & $> 6 \times 10^{-3}$ \\
J10423011$-$3340162, P  & TWA            &  0.54              & $-$17.85 &  3100         &  9            &    4.64            & $-14.3^{+0.5}_{-0.8}$ & 80$^{+40}_{-30}$    & 16$^{+5}_{-6} \times 10^{-4}$ \\
J17173128$-$6657055, Q  & BPMG           &  2.83              & $-$18.69 &  3600         &  4*           &    0.18            & $-$16.5               & $< 380$             & $> 1.1 \times 10^{-3}$ \\
J17292067$-$5014529, R  & BPMG           &  0.20              & $-$18.86 &  3700         &  4*           &    0.41            & $-$17.7               & $< 590$             & $> 1.2 \times 10^{-4}$ \\
J20450949$-$3120266, S  & BPMG           &  4.58              & $-$17.38 &  3400         & 16            &    6.67            & $-13.4^{+0.5}_{-0.4}$ & 50$^{+30}_{-10}$    & 4$^{+6}_{-3} \times 10^{-4}$ \\
\bottomrule
\end{tabular}
\caption{Best-fit parameters to SED models for the sample of objects with evidence of IR excess. The number of flux points at wavelengths longer than $K$ are provided in column 3.  *No data available to constrain the disc at wavelengths longer than the peak wavelength of the black-body fit --- $2\sigma$ upper/lower limits are provided for $T_{\rm d}$ and $f_{\rm d}$ for these (see $\S$\ref{S_Sensitivity}).  Error bars for the $\eta$ values are $\sim$ 0.01. {\bf Two temperature models are fitted for targets D, I and L. $^{\rm a}$ $3\sigma$ upper-limits from $MIPS$ 24, 70 and 160\,$\mu$m and Herschel 70, 100 and 160\,$\mu$m; these are used only as a constraint in the SED modelling such that model fluxes at these wavelengths must lie below the upper limits. $^{\rm b}$ $3\sigma$ upper-limits from Herschel 70, 100 and 160\,$\mu$m.}}
\label{T_SED_Params}
\end{table*}}

\section{Sensitivity limits for detecting debris discs}\label{S_Sensitivity}

A working definition for debris discs is that they have $f_{\rm
  d}/f_{*}\leq 0.01$ \citep{2000a_Lagrange}. Optically thick primordial discs will generally have much higher values than this. Given that the objects in this work are relatively faint M-dwarfs, there will be a limiting $W1-W4$ for which $f_{\rm d}/f_{*} \leq 0.01$.

To explore the relationship between $f_{\rm d}/f_{*}$ and $W1-W4$, a
simulation was made to artifically increase the $W4$ flux from a
disc-less M-dwarf star until $f_{\rm d}/f_{*} = 0.01$. The disc-less
stars used in the simulation are 2MASSJ05531299$-$4505119 (M0.5V, ABDMG),
BD+36\,2322 (M4V, ABDMG) and DEN\,1048$-$3956 (M8.5, field brown
dwarf -- see \citealt{2012a_Avenhaus}). Photospheres were modelled
using the same procedure described in $\S$\ref{S_SED}. The disc
temperature was fixed at either 100\,K or 300\,K and the threshold $W4$ magnitude and disc area are outputs from the model.

\begin{figure}
 \begin{center}
\hspace*{+0.4cm}\includegraphics[width=0.47\textwidth]{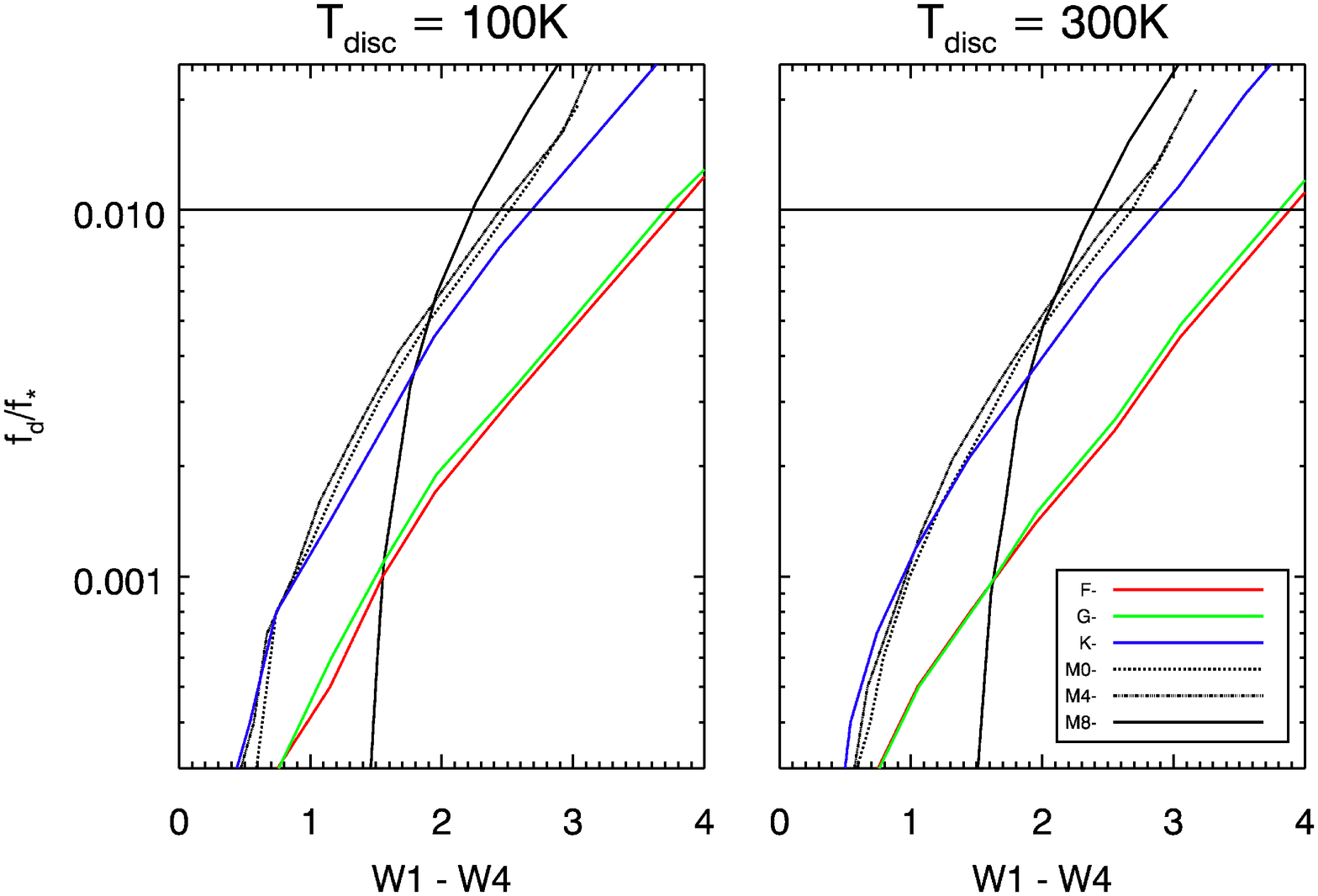}
 \flushleft
 \caption{Sensitivity limits for detecting debris discs in stars of
   various spectral types, with assumed disc temperatures of 100\,K (left) and 300\,K (right).}
  \label{F_Sensitivity}
 \end{center}
\end{figure}

For comparison, the same simulation was carried out for an F5, a G3 and a
K7 star (HIP~19183, HIP~113579 and HIP~31878, respectively, all members
of the ABDMG). The simulations show (see Figure~\ref{F_Sensitivity}), for the
assumed disc temperatures chosen, that $W1-W4 > 1.0$ corresponds to
$f_{\rm d}/f_{*} > 10^{-3}$ for K- to mid-M-dwarfs (and $> 3\times
10^{-4}$ for F- and G-dwarfs). Sensitivity to (debris) discs with less IR flux than this, using $W1-W4$, is compromised because smaller $W4$ excesses become indistinguishable from the scatter in $W1-W4$ values seen in disc-less field stars.

The $W1-W4$ colour at which $f_{\rm d}/f_{*} \geq 0.01$ is $\approx
2.5$ for early to mid-M spectral-types and $\approx 2.2$ for late
M-dwarfs.  Given that M8/9 represents the faintest object in the MG
sample, this suggests that debris discs will have $W1-W4$ values
somewhere between these upper limits and the $W1-W4$ values of
disc-less field stars. Primordial discs should have $W1-W4>2.2$. 
For F and G stars the sensitivity improves, and debris
disc objects should have $W1-W4 < 3.7$. Observations of K stars are
only slightly more sensitive than M-dwarfs and debris discs should have
$W1-W4 < 2.8$. There were only small differences of $W1-W4 \sim 0.1$
mag in the results depending on whether a disc temperature of 100\,K or
300\,K was adopted. These simple models do not take into account disc
gaps or disc orientation, which will be an additional source of scatter.

\begin{figure}
 \begin{center}
\hspace*{+1.1cm}\includegraphics[scale=0.24]{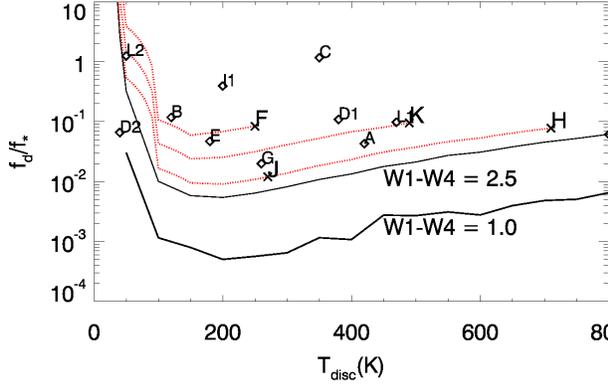}
 \flushleft
\caption{Diagram showing the expected relationship between $f_{\rm d}/f_{*}$ and disc temperature at a given $W1-W4$ for an M0 star. For comparison with Figure~\ref{F_Sensitivity} we have plotted the $f_{\rm d}/f_{*}$ vs $T_{\rm d}$ at constant $W1-W4 = 1.0$ and 2.5. Diamond symbols indicate our 6 objects which have a measured $T_{\rm d}$ and $f_{\rm d}/f_{*}$. For the other 6 objects only with an upper limit to the disc temperature we plot lines of constant $W1-W4$ which terminate at their 2-sigma $T_{\rm d}$ upper limits, as described in $\S$\ref{S_SED}. We note that none of these objects could have $f_{\rm d}/f_{*} < 0.01$. For targets D, I and L with two-temperature fits we split the flux from each disc into their two components and label them as `1' and `2'.}
  \label{F_M0_Map}
 \end{center}
\end{figure}

Figure~\ref{F_M0_Map} is an alternative way of looking at this
sensitivity to debris discs, showing the relationship between disc
temperature and $f_{\rm d}/f_{*}$ at a fixed M0 spectral-type for
$W1-W4 = 1.0$ and $2.5$ respectively. We find that debris discs with
$f_{\rm d}/f_{*}<0.01$ must have $T_{\rm d}>70$\,K if they are to have $W1-W4 > 1.0$. The debris disk threshold of $f_{\rm d}/f_{*}<0.01$ corresponds to $W1-W4 < 2.5$ for $100<T_{\rm d}<300$\,K and $W1-W4 \lesssim 2.0$ for $300<T_{\rm d}<500$\,K.

\section{Nature of the IR excess}\label{S_Nature}
\subsection{Disc criteria}\label{S_Disc_Criteria}

This section focuses on objects that have $W1-W4 > 1.0$. In $\S$\ref{S_Other} objects with $W1-W4 < 1.0$ and $E_{W4} > 3.0$ are discussed. A variety of indicators have been used in the literature to try to separate primordial, transitional and debris discs and we examine these with our sample. The final disc designation for each target is listed in Table~\ref{T_Disc_Type}. 

Figure~\ref{F_Colours} shows four separate colour-colour diagrams for the entire sample used in this analysis. Large $K-W1$ and $K-W2$ colours are indicative of large amounts of near-IR flux, consistent with what is expected from primordial discs. In Table~\ref{T_Disc_Type} we assign a disc type of either `P' (primordial) or `T' (transitional) based on an object's position (see Figure~\ref{F_Colours}) in the $K-W2$ versus $K-W1$ (top-left), $K-W3$ versus $K-W2$ (top-right) and $W3-W4$ versus $W1-W2$ (bottom-left) diagrams compared with the distribution of primordial and transitional discs observed in Taurus. The bottom-right panel in Figure~\ref{F_Colours} ($W1-W4$ versus $K-W1$) is the most effective way of separating debris discs from primordial discs, given only $2MASS$ and $WISE$ data, as it can potentially probe both hot discs ($K-W1$) and cooler discs ($W1-W4$). Figure~\ref{F_Colours} shows that both primordial and transitional discs in Taurus have colours of $3 < W1-W4 < 6$, which is consistent with the simulations in $\S$\ref{S_Sensitivity}. Overall it appears that transitional discs have slightly larger $W1-W4$ values ($\sim 4.5$ to 6.0) although not enough to strongly distinguish them from primordial discs. Choosing a disc type based on colour-colour diagrams is ambiguous, although the evidence suggests the majority of our targets with $W1-W4 > 1.0$ are primordial/transitional.

We expect debris discs to be class III sources. If they are class II sources then it is likely that any NIR excess is from a primordial disc. Because discs contribute large amounts of IR flux compared to the star, the flux gradient observed in SEDs is shallower than for objects without discs. The slope of the SED at IR wavelengths is characterised by $\alpha$, where $\alpha = {\rm d}\log\lambda F_{\lambda}/{\rm d}\log\lambda$. In identifying the disc frequencies amongst brown-dwarfs in the Upper Scorpius OB association \cite{2012b_Riaz} suggest that objects with $\alpha_{K-W3} \geq -0.2, -1.8 < \alpha_{K-W3} < -0.2$ and $\alpha_{K-W3} \leq -2.0$ correspond to class I/II/III sources, respectively. This method was calibrated for objects as early as M3.5 and should be applicable to most M-dwarfs in this work. There is an overlap of class II/III sources between $-2.0 < \alpha < -1.8$. Should an object have $-2.0 < \alpha_{K-W3} < -1.8$ it is assumed to be class III.

We choose the disc flux fractions as the most important indicator of disc type
and this indicator has been recommended in several previous works
(\citealt{2000a_Lagrange, 2008a_Wyatt, 2014a_Matthews}). We have values
or upper limits which make a decision between debris disc and
transitional/primordial disc rather clear. Debris discs generally have
$f_{\rm d}/f_{*} < 0.01$, whereas \cite{2012a_Cieza} estimate
primordial discs to have $f_{\rm d}/f_{*} \gtrsim 0.1$. In this work objects
with $f_{\rm d}/f_{*} > 0.1$ are classed as primordial, $0.01 < f_{\rm
  d}/f_{*} < 0.1$ are either primordial or transitional and $f_{\rm
  d}/f_{*} < 10^{-2}$ are debris discs. In the case of target B, the most probable disc-type based on $f_{\rm d}/f_{*}$ is primordial, however, this is less than 1 error bar from the primordial and primordial/transitional boundary and we flag this object accordingly in Table~\ref{T_Disc_Type}. If we were to only use the hotter component in the two-temperature models our disc classification would remain the same. The addition of an extra, cooler disc increases our $f_{\rm d}/f_{*}$, but in no cases is this by enough to change the type of disc we infer.

Figure~\ref{F_M0_Map} illustrates this test for both stars with
estimated disc temperatures and for those with upper limits to disc
temperatures. For the latter, their $W1-W4$ colour defines a locus in the $f_{\rm d}/f_{*}$ versus
$T_{\rm d}$ diagram which terminates at their 2$\sigma$ disc temperature
upper limits. Assuming that all 12 of our targets with $W1-W4 > 1.0$
are similar to M0 stars, we find that none of them have $f_{\rm
  d}/f_{*} < 0.01$, and this remains true even when a more precise
spectral-type is chosen (see Table~\ref{T_SED_Params}). For targets that only have upper limits for $f_{\rm d}/f_{*}$ we report the disc type as the type corresponding to the upper limit, however, we place in parenthesis other possible disc types. Five excesses appear to be due
to primordial discs, whilst the remainder are either primordial or transitional discs.

In Table~\ref{T_Disc_Type} we provide the results of our deliberations
of disc type using each method described in this section. The
$f_{\rm d}/f_{*}$ classification assumes primacy, followed by consideration
of the other indicators where this is ambiguous. The final totals are 8
primordial discs, 4 transitional discs and no debris discs.

{\tiny
\begin{table*}
\centering
\begin{tabular}{lrrrrrl}
\toprule
\toprule
Label/Common identifier & $W1-W4$ & Excess & $\alpha$ & $f_{\rm d}/f_{*}$ & Final & Additional fluxes ($\times 10^{-15}$ erg/s/cm$^{2}$/\AA) \\
\toprule
\multicolumn{7}{c}{$W1-W4 > 1.0$} \\
\toprule
A, ES Cha            & 3.39    & PPP(P)   & II       & T/P     & P   & 0.41 (3.6) 0.21 (4.5) 0.10 (5.8) 0.05 (8) \\
                     &         &          &          &         &     & $< 0.004$ (M24) $< 0.001$ (M70) $< 0.0009$ (H70) \\
                     &         &          &          &         &     & $< 0.0004$ (H100) $< 0.002$ (M160) \\
                     &         &          &          &         &     & $< 0.0003$ (H160)$^{\rm a}$ \\
B, EK Cha            & 4.53    & TTT(P/T) & III      & P(P/T?) & P?  & 0.96 (3.6) 0.43 (4.5) 0.19 (5.8) 0.08 (8) \\
                     &         &          &          &         &     & 0.03 (M24) 0.006 (M70) 0.007 (H70) \\
                     &         &          &          &         &     & 0.006 (H100) 0.002 (M160) 0.002 (H160)$^{\rm a}$ \\
C, ET Cha            & 5.15    & PPP(P)   & II       & P       & P   & 2.93 (3.6) 1.87 (4.5) 1.15 (5.8) 0.76 (8) \\
                     &         &          &          &         &     & 0.74 (9) 0.30 (18) 0.12 (M24) 0.16 (25) \\
                     &         &          &          &         &     & 0.01 (M70) 0.01 (H70) 0.004 (H100) \\
                     &         &          &          &         &     & 0.0009 (M160) 0.0008 (H160)$^{\rm a}$ \\
D, RECX 16           & 3.94    & PPP(P)   & II       & P       & P   & 0.26 (3.6) 0.14 (4.5) 0.07 (5.8) 0.04 (8) \\
                     &         &          &          &         &     & 0.004 (M24) 0.002 (H70) 0.0007 (H160)$^{\rm a}$ \\
E, EN Cha            & 3.61    & TPP(P/T) & II       & T/P     & T   & 1.67 (3.6) 0.82 (4.5) 0.40 (5.8) 0.20 (8) \\
                     &         &          &          &         &     & 0.02 (M24) 0.003 (M70) 0.002 (H70) \\
                     &         &          &          &         &     & 0.0006 (H100)$^{\rm a}$ \\
F, $\epsilon$~Cha 13 & 4.95    & TTT(P/T) & III      & T/P(P?) & T?  & 0.12 (18) \\
G, TWA 30            & 3.64    & TPT(P/T) & III      & T/P     & T   & 0.0009 (H70) 0.0003 (H100) 0.00002 (H160)$^{\rm b}$ \\
H, $\epsilon$~Cha 17 & 3.12    & PPP(P)   & III      & T/P(P?) & P?  & \\
I, $\epsilon$~Cha 1  & 6.38    & PPP(P/T) & II       & P       & P   & 1.50 (9) 1.13 (12) 1.09 (18) 0.84 (25) \\
                     &         &          &          &         &     & 0.12 (60) 0.04 (90) \\
J, T Cha B           & 3.03    & TTP(P/T) & III      & T/P(P?) & T?  & \\
K, $\epsilon$~Cha 10 & 4.01    & PPP(P)   & II       & T/P(P?) & P?  & \\
L, 2M 1207           & 3.53    & PPP(P)   & II       & P       & P   & 0.04 (H70) 0.02 (H100) 0.005 (H160) \\
                     &         &          &          &         &     & 0.001 (250) 0.0003 (350) 0.00008 (500)$^{\rm c}$ \\
\toprule
\multicolumn{7}{c}{$W1-W4 < 1.0$ and $E_{W4} > 3.0$} \\
\toprule
M, 2MASS J08173843$-$8243298 & 0.43 & T/D & III      & D(P/T?) & D? & 0.76 (9) \\
N, EI Cha            & 0.75    &      T/D & III      & D(T?P?) & D?  & 2.75 (3.6) 1.14 (4.5) 0.47 (5.8) 0.14 (8.0) \\
                     &         &          &          &         &     & 0.03 (12) 0.003 (M24) 0.003 (25) \\
                     &         &          &          &         &     & $< 0.0006$ (H70) \\
                     &         &          &          &         &     & $< 0.0003$ (H100) $< 0.0002$ (H160)$^{\rm a}$ \\  
O, TWA 22AB          & 0.26    &      T/D & III      & D(T?P?) & D?  & 0.36 (9) 0.053 (12) 0.006 (M24) 0.006 (25) \\
P, TWA 7             & 0.79    &      T/D & III      & D       & D   & 0.040 (12) 0.046 (M24) 0.005 (M70) \\
                     &         &          &          &         &     & 0.00003 (450) 0.000004 (850)$^{\rm d}$ \\
Q, HD 155555C        & 0.59    &      T/D & III      & D(T?P?) & D?  & \\
R, GSC 08350$-$01924 & 0.23    &      T/D & III      & D(T?P?) & D?  & \\
S, AU Mic            & 0.36    &      T/D & III      & D       & D   & 1.39 (9) 1.59 (12) 0.22 (18) 0.076 (M24) \\
                     &         &          &          &         &     & 0.14 (25) 0.01 (M70) 0.01 (H70) \\
                     &         &          &          &         &     & 0.002 (M160) \\
                     &         &          &          &         &     & 0.007 (H160)$^{\rm e}$ \\
\bottomrule
\end{tabular}
\caption{Disc types derived from various methods described in this work. The labels in column 1 refer to the 2MASS sources in Table~\ref{T_SED_Params}. The excess column is an approximation of the disc-type based on the $K-W2$ vs $K-W1$, $K-W3$ vs $K-W2$, $W3-W4$ vs $W1-W2$ and $W1-W4$ vs $K-W1$ (in parentheses) colour-colour diagrams (respectively) of Figure~\ref{F_Colours}. For objects with $W1-W4 < 1.0$, only the $W1-W4$ vs $K-W1$ disc designation is prescribed. The entries in column 4 categorise an object as a class II or class III source based on the $\alpha$ slopes calculated in $\S$\ref{S_Identifying_IR_Excess}. {\bf The disc types derived from disc flux fractions are provided in column 5 where designations in parentheses are alternative possible disc types as described in $\S$\ref{S_Disc_Criteria}}. The final derived disc type is given in column 6. P = primordial disc, T = transitional disc, D = debris disc. The final column lists any additional fluxes where values in parentheses are the wavelengths (in $\mu$m) for each data point: 3.6, 4.5, 5.8 and 8 are from $IRAC$; 12, 25, 60 and 100 are from $IRAS$; M24, M70 and M160 are from $MIPS${\bf ; H70, H100 and H160 are from $Herschel$ $PACS$; 250, 350 and 500 are from $Herschel$ $SPIRE$; 450 and 850 are from $SCUBA2$ and 9, 18 and 90 are from $AKARI$. Additional fluxes prefixed with `$<$' are 3$\sigma$ upper limits. Fluxes from $Herschel$ or $SCUBA2$ provided in: a) \protect\cite{2015a_Riviere-Marichalar}; b) \protect\cite{2015a_Liu}; c) \protect\cite{2012b_Riaz}; d) \protect\cite{2007a_Matthews}; e) \protect\cite{2015a_Matthews}}.}
\label{T_Disc_Type}
\end{table*}}

\subsection{Other disc candidates identified with $W4$ excess}\label{S_Other}

In this section we analyse the 7 objects with $W1-W4 < 1.0$ and $E_{W4}
> 3.0$ that are listed towards the bottom of Table~\ref{T_Disc_Type}. There is one member of $\eta$~Cha (target N=EI Cha), one TWA member (target P=TWA~7) and five BPMG members (targets M=2MASSJ08173843$-$8243298, hereafter 2MJ0817; O=TWA~22AB; Q=HD~155555C; R=2MASSJ17292067$-$5014529, hereafter 2MJ1729 and S=AU~Mic).

SEDs were modelled for these seven objects using the same procedure
described in $\S$\ref{S_SED} and are displayed in
Figure~\ref{F_EW4}, with results listed in Table~\ref{T_SED_Params}.
 Only three of these
objects had estimated values of $T_{\rm d}$ and $f_{\rm d}/f_{*}$, all of which were
less than 0.01 (2M~J0817, TWA~7 and AU~Mic, see \citealt{2016a_Cotten}). For 2M~J0817, the flux fraction is within one error bar of 0.01, the boundary between debris and primordial/transitional discs. AU Mic and TWA~7 are already well-known to have a debris disc that
has been imaged in scattered light (\citealt{2004a_Kalas}; \citealt{2016a_Choquet}).

TWA~22AB, 2M~J1729 and HD~155555C did not have sufficient data at far-IR
wavelengths to constrain a disc fit and no information on their disc
type or $f_{\rm d}/f_{*}$ could be found in previous works, however \cite{2009a_Schutz} claim a null detection of IR excess for HD~155555C based on mid-IR spectroscopic measurements between $8-13\,\mu$m. Disc
temperature upper limits of 80, 120 and 180\,K, and corresponding $f_{\rm d}/f_{*}$ lower limits of $6.0\times10^{-3}$, $1.1\times10^{-3}$ and $1.2\times10^{-4}$, respectively, were estimated, so these were
classed here as candidate debris discs which require further observation.
EI~Cha also had unconstrained disc parameters in this
work; EI~Cha has previously been flagged as a possible debris disk source with $f_{\rm d}/f_{*}=3\times 10^{-5}$ (based on far-IR Herschel photometry) by \cite{2008a_Gautier}, however in our work we only calculate a lower-limit to $f_{\rm d}/f_{*}$ because the Herschel data for EI~Cha are upper limits.

In summary, this set of seven candidates, with small but significant
$W4$ excesses, consists of three ``recovered''
objects whose status as debris disc objects was already known or
suspected, together with a new probable debris disc source
(2M~J0817) and three new candidate debris disc sources
(targets TWA~22AB, HD~155555C and 2M~J1729) that are all likely
members of the BPMG.

\subsection{Comparison with the literature}

Many of the 19 stars we have identified as harbouring discs have been
previously identified or studied. Appendix~B gives a star-by-star
comparison with literature values and classifications. There are
some differences between our estimates of $f_{\rm d}/f_{*}$ and those in the
literature, but not by enough to change our disc classifications. For the 12 objects with $W1-W4 > 1.0$, in all but two cases there is broad agreement and agreement on the
nature of the disc: for $\epsilon$ Cha~17 and 2M~J1207 (targets H and L)
 we find primordial discs whereas in the literature these were reported
as transitional and debris discs, respectively
(\citealt{2011a_Manoj} and \citealt{2012a_Riaz}).
Four of the seven objects with $W1-W4 < 1.0$, but significant $W4$ excess have been reported in previous work as either confirmed debris discs or debris disc candidates, which is in broad agreement with our classification scheme.

\section{Discussion}\label{S_Discussion}
\subsection{Did we find \textit{any} new debris discs?}

We have identified IR (22\,$\mu$m) excesses from 19/100 M-dwarfs that are
likely MG members. These excesses are bimodal in nature -- 12 sources
have $W1-W4 > 3$, which simple disc models suggest corresponds to
$f_{\rm d}/f_{*} > 0.02$ for these spectral-types; 7 sources have $W1-W4 < 1$ and
although we are unable to adequately constrain the disc temperatures,
it is likely that these have $f_{\rm d}/f_{*} \lesssim 3\times
10^{-3}$. The former objects are all almost certainly primordial, gas-rich discs, although
some may have transitional inner holes as recorded by our
classifications in Table~\ref{T_Disc_Summary}. The latter do not exhibit 
excesses at shorter wavelengths or signs of accretion activity and should be
regarded as debris disc {\it candidates}, although some confidence in
our methods is drawn from the fact that these include 2 objects with
previously known debris discs that have been imaged in scattered light
and another object identified as a debris disc in other studies. The
four remaining candidates need additional sensitive measurements at
mid- and far-IR wavelengths in order to confirm their nature.

Table~\ref{T_Disc_Summary} provides the number of primordial,
transitional or debris discs for each MG containing at least one
disc. All of the IR excess objects were found in MGs with age $< 30$\,Myr.
The total fraction of M-dwarf MG members with debris
discs identified in this work is 7/100. Given that identification of
debris discs for objects with $W1-W4 < 1$ is incomplete then this
fraction is of course a lower limit and likely to be age-dependent. No
debris disc candidates are found with
$1.0 < W1-W4 < 3.0$, implying that $<3$ per cent of young M-dwarfs have
very dusty debris discs with $f_{\rm d}/f_{*} \gtrsim 10^{-3}$.

{\tiny
\begin{table}
\begin{center}
\setlength{\tabcolsep}{0.0cm}
\begin{tabular}{p{2.0cm}p{2.0cm}p{1.0cm}p{1.0cm}p{1.0cm}}
\toprule
\toprule
MG Name		& $N_{\rm sample}$	& P	& T	& D	\\
\toprule
$\epsilon$ Cha	& 8			& 3		& 2		& 0		\\
TWA		& 11			& 1		& 1		& 1		\\
$\eta$ Cha	& 8			& 4		& 1		& 1		\\
BPMG		& 28			& 0		& 0		& 5		\\
\midrule
Total		& 55			& 8		& 4		& 7		\\
\bottomrule
\end{tabular}
\end{center}
\caption{The number of primordial (P), transitional (T) or debris (D) discs for each MG that contains at least one type of disc. $N_{\rm sample}$ is the number of M-dwarfs in the MG which satisfy all criteria described in $\S$\ref{S_Target_Selection}.}
\label{T_Disc_Summary}
\end{table}}

\subsection{Evidence for debris disc evolution in M-dwarfs?}

All of the discs we have found, primordial, transitional or debris,
are in MGs at the age of BPMG ($24 \pm 4\,$Myr) or younger. This is
consistent with the claimed lifetimes of primordial discs in M-dwarfs
(2--3\,Myr, \citealt{2011a_Williams}), but may also be evidence for a
decline of debris disc frequency with age. Dividing the sample into MGs
younger and older than 30\,Myr, there are 7/55 debris disc candidates
in the younger group and 0/45 in the older group. This corresponds to
debris disc frequencies of $12 \pm 5$ per cent in the younger group and 
$<7$ per cent (at 95 per cent confidence) in the older group. Assuming that the younger and older
samples were observed with similar sensitivities, then a $2\times 2$
chi-squared contingency test suggests a difference in these frequencies
at 98.5 per cent confidence.

\cite{2007a_Gautier} found no evidence for debris discs in a sample of
62 field M dwarfs that were estimated to be older than 1 Gyr.
\cite{2009a_Lestrade} identified only one object (GJ842.2) from a
sample of 19 M-dwarfs younger than 200\,Myr as having a cold debris
disc based on sub-mm observations, deriving a debris disc frequency of
$5.3^{+10.5}_{-5.0}$ per cent. Following a deep $Spitzer$ $MIPS$ survey,
\cite{2008a_Forbrich} found evidence for debris discs with a 24\,$\mu$m
excess for 11 M-dwarfs in NGC~2547 (age $35 \pm 3$\,Myr,
\citealt{2005a_Jeffries}), providing a lower limit to the debris disc
frequency of 4.9 per cent. Our measurement of a debris disc frequency
(for MG members with ages of 30--150\,Myr) of $<7$ per cent is consistent with all of this previous work.

\subsection{Do M-dwarfs dissipate their discs faster than solar-type stars?}\label{S_Age_Comparison}

There have been suggestions in the literature that M-dwarfs may
disperse their debris disks more rapidly than higher mass stars, due to
drag caused by stronger stellar winds (\citealt{2002a_Wood, 2005a_Plavchan}), by earlier photoevaporation due to the stonger EUV and FUV radiation fields (\citealt{2014a_Galvan-Madrid}) or by the stripping of planetestimals from less massive M-dwarfs due to stellar encounters \citep{2009a_Lestrade}, but these arguments also depend on a straightforward linear scaling between disc mass and stellar mass. The empirical situation is far from clear due to the small number of identified M-dwarf debris discs and the difficulties of comparing samples with threshold sensitivities that vary with brightness and spectral-type.

A straightforward comparison between FGK-stars in MGs and the M-dwarfs studied
in this work is difficult because of small numbers and whilst $WISE$
can detect debris discs amongst FGK-dwarfs with $f_{\rm d}/f_{*} >
3\times 10^{-4}$ (corresponding to $W1-W4 \geq 1.0$ -- see
Figure~\ref{F_Sensitivity}), data of the same quality is only capable of
detecting M-dwarf debris discs if they have $f_{\rm d}/f_{*} > 10^{-3}$. The
only comparable study was performed by \cite{2016a_Moor} who surveyed
29 FGK members of MGs with ages of 20--50\,Myr, finding 6 debris discs
with $10^{-4} < f_{\rm d}/f_{*} < 10^{-3}$. The frequency of discs in the older
FGK stars with age 30--50\,Myr was 5/25. A $2 \times 2$ chi-squared
contingency test suggests that this is incompatible (at $> 99$ per cent
confidence) with the lack of debris discs we have found in M-dwarfs at
similar ages, however it could be that the survey for FGK debris discs
is more sensitive and thus more complete, with several M-dwarf debris
discs with $10^{-4}<f_{\rm d}/f_{*}<10^{-3}$ awaiting discovery.

It has also been suggested that although dust may be dispersed more
rapidly around M-dwarfs there could be a delay in the production of
dust from planetestimal collisions with respect to higher mass stars,
with the time of maximum dust visibility $t_{\rm max} \propto M_{*}^{-1}$
\citep{2008a_Kenyon}. Empirically, the strength of debris disc emission
and the debris disc frequency both appear to peak at $\sim$20--30\,Myr
in FGK stars but occur earlier at 10--20\,Myr for A/B-type stars
(\citealt{2008a_Currie, 2009a_Hernandez, 2012a_Smith}). There is
no evidence from our survey that this trend continues towards lower masses
-- our data would suggest that any peak in dust visibility for M-dwarfs
occurs at $\leq 30$\,Myr.

\section{Summary}\label{S_Summary}
We have performed a survey using $2MASS$ and $WISE$ for discs around a
sample of 100 candidate M-dwarf members of young MGs (with ages of
5--150\,Myr). Potential discs were identified on the basis of
significant $W1-W4$ excesses and their nature investigated using a
variety of diagnostics including simple models for the SEDs and
longer wavelength data where available.

Tests of the sensitivity of $WISE$ data to dust discs suggest that
debris discs should have been readily identified around M-dwarfs if $W1-W4 > 1.0$,
corresponding to a disc flux fraction $f_{\rm d}/f_{*}>10^{-3}$.
For stars with $W1-W4 < 1.0$, significant excess can be detectable, but
becomes harder to separate from the photospheres of (presumed disc-less)
M-dwarf field stars. The main results of our search are:

1. We found 12 objects with clear IR excesses and $W1-W4 > 3.0$, all were
younger than 30\,Myr, all have $f_{\rm d}/f_{*} > 0.02$ and all have
SEDs consistent with primordial discs or transitional discs with inner holes.

2. No M-dwarfs were found with $1.0 < W1-W4 < 3.0$, indicating that $< 3$
per cent (at 95 per cent confidence) of young M-dwarfs have very dusty debris
discs with $f_{\rm d}/f_{*} \gtrsim 10^{-3}$.

3. Seven objects were identified with $W1-W4 < 1.0$ but with a significant
$W4$ excess over that expected from the photosphere, corresponding to
$f_{\rm d}/f_{*} \lesssim 3 \times 10^{-3}$, although the disc flux fractions and disc
temperatures are poorly constrained. These IR excesses are assumed to
be due to debris discs and all these stars are younger than 30\,Myr. 
Three of the debris discs have
been previously identified in the literature and in two cases the discs have
been imaged in scattered light. The four other sources are new debris disc
candidates and all are candidate members of the BPMG.

4. The frequency of debris discs in young M-dwarfs with age $<30\,$Myr
is at least $13 \pm 5$ per cent, but this falls to $<7$ per cent in
older MG M-dwarfs at equivalent observational sensitivity. This
provides evidence for some evolution of debris disc properties and
argues against the peak of debris disc activity occurring later in
lower-mass stars.

\section{Acknowledgements}

ASB acknowledges the financial support from the United Kingdom Science and Technology Funding Council (STFC) and the Consejo Nacional de Ciencia y Tecnolog\'ia (M\'exico). RDJ acknowledges financial support from the STFC. This publication makes use of data products from the Wide-field Infrared Survey Explorer, which is a joint project of the University of California, Los Angeles, and the Jet Propulsion Laboratory/California Institute of Technology, funded by the National Aeronautics and Space Administration. This research has made use of the NASA/IPAC Infrared Science Archive, which is operated by the Jet Propulsion Laboratory, California Institute of Technology, under contract with the National Aeronautics and Space Administration. This work is based [in part] on observations made with the Spitzer Space Telescope, which is operated by the Jet Propulsion Laboratory, California Institute of Technology under a contract with NASA. This publication makes use of data products from the Two Micron All Sky Survey, which is a joint project of the University of Massachusetts and the Infrared Processing and Analysis Center/California Institute of Technology, funded by the National Aeronautics and Space Administration and the National Science Foundation. This research is based on observations with AKARI, a JAXA project with the participation of ESA. This research has made use of the SIMBAD database, operated at CDS, Strasbourg, France. We thank an anonymous referee for their comments and suggestions which improved this manuscript.

\bibliography{BJ_11_11_16_response}

\begin{thebibliography}{90}
\expandafter\ifx\csname natexlab\endcsname\relax\def\natexlab#1{#1}\fi

\bibitem[{{Allard}(2014)}]{2014a_Allard}
{Allard} F., 2014, in IAU Symposium, Vol. 299, IAU Symposium, {Booth} M.,
  {Matthews} B.~C., {Graham} J.~R., eds., pp. 271--272

\bibitem[{{Augereau} \& {Beust}(2006)}]{2006a_Augereau}
{Augereau} J.-C., {Beust} H., 2006, \aap, 455, 987

\bibitem[{{Avenhaus}, {Schmid} \& {Meyer}(2012){Avenhaus}, {Schmid}, \&
  {Meyer}}]{2012a_Avenhaus}
{Avenhaus} H., {Schmid} H.~M., {Meyer} M.~R., 2012, \aap, 548, A105

\bibitem[{{Backman} \& {Paresce}(1993)}]{1993a_Backman}
{Backman} D.~E., {Paresce} F., 1993, in Protostars and Planets III, {Levy}
  E.~H., {Lunine} J.~I., eds., pp. 1253--1304

\bibitem[{{Barman} {et~al}\mbox{.}(2011){Barman}, {Macintosh}, {Konopacky}, \&
  {Marois}}]{2011a_Barman}
{Barman} T.~S., {Macintosh} B., {Konopacky} Q.~M., {Marois} C., 2011, \apjl,
  735, L39

\bibitem[{{Barrado y Navascu{\'e}s} {et~al}\mbox{.}(1999){Barrado y
  Navascu{\'e}s}, {Stauffer}, {Song}, \&
  {Caillault}}]{1999a_Barrado_y_Navascues}
{Barrado y Navascu{\'e}s} D., {Stauffer} J.~R., {Song} I., {Caillault} J.-P.,
  1999, \apjl, 520, L123

\bibitem[{{Bell}, {Mamajek} \& {Naylor}(2015){Bell}, {Mamajek}, \&
  {Naylor}}]{2015a_Bell}
{Bell} C.~P.~M., {Mamajek} E.~E., {Naylor} T., 2015, \mnras, 454, 593

\bibitem[{{Binks} \& {Jeffries}(2014)}]{2014a_Binks}
{Binks} A.~S., {Jeffries} R.~D., 2014, \mnras, 438, L11

\bibitem[{{Binks} \& {Jeffries}(2016)}]{2016a_Binks}
{Binks} A.~S., {Jeffries} R.~D., 2016, \mnras, 455, 3345

\bibitem[{{Binks}, {Jeffries} \& {Maxted}(2015){Binks}, {Jeffries}, \&
  {Maxted}}]{2015a_Binks}
{Binks} A.~S., {Jeffries} R.~D., {Maxted} P.~F.~L., 2015, \mnras, 452, 173

\bibitem[{{Chabrier} \& {Baraffe}(2000)}]{2000a_Chabrier}
{Chabrier} G., {Baraffe} I., 2000, \araa, 38, 337

\bibitem[{{Choquet} {et~al}\mbox{.}(2016){Choquet}, {Perrin}, {Chen},
  {Soummer}, {Pueyo}, {Hagan}, {Gofas-Salas}, {Rajan}, {Golimowski}, {Hines},
  {Schneider}, {Mazoyer}, {Augereau}, {Debes}, {Stark}, {Wolff}, {N'Diaye}, \&
  {Hsiao}}]{2016a_Choquet}
{Choquet} {\'E}. {et~al.}, 2016, \apjl, 817, L2

\bibitem[{{Cieza} {et~al}\mbox{.}(2013){Cieza}, {Olofsson}, {Harvey}, {Evans},
  {Najita}, {Henning}, {Mer{\'{\i}}n}, {Liebhart}, {G{\"u}del}, {Augereau}, \&
  {Pinte}}]{2013a_Cieza}
{Cieza} L.~A. {et~al.}, 2013, \apj, 762, 100

\bibitem[{{Cieza} {et~al}\mbox{.}(2012){Cieza}, {Schreiber}, {Romero},
  {Williams}, {Rebassa-Mansergas}, \& {Mer{\'{\i}}n}}]{2012a_Cieza}
{Cieza} L.~A., {Schreiber} M.~R., {Romero} G.~A., {Williams} J.~P.,
  {Rebassa-Mansergas} A., {Mer{\'{\i}}n} B., 2012, \apj, 750, 157

\bibitem[{{Cohen}, {Wheaton} \& {Megeath}(2003){Cohen}, {Wheaton}, \&
  {Megeath}}]{2003a_Cohen}
{Cohen} M., {Wheaton} W.~A., {Megeath} S.~T., 2003, \aj, 126, 1090

\bibitem[{{Cotten} \& {Song}(2016)}]{2016a_Cotten}
{Cotten} T.~H., {Song} I., 2016, \apjs, 225, 15

\bibitem[{{Currie} {et~al}\mbox{.}(2008){Currie}, {Kenyon}, {Balog}, {Rieke},
  {Bragg}, \& {Bromley}}]{2008a_Currie}
{Currie} T., {Kenyon} S.~J., {Balog} Z., {Rieke} G., {Bragg} A., {Bromley} B.,
  2008, \apj, 672, 558

\bibitem[{{Currie} \& {Sicilia-Aguilar}(2011)}]{2011a_Currie}
{Currie} T., {Sicilia-Aguilar} A., 2011, \apj, 732, 24

\bibitem[{{Cutri} {et~al}\mbox{.}(2003){Cutri}, {Skrutskie}, {van Dyk},
  {Beichman}, {Carpenter}, {Chester}, {Cambresy}, {Evans}, {Fowler}, {Gizis},
  {Howard}, {Huchra}, {Jarrett}, {Kopan}, {Kirkpatrick}, {Light}, {Marsh},
  {McCallon}, {Schneider}, {Stiening}, {Sykes}, {Weinberg}, {Wheaton},
  {Wheelock}, \& {Zacarias}}]{2003a_Cutri}
{Cutri} R.~M. {et~al.}, 2003, {2MASS All Sky Catalog of point sources.}

\bibitem[{{De Silva} {et~al}\mbox{.}(2013){De Silva}, {D'Orazi}, {Melo},
  {Torres}, {Gieles}, {Quast}, \& {Sterzik}}]{2013a_De_Silva}
{De Silva} G.~M., {D'Orazi} V., {Melo} C., {Torres} C.~A.~O., {Gieles} M.,
  {Quast} G.~R., {Sterzik} M., 2013, \mnras, 431, 1005

\bibitem[{{Ducourant} {et~al}\mbox{.}(2014){Ducourant}, {Teixeira}, {Galli},
  {Le Campion}, {Krone-Martins}, {Zuckerman}, {Chauvin}, \&
  {Song}}]{2014a_Ducourant}
{Ducourant} C., {Teixeira} R., {Galli} P.~A.~B., {Le Campion} J.~F.,
  {Krone-Martins} A., {Zuckerman} B., {Chauvin} G., {Song} I., 2014, \aap, 563,
  A121

\bibitem[{{Elliott} {et~al}\mbox{.}(2016){Elliott}, {Bayo}, {Melo}, {Torres},
  {Sterzik}, {Quast}, {Montes}, \& {Brahm}}]{2016a_Elliott}
{Elliott} P., {Bayo} A., {Melo} C.~H.~F., {Torres} C.~A.~O., {Sterzik} M.~F.,
  {Quast} G.~R., {Montes} D., {Brahm} R., 2016, \aap, 590, A13

\bibitem[{{Esplin}, {Luhman} \& {Mamajek}(2014){Esplin}, {Luhman}, \&
  {Mamajek}}]{2014a_Esplin}
{Esplin} T.~L., {Luhman} K.~L., {Mamajek} E.~E., 2014, \apj, 784, 126

\bibitem[{{Fang} {et~al}\mbox{.}(2013){Fang}, {van Boekel}, {Bouwman},
  {Henning}, {Lawson}, \& {Sicilia-Aguilar}}]{2013a_Fang}
{Fang} M., {van Boekel} R., {Bouwman} J., {Henning} T., {Lawson} W.~A.,
  {Sicilia-Aguilar} A., 2013, \aap, 549, A15

\bibitem[{{Forbrich} {et~al}\mbox{.}(2008){Forbrich}, {Lada}, {Muench}, \&
  {Teixeira}}]{2008a_Forbrich}
{Forbrich} J., {Lada} C.~J., {Muench} A.~A., {Teixeira} P.~S., 2008, \apj, 687,
  1107

\bibitem[{{Gagn{\'e}} {et~al}\mbox{.}(2014){Gagn{\'e}}, {Lafreni{\`e}re},
  {Doyon}, {Malo}, \& {Artigau}}]{2014a_Gagne}
{Gagn{\'e}} J., {Lafreni{\`e}re} D., {Doyon} R., {Malo} L., {Artigau} {\'E}.,
  2014, ArXiv e-prints

\bibitem[{{Gagn{\'e}} {et~al}\mbox{.}(2015){Gagn{\'e}}, {Lafreni{\`e}re},
  {Doyon}, {Malo}, \& {Artigau}}]{2015a_Gagne}
{Gagn{\'e}} J., {Lafreni{\`e}re} D., {Doyon} R., {Malo} L., {Artigau} {\'E}.,
  2015, \apj, 798, 73

\bibitem[{{Galv{\'a}n-Madrid} {et~al}\mbox{.}(2014){Galv{\'a}n-Madrid}, {Liu},
  {Manara}, {Forbrich}, {Pascucci}, {Carrasco-Gonz{\'a}lez}, {Goddi},
  {Hasegawa}, {Takami}, \& {Testi}}]{2014a_Galvan-Madrid}
{Galv{\'a}n-Madrid} R. {et~al.}, 2014, \aap, 570, L9

\bibitem[{{Gautier} {et~al}\mbox{.}(2008){Gautier}, {Rebull}, {Stapelfeldt}, \&
  {Mainzer}}]{2008a_Gautier}
{Gautier}, III T.~N., {Rebull} L.~M., {Stapelfeldt} K.~R., {Mainzer} A., 2008,
  \apj, 683, 813

\bibitem[{{Gautier} {et~al}\mbox{.}(2007){Gautier}, {Rieke}, {Stansberry},
  {Bryden}, {Stapelfeldt}, {Werner}, {Beichman}, {Chen}, {Su}, {Trilling},
  {Patten}, \& {Roellig}}]{2007a_Gautier}
{Gautier}, III T.~N. {et~al.}, 2007, \apj, 667, 527

\bibitem[{{Gliese} \& {Jahreiss}(1991)}]{1991a_Gliese}
{Gliese} W., {Jahreiss} H., 1991, NASA STI/Recon Technical Report A, 92, 33932

\bibitem[{{Haisch}, {Lada} \& {Lada}(2001){Haisch}, {Lada}, \&
  {Lada}}]{2001a_Haisch}
{Haisch}, Jr. K.~E., {Lada} E.~A., {Lada} C.~J., 2001, \apjl, 553, L153

\bibitem[{{Helou} \& {Walker}(1988)}]{1988a_Helou}
{Helou} G., {Walker} D.~W., eds., 1988, {Infrared astronomical satellite (IRAS)
  catalogs and atlases. Volume 7: The small scale structure catalog}, {Helou}
  G., {Walker} D.~W., eds., Vol.~7

\bibitem[{{Hern{\'a}ndez} {et~al}\mbox{.}(2009){Hern{\'a}ndez}, {Calvet},
  {Hartmann}, {Muzerolle}, {Gutermuth}, \& {Stauffer}}]{2009a_Hernandez}
{Hern{\'a}ndez} J., {Calvet} N., {Hartmann} L., {Muzerolle} J., {Gutermuth} R.,
  {Stauffer} J., 2009, \apj, 707, 705

\bibitem[{{Hern{\'a}ndez} {et~al}\mbox{.}(2008){Hern{\'a}ndez}, {Hartmann},
  {Calvet}, {Jeffries}, {Gutermuth}, {Muzerolle}, \&
  {Stauffer}}]{2008a_Hernandez}
{Hern{\'a}ndez} J., {Hartmann} L., {Calvet} N., {Jeffries} R.~D., {Gutermuth}
  R., {Muzerolle} J., {Stauffer} J., 2008, \apj, 686, 1195

\bibitem[{{Hu{\'e}lamo} {et~al}\mbox{.}(2011){Hu{\'e}lamo}, {Lacour},
  {Tuthill}, {Ireland}, {Kraus}, \& {Chauvin}}]{2011a_Huelamo}
{Hu{\'e}lamo} N., {Lacour} S., {Tuthill} P., {Ireland} M., {Kraus} A.,
  {Chauvin} G., 2011, \aap, 528, L7

\bibitem[{{Jarrett} {et~al}\mbox{.}(2011){Jarrett}, {Cohen}, {Masci}, {Wright},
  {Stern}, {Benford}, {Blain}, {Carey}, {Cutri}, {Eisenhardt}, {Lonsdale},
  {Mainzer}, {Marsh}, {Padgett}, {Petty}, {Ressler}, {Skrutskie}, {Stanford},
  {Surace}, {Tsai}, {Wheelock}, \& {Yan}}]{2011a_Jarrett}
{Jarrett} T.~H. {et~al.}, 2011, \apj, 735, 112

\bibitem[{{Jeffries} \& {Oliveira}(2005)}]{2005a_Jeffries}
{Jeffries} R.~D., {Oliveira} J.~M., 2005, \mnras, 358, 13

\bibitem[{{Kalas}(2004)}]{2004a_Kalas}
{Kalas} P., 2004, in Bulletin of the American Astronomical Society, Vol.~36,
  American Astronomical Society Meeting Abstracts 204, p. 690

\bibitem[{{Kastner} {et~al}\mbox{.}(2012){Kastner}, {Thompson}, {Montez},
  {Murphy}, {Bessell}, \& {Sacco}}]{2012a_Kastner}
{Kastner} J.~H., {Thompson} E.~A., {Montez} R., {Murphy} S.~J., {Bessell}
  M.~S., {Sacco} G.~G., 2012, \apjl, 747, L23

\bibitem[{{Kennedy} \& {Wyatt}(2012)}]{2012a_Kennedy}
{Kennedy} G.~M., {Wyatt} M.~C., 2012, \mnras, 426, 91

\bibitem[{{Kenyon} \& {Bromley}(2005)}]{2005a_Kenyon}
{Kenyon} S.~J., {Bromley} B.~C., 2005, \aj, 130, 269

\bibitem[{{Kenyon} \& {Bromley}(2008)}]{2008a_Kenyon}
{Kenyon} S.~J., {Bromley} B.~C., 2008, \apjs, 179, 451

\bibitem[{{Kiss} {et~al}\mbox{.}(2011){Kiss}, {Mo{\'o}r}, {Szalai},
  {Kov{\'a}cs}, {Bayliss}, {Gilmore}, {Bienaym{\'e}}, {Binney},
  {Bland-Hawthorn}, {Campbell}, {Freeman}, {Fulbright}, {Gibson}, {Grebel},
  {Helmi}, {Munari}, {Navarro}, {Parker}, {Reid}, {Seabroke}, {Siebert},
  {Siviero}, {Steinmetz}, {Watson}, {Williams}, {Wyse}, \&
  {Zwitter}}]{2011a_Kiss}
{Kiss} L.~L. {et~al.}, 2011, \mnras, 411, 117

\bibitem[{{Kraus} {et~al}\mbox{.}(2014){Kraus}, {Shkolnik}, {Allers}, \&
  {Liu}}]{2014a_Kraus}
{Kraus} A.~L., {Shkolnik} E.~L., {Allers} K.~N., {Liu} M.~C., 2014, \aj, 147,
  146

\bibitem[{{Lagrange}, {Backman} \& {Artymowicz}(2000){Lagrange}, {Backman}, \&
  {Artymowicz}}]{2000a_Lagrange}
{Lagrange} A.-M., {Backman} D.~E., {Artymowicz} P., 2000, Protostars and
  Planets IV, 639

\bibitem[{{Lawson} \& {Feigelson}(2001)}]{2001a_Lawson}
{Lawson} W., {Feigelson} E.~D., 2001, in Astronomical Society of the Pacific
  Conference Series, Vol. 243, From Darkness to Light: Origin and Evolution of
  Young Stellar Clusters, {Montmerle} T., {Andr{\'e}} P., eds., p. 591

\bibitem[{{Lestrade} {et~al}\mbox{.}(2009){Lestrade}, {Wyatt}, {Bertoldi},
  {Menten}, \& {Labaigt}}]{2009a_Lestrade}
{Lestrade} J.-F., {Wyatt} M.~C., {Bertoldi} F., {Menten} K.~M., {Labaigt} G.,
  2009, \aap, 506, 1455

\bibitem[{{Low} {et~al}\mbox{.}(2005){Low}, {Smith}, {Werner}, {Chen},
  {Krause}, {Jura}, \& {Hines}}]{2005a_Low}
{Low} F.~J., {Smith} P.~S., {Werner} M., {Chen} C., {Krause} V., {Jura} M.,
  {Hines} D.~C., 2005, \apj, 631, 1170

\bibitem[{{Luhman}(2004)}]{2004a_Luhman}
{Luhman} K.~L., 2004, \apj, 616, 1033

\bibitem[{{Luhman}(2007)}]{2007a_Luhman}
{Luhman} K.~L., 2007, \apjs, 173, 104

\bibitem[{{MacGregor}(2014)}]{2014a_MacGregor}
{MacGregor} M., 2014, in IAU Symposium, Vol. 299, IAU Symposium, {Booth} M.,
  {Matthews} B.~C., {Graham} J.~R., eds., pp. 313--317

\bibitem[{{Malo} {et~al}\mbox{.}(2014{\natexlab{a}}){Malo}, {Artigau}, {Doyon},
  {Lafreni{\`e}re}, {Albert}, \& {Gagn{\'e}}}]{2014b_Malo}
{Malo} L., {Artigau} {\'E}., {Doyon} R., {Lafreni{\`e}re} D., {Albert} L.,
  {Gagn{\'e}} J., 2014{\natexlab{a}}, \apj, 788, 81

\bibitem[{{Malo} {et~al}\mbox{.}(2014{\natexlab{b}}){Malo}, {Artigau}, {Doyon},
  {Lafreni{\`e}re}, {Albert}, \& {Gagn{\'e}}}]{2014a_Malo}
{Malo} L., {Artigau} {\'E}., {Doyon} R., {Lafreni{\`e}re} D., {Albert} L.,
  {Gagn{\'e}} J., 2014{\natexlab{b}}, \apj, 788, 81

\bibitem[{{Malo} {et~al}\mbox{.}(2013){Malo}, {Doyon}, {Lafreni{\`e}re},
  {Artigau}, {Gagn{\'e}}, {Baron}, \& {Riedel}}]{2013a_Malo}
{Malo} L., {Doyon} R., {Lafreni{\`e}re} D., {Artigau} {\'E}., {Gagn{\'e}} J.,
  {Baron} F., {Riedel} A., 2013, \apj, 762, 88

\bibitem[{{Mamajek}(2005)}]{2005a_Mamajek}
{Mamajek} E.~E., 2005, \apj, 634, 1385

\bibitem[{{Manoj} {et~al}\mbox{.}(2011){Manoj}, {Kim}, {Furlan}, {McClure},
  {Luhman}, {Watson}, {Espaillat}, {Calvet}, {Najita}, {D'Alessio}, {Adame},
  {Sargent}, {Forrest}, {Bohac}, {Green}, \& {Arnold}}]{2011a_Manoj}
{Manoj} P. {et~al.}, 2011, \apjs, 193, 11

\bibitem[{{Matthews} {et~al}\mbox{.}(2014){Matthews}, {Krivov}, {Wyatt},
  {Bryden}, \& {Eiroa}}]{2014a_Matthews}
{Matthews} B.~C., {Krivov} A.~V., {Wyatt} M.~C., {Bryden} G., {Eiroa} C., 2014,
  Protostars and Planets VI, 521

\bibitem[{{Mo{\'o}r} {et~al}\mbox{.}(2016){Mo{\'o}r}, {K{\'o}sp{\'a}l},
  {{\'A}brah{\'a}m}, {Balog}, {Csengeri}, {Henning}, {Juh{\'a}sz}, \&
  {Kiss}}]{2016a_Moor}
{Mo{\'o}r} A., {K{\'o}sp{\'a}l} {\'A}., {{\'A}brah{\'a}m} P., {Balog} Z.,
  {Csengeri} T., {Henning} T., {Juh{\'a}sz} A., {Kiss} C., 2016, \apj, 826, 123

\bibitem[{{Murphy}, {Lawson} \& {Bessell}(2013){Murphy}, {Lawson}, \&
  {Bessell}}]{2013a_Murphy}
{Murphy} S.~J., {Lawson} W.~A., {Bessell} M.~S., 2013, \mnras, 435, 1325

\bibitem[{{Nakajima} \& {Morino}(2012)}]{2012a_Nakajima}
{Nakajima} T., {Morino} J.-I., 2012, \aj, 143, 2

\bibitem[{{Patel}, {Metchev} \& {Heinze}(2014){Patel}, {Metchev}, \&
  {Heinze}}]{2014a_Patel}
{Patel} R.~I., {Metchev} S.~A., {Heinze} A., 2014, \apjs, 212, 10

\bibitem[{{Pecaut} \& {Mamajek}(2013)}]{2013a_Pecaut}
{Pecaut} M.~J., {Mamajek} E.~E., 2013, \apjs, 208, 9

\bibitem[{{Plavchan}, {Jura} \& {Lipscy}(2005){Plavchan}, {Jura}, \&
  {Lipscy}}]{2005a_Plavchan}
{Plavchan} P., {Jura} M., {Lipscy} S.~J., 2005, \apj, 631, 1161

\bibitem[{{Rhee}, {Song} \& {Zuckerman}(2007){Rhee}, {Song}, \&
  {Zuckerman}}]{2007a_Rhee}
{Rhee} J.~H., {Song} I., {Zuckerman} B., 2007, \apj, 671, 616

\bibitem[{{Riaz} \& {Gizis}(2012)}]{2012a_Riaz}
{Riaz} B., {Gizis} J.~E., 2012, \aap, 548, A54

\bibitem[{{Riaz} {et~al}\mbox{.}(2012){Riaz}, {Lodieu}, {Goodwin},
  {Stamatellos}, \& {Thompson}}]{2012b_Riaz}
{Riaz} B., {Lodieu} N., {Goodwin} S., {Stamatellos} D., {Thompson} M., 2012,
  \mnras, 420, 2497

\bibitem[{{Riviere-Marichalar} {et~al}\mbox{.}(2014){Riviere-Marichalar},
  {Barrado}, {Montesinos}, {Duch{\^e}ne}, {Bouy}, {Pinte}, {Menard},
  {Donaldson}, {Eiroa}, {Krivov}, {Kamp}, {Mendigut{\'{\i}}a}, {Dent}, \&
  {Lillo-Box}}]{2014a_Riviere-Marichalar}
{Riviere-Marichalar} P. {et~al.}, 2014, \aap, 565, A68

\bibitem[{{Schlieder}, {L{\'e}pine} \& {Simon}(2010){Schlieder}, {L{\'e}pine},
  \& {Simon}}]{2010a_Schlieder}
{Schlieder} J.~E., {L{\'e}pine} S., {Simon} M., 2010, \aj, 140, 119

\bibitem[{{Schlieder}, {L{\'e}pine} \& {Simon}(2012){Schlieder}, {L{\'e}pine},
  \& {Simon}}]{2012a_Schlieder}
{Schlieder} J.~E., {L{\'e}pine} S., {Simon} M., 2012, \aj, 144, 109

\bibitem[{{Schneider}, {Melis} \& {Song}(2012){Schneider}, {Melis}, \&
  {Song}}]{2012a_Schneider}
{Schneider} A., {Melis} C., {Song} I., 2012, \apj, 754, 39

\bibitem[{{Shkolnik}, {Liu} \& {Reid}(2009){Shkolnik}, {Liu}, \&
  {Reid}}]{2009a_Shkolnik}
{Shkolnik} E., {Liu} M.~C., {Reid} I.~N., 2009, \apj, 699, 649

\bibitem[{{Shkolnik} {et~al}\mbox{.}(2012){Shkolnik}, {Anglada-Escud{\'e}},
  {Liu}, {Bowler}, {Weinberger}, {Boss}, {Reid}, \& {Tamura}}]{2012a_Shkolnik}
{Shkolnik} E.~L., {Anglada-Escud{\'e}} G., {Liu} M.~C., {Bowler} B.~P.,
  {Weinberger} A.~J., {Boss} A.~P., {Reid} I.~N., {Tamura} M., 2012, \apj, 758,
  56

\bibitem[{{Sicilia-Aguilar} {et~al}\mbox{.}(2009){Sicilia-Aguilar}, {Bouwman},
  {Juh{\'a}sz}, {Henning}, {Roccatagliata}, {Lawson}, {Acke}, {Feigelson},
  {Tielens}, {Decin}, \& {Meeus}}]{2009a_Sicilia-Aguilar}
{Sicilia-Aguilar} A. {et~al.}, 2009, \apj, 701, 1188

\bibitem[{{Siegler} {et~al}\mbox{.}(2007){Siegler}, {Muzerolle}, {Young},
  {Rieke}, {Mamajek}, {Trilling}, {Gorlova}, \& {Su}}]{2007a_Siegler}
{Siegler} N., {Muzerolle} J., {Young} E.~T., {Rieke} G.~H., {Mamajek} E.~E.,
  {Trilling} D.~E., {Gorlova} N., {Su} K.~Y.~L., 2007, \apj, 654, 580

\bibitem[{{Simon} {et~al}\mbox{.}(2012){Simon}, {Schlieder}, {Constantin}, \&
  {Silverstein}}]{2012a_Simon}
{Simon} M., {Schlieder} J.~E., {Constantin} A.-M., {Silverstein} M., 2012,
  \apj, 751, 114

\bibitem[{{Smith} \& {Jeffries}(2012)}]{2012a_Smith}
{Smith} R., {Jeffries} R.~D., 2012, \mnras, 420, 2884

\bibitem[{{Tanab{\'e}} {et~al}\mbox{.}(2008){Tanab{\'e}}, {Sakon}, {Cohen},
  {Wada}, {Ita}, {Ohyama}, {Oyabu}, {Uemizu}, {Takagi}, {Ishihara}, {Kim},
  {Ueno}, {Matsuhara}, \& {Onaka}}]{2008a_Tanabe}
{Tanab{\'e}} T. {et~al.}, 2008, \pasj, 60, 375

\bibitem[{{Teixeira} {et~al}\mbox{.}(2009){Teixeira}, {Ducourant}, {Chauvin},
  {Krone-Martins}, {Bonnefoy}, \& {Song}}]{2009a_Teixeira}
{Teixeira} R., {Ducourant} C., {Chauvin} G., {Krone-Martins} A., {Bonnefoy} M.,
  {Song} I., 2009, \aap, 503, 281

\bibitem[{{Th{\'e}bault} \& {Wu}(2008)}]{2008a_Thebault}
{Th{\'e}bault} P., {Wu} Y., 2008, \aap, 481, 713

\bibitem[{{Theissen} \& {West}(2014)}]{2014a_Theissen}
{Theissen} C.~A., {West} A.~A., 2014, \apj, 794, 146

\bibitem[{{Trilling} {et~al}\mbox{.}(2008){Trilling}, {Bryden}, {Beichman},
  {Rieke}, {Su}, {Stansberry}, {Blaylock}, {Stapelfeldt}, {Beeman}, \&
  {Haller}}]{2008a_Trilling}
{Trilling} D.~E. {et~al.}, 2008, \apj, 674, 1086

\bibitem[{{Wahhaj} {et~al}\mbox{.}(2010){Wahhaj}, {Cieza}, {Koerner},
  {Stapelfeldt}, {Padgett}, {Case}, {Keller}, {Mer{\'{\i}}n}, {Evans},
  {Harvey}, {Sargent}, {van Dishoeck}, {Allen}, {Blake}, {Brooke}, {Chapman},
  {Mundy}, \& {Myers}}]{2010a_Wahhaj}
{Wahhaj} Z. {et~al.}, 2010, \apj, 724, 835

\bibitem[{{Williams} \& {Cieza}(2011)}]{2011a_Williams}
{Williams} J.~P., {Cieza} L.~A., 2011, \araa, 49, 67

\bibitem[{{Wood} {et~al}\mbox{.}(2002){Wood}, {Lada}, {Bjorkman}, {Kenyon},
  {Whitney}, \& {Wolff}}]{2002a_Wood}
{Wood} K., {Lada} C.~J., {Bjorkman} J.~E., {Kenyon} S.~J., {Whitney} B.,
  {Wolff} M.~J., 2002, \apj, 567, 1183

\bibitem[{{Wright} {et~al}\mbox{.}(2010){Wright}, {Eisenhardt}, {Mainzer},
  {Ressler}, {Cutri}, {Jarrett}, {Kirkpatrick}, {Padgett}, {McMillan},
  {Skrutskie}, {Stanford}, {Cohen}, {Walker}, {Mather}, {Leisawitz}, {Gautier},
  {McLean}, {Benford}, {Lonsdale}, {Blain}, {Mendez}, {Irace}, {Duval}, {Liu},
  {Royer}, {Heinrichsen}, {Howard}, {Shannon}, {Kendall}, {Walsh}, {Larsen},
  {Cardon}, {Schick}, {Schwalm}, {Abid}, {Fabinsky}, {Naes}, \&
  {Tsai}}]{2010a_Wright}
{Wright} E.~L. {et~al.}, 2010, \aj, 140, 1868

\bibitem[{{Wyatt}(2008)}]{2008a_Wyatt}
{Wyatt} M.~C., 2008, \araa, 46, 339

\bibitem[{{Zuckerman} \& {Song}(2004)}]{2004a_Zuckerman}
{Zuckerman} B., {Song} I., 2004, \araa, 42, 685

\bibitem[{{Zuckerman}, {Song} \& {Bessell}(2004){Zuckerman}, {Song}, \&
  {Bessell}}]{2004b_Zuckerman}
{Zuckerman} B., {Song} I., {Bessell} M.~S., 2004, \apjl, 613, L65

\bibitem[{{Zuckerman} {et~al}\mbox{.}(2001){Zuckerman}, {Song}, {Bessell}, \&
  {Webb}}]{2001a_Zuckerman}
{Zuckerman} B., {Song} I., {Bessell} M.~S., {Webb} R.~A., 2001, \apjl, 562, L87

\end{thebibliography}

\appendix

\section{Objects with $W1-W4 > 1.0$ and SNR $> 5.0$, but rejected from the sample}\label{A_1}

{\tiny
\begin{table*}
\begin{tabular}{lrrrrrrr}
\hline
\hline
Name                                           & $W1$                      & $W2$                      & $W3$                      & $W4$                                & $f_{\rm d}/f_{*}$ & Disc Type & Reason \\ 
(WISE- )                                       & (mag)                     & (mag)                     & (mag)                     & (mag, SNR)                          &                   &           & \\
\hline 
J010335.75$-$551556.6, $A_{X}$         &      9.03 $\pm$      0.02 &      8.79 $\pm$      0.02 &      8.56 $\pm$      0.02 &      7.86 $\pm$      0.15,      7.4 & $> 2 \times 10^{-4}$      & D? & 1     \\ 
J044356.87+372302.7, $B_{X}$           &      8.65 $\pm$      0.02 &      8.56 $\pm$      0.02 &      8.35 $\pm$      0.02 &      7.55 $\pm$      0.16,      6.9 & $> 1 \times 10^{-4}$      & D? & 1, 2  \\ 
J101209.04$-$312445.3, $C_{X}^{\rm a}$ &      7.81 $\pm$      0.02 &      7.54 $\pm$      0.02 &      6.14 $\pm$      0.01 &      4.80 $\pm$      0.03,     39.2 & $0.021^{+0.007}_{-0.003}$ & T  & 2     \\ 
J111027.80$-$373152.0, $D_{X}^{\rm b}$ &      6.60 $\pm$      0.04 &      6.34 $\pm$      0.02 &      3.88 $\pm$      0.02 &      1.73 $\pm$      0.01,     77.5 & $0.092^{+0.004}_{-0.010}$ & T  & 2     \\
J120144.32$-$781926.7, $E_{X}$         &     10.16 $\pm$      0.02 &      9.74 $\pm$      0.02 &      8.35 $\pm$      0.02 &      6.70 $\pm$      0.05,     21.4 & $> 4 \times 10^{-3}$      & P  & 1     \\ 
\hline
\end{tabular}
\caption{$^{\rm a}AKARI$ flux (all fluxes quoted as $\times 10^{-14}$ erg/s/cm${^2}$/\AA) at $9\,{\mu}$m = 0.46. $^{\rm b}AKARI$ flux at $9, 18$ and $90\,{\mu}$m = 2.60, 1.28 and 0.02 and $IRAS$ flux at 12, 25 and 60$\,{\mu}$m =2.06, 0.84 and 0.08. Reasons for rejection: 1. Object has a projected companion with $\Delta K_{\rm s} \leq 5$\,mag within 16'' in 2MASS. 2. The mean photometry of all frames in the L1b catalog for $W1$ and/or $W2$ is $> 2\sigma$ from the value quoted in $AllWISE$ (see $\S$\ref{S_Selection_Criteria}).}
\label{A_WISE_Phot}
\end{table*}}

In addition to the 19 objects identified with a significant IR excess, there are five objects that have $W1-W4 > 1.0$ and SNR $> 5.0$, however, they fail criteria (see $\S$\ref{S_Selection_Criteria}) to be included in the final sample. We have included these objects for completeness and interest but note that the SED fits may not be reliable because these objects failed some of the criteria in $\S$\ref{S_Selection_Criteria} and may have unreliable WISE photometry. These are listed in Table~\ref{A_WISE_Phot} with SED fitting parameters and the reasons for their rejection.

Two of these objects, $A_{x}$ and $B_{x}$, which are probable members of Tuc-Hor (\citealt{2013a_Delorme, 2013a_Malo}) and BPMG (\citealt{2013a_Malo}), respectively, are debris disc candidates (based on $W1-W4$ colour) but had no IR data beyond 25\,$\mu$m to constrain an SED fit and neither have been identified as having discs in previous work. Two objects, $C_{x}$ and $D_{x}$ are both probable members of TWA (\citealt{2014a_Malo}) and have $f_{\rm d}/f_{*}$ consistent with transitional discs. For object $D_{x}$ \cite{2013a_Riviere-Marichalar} measure $f_{\rm d}/f_{*} = 0.098$, within $1\,\sigma$ of our measurement and identify a disc `gap' characteristic of a transitional disc. Object $E_{x}$ is an $\epsilon$~Cha member (\citealt{2003a_Feigelson}) which has a mid-IR SED measured in \cite{2013a_Fang}, who were also unable to measure an absolute $f_{\rm d}/f_{*}$. Based on the $W1-W4$ colour we suggest $E_{x}$ is a protoplanetary disc candidate.

\section{Individual objects with significant mid-IR excess}\label{S_Indiv}

\noindent\textbf{Target A, ES~Cha} - The $f_{\rm d}/f_{*}$ value of 0.043 is $\sim 3\sigma$ from the debris disc threshold of 0.01 and is consistent to within $1\sigma$ of the $f_{\rm d}/f_{*}$ value of 0.04 in \cite{2008a_Gautier}.

\noindent\textbf{Target B, EK~Cha} - This target is reported as a primordial disc in $\eta$~Cha in \cite{2014a_Riviere-Marichalar} which is in agreement with our prescription. We measure $f_{\rm d}/f_{*} = 0.118$, within 2$\sigma$ of the $f_{\rm d}/f_{*}$ in \cite{2008a_Gautier}, who calculate a value of 0.06. \cite{2016a_Cotten} measure $T_{\rm d} = 140\,$K and $f_{\rm d}/f_{*} = 0.0876$, which would not change the outcome of the disc type based on the analysis in this work.

\noindent\textbf{Target C, ET~Cha} - The flux fraction calculated in this work is 1.170, $\sim$ triple the value calculated in \cite{2008a_Gautier}. \cite{2001a_Lawson} categorise this object as a CTTS in $\eta$~Cha with a primordial disc, in agreement with our judgement.

\noindent\textbf{Target D, RECX~16} - Our SED model yields a flux fraction of 0.175, which is larger than the value of 0.04 calculated in \cite{2008a_Gautier}. \cite{2011a_Currie} designate this object as a primordial disc, however \cite{2007a_Rhee} identify warm dusty material in the disc and suggest it could be either a debris or primordial disc.

\noindent\textbf{Target E, EN~Cha} - The flux fraction of $0.047^{+0.028}_{-0.016}$ calculated in this work is within $1\sigma$ of the value of 0.04 calculated in \cite{2008a_Gautier}. \cite{2009a_Sicilia-Aguilar} and \cite{2011a_Currie} identify that the star is a CTTS with a transitional disc, consistent with our classification.

\noindent\textbf{Target F, $\epsilon$ Cha 13} - \cite{2007a_Luhman} identify this object as a WTTS and \cite{2011a_Manoj} suggest the object has a transitional disc with an inner hole, consistent with our findings.

\noindent\textbf{Target G, TWA 30} - \cite{2012a_Schneider} classify it as a debris disc based on photometric IR excess, however, our $f_{\rm d}/f_{*}$ value of $0.020^{+0.016}_{-0.006}$ is inconsistent with this and we class this objects as transitional. \cite{2016a_Cotten} have measured $f_{\rm d}/f_{*} = 0.0215$ and $T_{\rm d} = 240\,$K, which are both consistent with our measurements to within 1$\sigma$.

\noindent\textbf{Target H, $\epsilon$ Cha 17} - Based on $Spitzer$ $IRS$ spectra, \cite{2011a_Manoj} measure a class II transitional disc, whereas we suggest a primordial disc.

\noindent\textbf{Target I, $\epsilon$ Cha 1} - Our $f_{\rm d}/f_{*}$ value of 0.455 is larger than the value of 0.280 calculated in \cite{2010a_Wahhaj}. \cite{2013a_Murphy} suggest that although \cite{2013a_Malo} characterise it as a BPMG member, its kinematics are better matched with $\epsilon$~Cha ($\epsilon$~Cha was not considered in the Malo et al. analysis). \cite{2012a_Simon} identify this target as a debris disc, however, given the large flux fraction we classify this as a primordial disc source.

\noindent\textbf{Target J, T Cha B} - Both the works of \cite{2011a_Huelamo} and \cite{2012a_Kastner} identify a transitional disc around this source, in agreement with this work.

\noindent\textbf{Target K, $\epsilon$ Cha 10} - The best-fit SED models in \cite{2013a_Fang} are for a thick primordial disc, which we also find.

\noindent\textbf{Target L, 2MASSJ12073346-3932539} - Literature searches for any additional evidence for a disc revealed that this target was the brown dwarf 2M1207, which is host to the giant exoplanet candidate 2M1207b (\citealt{2005a_Mamajek, 2011a_Barman}). \cite{2012a_Riaz} suggest that 2M1207 has a transition disc with an inner disc evacuation due to grain growth/dust settling, however our measurement of $f_{\rm d}/f_{*} > 1.0$ clearly indicates a primordial disc.

\noindent\textbf{Target M, 2MASSJ08173943$-$8243298} - This target is reported as a likely new member of the BPMG in \cite{2014a_Malo}. No previous evidence of disc presence has been found in the literature.

\noindent\textbf{Target N, EI~Cha} - The flux fraction measured by \cite{2008a_Gautier} and \cite{2013a_Cieza} of $3 \times 10^{-4}$ and $1.6 \times 10^{-4}$, respectively, using far-IR data to constrain the SED fit, are slightly below our measured lower limit. We use equation 3 in \cite{1993a_Backman} to calculate $T_{\rm d} \approx 30\,$K, which is consistent with our disc temperature upper limit. 

\noindent\textbf{Target O, TWA22~AB} - TWA22 was initially regarded as a member of the TWA, however subsequent kinematic analyses (e.g., \citealt{2009a_Teixeira, 2013a_Malo}) support a higher probability of membership to the BPMG. No previous work has reported a disc around TWA22.

\noindent\textbf{Target P, TWA~7} - The flux fractions of $2 \times 10^{-3}$ and $2.5 \times 10^{-3}$ measured by \cite{2005a_Low} and \cite{2016a_Cotten} are in agreement with our measurement to 1 and 2$\sigma$, respectively. \cite{2005a_Low} measure a disc temperature of 80\,K, whereas \cite{2016a_Cotten} model a two-temperature disc with 65 and 220\,K components. Both of these are consistent with our $T_{\rm d}$ measurement and the possibility of a debris disc.

\noindent\textbf{Target Q, HD~155555C} - \cite{2009a_Schutz} find no evidence for IR excess based on $8-13\,\mu$m spectroscopy, however, this is not necessarily inconsistent with our detection of a small excess at 22\,$\mu$m.

\noindent\textbf{Target R, 2MASSJ17292067$-$5014529} - Several works support membership to the BPMG (e.g., \citealt{2011a_Kiss, 2014a_Malo, 2016a_Elliott}), however, no previous work highlights this object as a disc candidate.

\noindent\textbf{Target S, AU Mic} - This target is well-known to have a visually-resolved debris disc (e.g., \citealt{2004a_Kalas, 2006a_Augereau, 2014a_MacGregor}). The most recent measurements $f_{\rm d}/f_{*} = 7 \times 10^{-4}$ and $T_{\rm d} = 65\,$K by \cite{2016a_Cotten} are both within $1\sigma$ of our findings.

\end{document}